\documentclass[paper]{JHEP3}
\usepackage{epsfig}
\usepackage{amsmath}
\usepackage{amssymb}
\usepackage{amsbsy}
\usepackage{enumerate}
\usepackage{bbm}
\usepackage{url}
\usepackage{wasysym}
\usepackage[nosort]{cite}

\usepackage[hang,sf]{subfigure}

\def\beq{\begin{equation}}
\def\beqn{\begin{eqnarray}}
\def\eeq{\end{equation}}
\def\eeqn{\end{eqnarray}}

\newcommand\HERWIG{{\tt HERWIG}}

\newcommand\PYTHIA{{\tt PYTHIA}}

\newcommand\SISCONE{{\tt SISCONE}}

\newcommand\FASTJET{{\textsc{Fastjet}}}
\newcommand\NLOPS{{\textsc{Nlops}}}

\newcommand\ptsupp{{\tt bornsuppfact}}

\def\({\left(} 
\def\){\right)}

\newcommand\sss{\mathchoice%
{\displaystyle}%
{\scriptstyle}%
{\scriptscriptstyle}%
{\scriptscriptstyle}%
}

\newdimen\hbigcirc
\newdimen\wbigcirc

\newdimen\figwidth

\figwidth=\textwidth

\newcommand\pt{p_{\sss\rm T}}
\newcommand\kt{k_{\sss\rm T}}
\newcommand\ktsupp{k_{\sss\rm T,supp}}
\newcommand\Et{E_{\sss\rm T}}

\newcommand\MCatNLO{{\tt MC@NLO}}

\newcommand     \MSB            {\ifmmode {\overline{\rm MS}} \else
                                 $\overline{\rm MS}$  \fi}

\newcommand\POWHEG{{\tt POWHEG}}
\newcommand\POWHEGBOX{{\tt POWHEG BOX}}

\newcount\minutes 
\newcount\scratch 
 
\def\timestamp{%
\scratch=\time 
\divide\scratch by 60 
\edef\hours{\the\scratch} 
\multiply\scratch by 60 
\minutes=\time 
\advance\minutes by -\scratch 
---$\,$\hours:\null 
\ifnum\minutes< 10 0\fi 
\the\minutes} 
 
\preprint{
DESY 10-233\\
SFB/CPP-10-128\\
IPPP/10/100\\
DCPT/10/200\\
MCnet/10/23}

\title{Jet pair production in \textsc{POWHEG}}
\author{Simone Alioli\\
  Deutsches Elektronen-Synchrotron DESY\\
  Platanenallee 6, D-15738 Zeuthen, Germany\\
  E-mail: \email{simone.alioli@desy.de}}
\author{Keith Hamilton\\
  INFN, Sezione di Milano-Bicocca,
  Piazza della Scienza 3, 20126 Milan, Italy\\
  E-mail: \email{Keith.Hamilton@mib.infn.it}}
\author{Paolo Nason\\
  INFN, Sezione di Milano-Bicocca,
  Piazza della Scienza 3, 20126 Milan, Italy\\
  E-mail: \email{Paolo.Nason@mib.infn.it}}
\author{Carlo Oleari\\
  Universit\`a di Milano-Bicocca and INFN, Sezione di Milano-Bicocca\\
  Piazza della Scienza 3, 20126 Milan, Italy\\
  E-mail: \email{Carlo.Oleari@mib.infn.it}}
\author{Emanuele Re\\
  Institute for Particle Physics Phenomenology, Department of Physics\\
  University of Durham, Durham, DH1 3LE, UK\\
  E-mail: \email{emanuele.re@durham.ac.uk}}
\abstract{
We present an implementation of the next-to-leading order dijet production
process in
hadronic collisions in the framework of \POWHEG{}, which is a method to
implement NLO calculations within a shower Monte Carlo context. In
constructing the simulation, we have made use of the \POWHEGBOX{} toolkit,
which makes light of many of the most technical steps.  The majority of this
article is concerned with the study of the predictions of the Monte Carlo
simulation. In so doing, we validate our program for use in experimental
analyses, elaborating on some of the more subtle features which arise from
the interplay of the NLO and resummed components of the calculation. We
conclude our presentation by comparing predictions from the simulation
against a number of Tevatron and LHC jet-production results.
}

\begin{document}

\section{Introduction}

Dijet production is by far the most frequently occurring of all hard
scattering processes in hadronic collisions, as such it is fundamental that
it be thoroughly studied and understood. In keeping with this fact the
physics programmes associated with these reactions at hadron colliders are
rich and diverse. From a purely experimental perspective, dijet events have an
important practical role to play as a tool in various aspects of jet
measurement and calibration e.g.~the determination of the jet energy
resolution. Also, from the point of view of QCD, jet pair production in
hadronic collisions is particularly interesting in that it is directly
dependent on the gluon parton distribution functions at the leading order.
More generally, and perhaps more importantly, in providing an abundant source
of high momentum transfer events, the dijet production process acts as both a
background to, and sensitive probe of, physics beyond the Standard Model.

Indeed, the first measurements and results of new physics searches in this
channel, with relatively small amounts of early LHC data, have been publicly
documented by the ATLAS and CMS collaborations in recent
weeks~\cite{Collaboration:2010wv,Khachatryan:2010jd}. Already these studies
have shown perturbative QCD to hold well in new kinematic regimes and
extended bounds on an impressive number of new physics models, from composite
quarks~\cite{Baur:1987ga} to TeV scale string
theories~\cite{Cullen:2000ef,Anchordoqui:2008di}. It is also very clear from
these early data, obtained at relatively low energies and luminosities, that
in the coming years jet pair production cross sections will be measured with
unprecedented precision in the TeV range.



As with the related Tevatron measurements, the Standard Model predictions
used in these analyses are derived from fixed order,
next-to-leading-order~(NLO) calculations which are corrected for showering,
hadronization and underlying event effects estimated using leading-order
parton-shower Monte Carlo simulations. Also, in studies such as these,
leading-order, leading-log, parton-shower event generators are frequently
used in assessing several systematic effects e.g.~jet triggering efficiencies
and jet energy scale corrections.

Given the significant and wide ranging applications of the dijet production
process, and with the LHC now beginning to take data in earnest, the need for
refined theoretical modeling is important.  Although the level of
maturity and attention to detail in current analyses is remarkable, there is
still room for improvement. In particular, the way in which the Standard
Model prediction is obtained could be made more easily and more coherently
through the use of a parton shower simulation consistently including the NLO
corrections to jet pair production. An event generator of this nature should
also be beneficial in understanding other experimental systematics for which
parton shower simulations are relied upon, improving the description of jet
profiles through the incorporation of exact, higher-order, real emission
matrix elements. Equally, when considered as a background process, all of
these higher order QCD corrections will offer markedly better, more robust,
predictions than those of the leading-order event generators.

In recent times the construction of such NLO accurate event generators has
become viable through the invention of the \MCatNLO{}~\cite{Frixione:2002ik}
and \POWHEG{}~\cite{Nason:2004rx,Frixione:2007vw} methods. The effectiveness
of these approaches has been demonstrated successfully and studied in some
detail through their application to a substantial array of hadron collider
processes~\cite{Frixione:2010wd,Bahr:2008pv,Hoeche:2010pf,svn_powhegbox}. 
In this paper we report on our construction and validation of a
next-to-leading order parton shower simulation of dijet production according
to the \POWHEG{} formalism. To this end we have utilized the public
\POWHEGBOX{} package~\cite{Alioli:2010xd}, which automates the most complex
technical steps of the implementation, essentially reducing it to the task of
realising the real, virtual, Born, spin- and colour-correlated Born matrix
elements as computer code.

The article is structured as follows. In section~\ref{sec:calc} we
elaborate on the next-to-leading order cross sections underlying the
simulation, as employed within the \POWHEGBOX{}, and related technical details. In
section~\ref{sec:theory_bit} the validity of the underlying NLO calculation
is demonstrated through comparisons with an independent computer
code~\cite{Frixione:1997ks} and the implementation of the \POWHEG{} algorithm
is checked in a series of non-trivial self-consistency tests. In
section~\ref{sec:phenomenology} we present results from our program
in comparison with a number of Tevatron and LHC measurements. Finally, in
section~\ref{sec:conclusions} we give our conclusions.

\section{Construction of the \POWHEG{} implementation}
As stated in the introduction, we have made use of the \POWHEGBOX{}
development framework in building our next-to-leading order parton shower
simulation, expediting the process considerably.  Essentially, provided with
a set of analytic formulae for the real, virtual, spin- and colour-correlated
Born cross sections, all that is required to produce the corresponding
\POWHEG{} simulation are simple computer programs returning their respective
values when given a list of particles and their associated momenta. The
underlying \POWHEGBOX{} machinery regulates the NLO corrections automatically,
using the FKS subtraction formalism~\cite{Frixione:1995ms,Frixione:1997np},
and builds the relevant Sudakov form factors internally, combining them to
form the \POWHEG{} hardest-emission cross section and, ultimately,
an executable to generate the associated events.
These single-emission events can then be further evolved to the hadron
level by general-purpose parton-shower event generators. In this section we
elaborate on the theoretical ingredients and some important technical aspects
of the implementation.


\label{sec:calc}
\subsection{Next-to-leading order cross sections}
The next-to-leading order real and virtual matrix elements for dijet
production were first computed over twenty years
ago~\cite{Ellis:1986er}. Later a more general approach to the computation of
NLO jet observables was considered in ref.~\cite{Kunszt:1992tn}, wherein one
can find, in addition, expressions for the colour-correlated Born cross
sections. As input for the \POWHEGBOX{} we have taken the one-loop matrix
elements and colour-correlated Born cross sections from the latter
publication; the evaluation of the corresponding soft real emission integrals
is delegated to the \POWHEGBOX{}. Note that, due to parity conservation,
helicity considerations and the fact that the leading-order process comprises
of just four massless partons, there are no non-trivial spin correlations
among the associated amplitudes.  The real cross sections have been built from
the concise analytic expressions taken from refs.~\cite{Ellis:1986er}
and~\cite{Ellis:1986bv}.

\subsection{Scale choices}
\label{sec:scales_choice}
In the \POWHEG{} algorithm, each event is built by first producing what is
referred to as an {underlying Born configuration}, here a QCD $2\rightarrow2$
scattering, before proceeding to generate the hardest branching in the event.
We have elected to use the $\pt$ of the underlying-Born configuration as the
renormalization and factorization scale in obtaining the fixed-order NLO
predictions, effectively resumming virtual corrections to the associated
$t$-channel gluon propagator.  This same scale choice is used in generating
the underlying Born kinematics, $\Phi_{B}$, of the \POWHEG{} events
(according to the $\bar{B}\left(\Phi_{B}\right)$ function of
ref.~\cite{Nason:2004rx,Frixione:2007vw}), while the component of the
hardest-emission cross section, responsible for the subsequent generation of
the hardest branching kinematics, uses the transverse momentum of the
branching, both in the evaluation of the strong coupling constant and the
PDFs~\cite{Frixione:2007vw}. 
Unless otherwise stated, these scale choices are the default ones.

\subsection[Colour assignment in the large-$N_{c}$ limit]
{Colour assignment in the large-$\boldsymbol{N_{c}}$ limit}
In order to shower and hadronize the hardest-emission events, it is
necessary to assign a colour structure to the event, comprised of a number of
{colour connections}: lines charting the flow of colour from one particle to
another. To this end, we have adopted the approach proposed in
ref.~\cite{Frixione:2007vw}, whereby a colour structure is first assigned to
the underlying Born configuration probabilistically, according to the
relative weight each one contributes to the leading-order cross section in
the limit of a large number of colours, $N_{c}$, also known as planar
limit. When the hardest branching is generated, its colour structure is
assigned by assuming that the colour flow among the mother parton and its two
daughters is also trivially planar.

\begin{figure}
\begin{center}
\epsfig{file=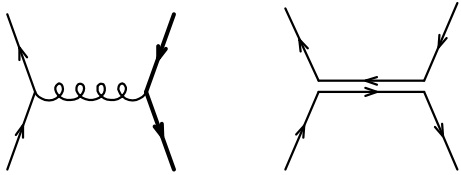,width=0.4\textwidth}
\end{center}
\caption{\label{fig:colourflqqQQ}
The Feynman diagram for non-identical quark scattering processes,
$q\bar{q}q'\bar{q}'$, with the corresponding planar colour
structure depicted on the right.}
\end{figure}

\begin{figure}
\begin{center}
\epsfig{file=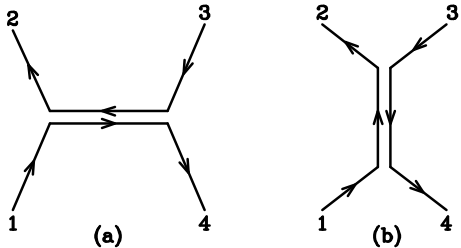,width=0.4\textwidth}
\end{center}
\caption{\label{fig:colourflqqqq}
The two planar colour flows for identical quark
scattering processes.}
\end{figure}

\begin{figure}
\begin{center}
\epsfig{file=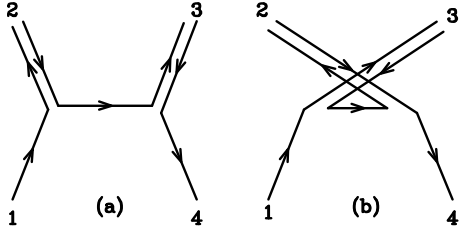,width=0.4\textwidth}
\end{center}
\caption{\label{fig:colourflqqgg}
Colour flows in the large $N_{C}$ limit for $q\bar{q}gg$ scattering processes.}
\end{figure}

\begin{figure}
\begin{center}
\epsfig{file=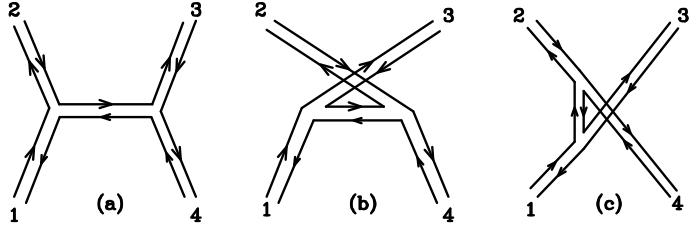,width=0.6\textwidth}
\end{center}
\caption{\label{fig:colourflgggg}
Colour connections for a $gggg$ amplitude.}
\end{figure}

To implement this prescription it is therefore necessary to also compute the
Born cross section, piecewise, in terms of the contributions made by each
individual colour structure. These component cross sections may be readily
computed using large-$N_{c}$ Feynman
rules~\cite{'tHooft:1973jz}. For the case at hand, they may
also be found in ref.~\cite{Marchesini:1987cf}.  In the case of a
$q\bar{q}q'\bar{q}'$ amplitude, only one Feynman diagram and one colour
structure are involved, with a gluon being exchanged between the $q'$ and the
$q$ line -- in the planar limit, the $q\bar{q}'$ and $\bar{q}q'$ pairs have
opposite colours, as shown in fig.~\ref{fig:colourflqqQQ}.
For identical quark scattering processes, $q\bar{q}q\bar{q}$, there are
 only two associated planar colour flows, as shown in
fig.~\ref{fig:colourflqqqq}.
The corresponding contributions to the leading-order cross section
can be easily deduced by comparing the $q\bar{q}q'\bar{q}'$ and
$q\bar{q}q\bar{q}$ squared amplitudes: omitting common factors,
they are given by
\begin{equation}
|{\cal M}_{(a)}|^2\propto \frac{s_{14}^2+s_{13}^2}{s_{12}^2}\,,\quad\quad
|{\cal M}_{(b)}|^2\propto \frac{s_{12}^2+s_{13}^2}{s_{14}^2}\,,
\end{equation}
and the respective colour structures are assigned, probabilistically, on
the basis of these values. The two colour flows associated with the
$q\bar{q}gg$ process are shown in fig.~\ref{fig:colourflqqgg}.
They are proportional to
\begin{equation}
|{\cal M}_{(a)}|^2\propto \left|\frac{s_{13}}{s_{12}}\right| 
\frac{s_{12}^2+s_{13}^2}{s_{14}^2}\,,\quad\quad 
|{\cal M}_{(b)}|^2\propto \left| \frac{s_{12}}{s_{13}} \right|
\frac{s_{12}^2+s_{13}^2}{s_{14}^2} 
\;.
\end{equation}
Thus, the colour structure is chosen with a probability proportional to
$s_{13}/s_{12}$ for (a), and $s_{12}/s_{13}$ for (b).
Finally, the $gggg$ amplitude has three colour structures,
depicted in fig.~\ref{fig:colourflgggg}.
Up to common kinematical factors and physical constants, their contributions
to the leading-order cross section are given by
\begin{equation}
|{\cal M}_{(a)}|^2\!\propto\!
\left(\frac{s_{12}}{s_{14}}+\frac{s_{14}}{s_{12}} + 1\right)^2\!\!\!,\quad
|{\cal M}_{(b)}|^2\!\propto\!
\left(\frac{s_{14}}{s_{13}}+\frac{s_{13}}{s_{14}} + 1\right)^2\!\!\!,\quad
|{\cal M}_{(c)}|^2\!\propto\!
\left(\frac{s_{12}}{s_{13}}+\frac{s_{13}}{s_{12}} + 1\right)^2\!\!\! .
\end{equation}

\subsection{Generation cut and suppression factor}
In the dijet process, as in the $V+j$ process~\cite{Alioli:2010qp}, the
leading-order contribution to the cross section is itself collinear and
soft divergent, mandating that a cut be placed on the transverse momentum
of the final-state partons in generating the underlying Born configuration.
Since this generation cut is unphysical, it is essential that in studying
the output of the simulation, the analysis cuts employed restrict the
transverse momenta of the leading jets to always be somewhat larger
than it, so as to render any related dependencies negligible.\footnote{For
a more involved discussion of this issue see ref.~\cite{Alioli:2010qp}.}

\begin{figure}
\begin{center}
\epsfig{file=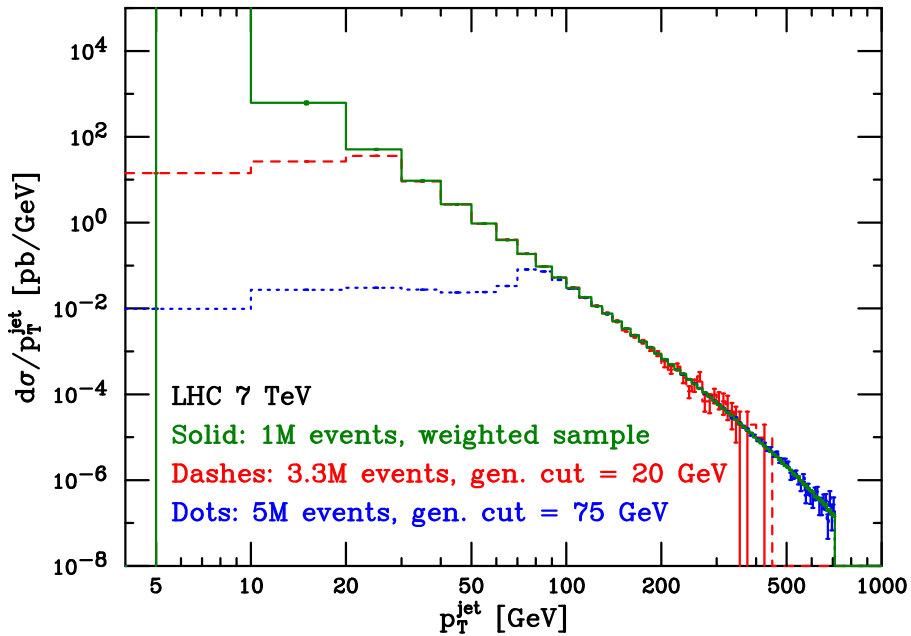,width=0.8\textwidth}
\end{center}
\caption{\label{fig:GenCuts}
Comparison of two \POWHEG{} unweighted samples for two different generation
cuts and a weighted sample with a generation cut at 1~GeV.}
\end{figure}

A further, somewhat related, technical point is that the cross section
falls very sharply, as the transverse momenta of the leading jets
increases, the rate being crudely proportional to $\kt^{-6}$, hence,
generating events with uniform weight generally fails to give a reasonable
yield in the high-$\kt$ regions of phase space. One approach to solving
this problem is to produce several independent samples of events, using
different values of the generation cut in each one, in order to populate
all the regions of interest. These samples may then be recombined by
weighting events discretely according to the cross section for the sample
from which they originated, having taken care to ensure that different samples
do not populate the same phase space. 
Alternatively, the \POWHEGBOX{} is capable of directly generating weighted
events samples, as described in ref.~\cite{Alioli:2010qp}. When the weighted
event mode is activated (by assigning a positive value to $\ktsupp$), events
are no longer distributed according to the differential cross section but
rather the differential cross section multiplied by 
\begin{equation}
\label{eq:suppfac}
{\cal S}\left(\kt\right)=\left(\frac{\kt^2}{\kt^2+\ktsupp^2}\right)^3,
\end{equation}
in the case of dijet production.\footnote{In the \POWHEGBOX{}, the value of
  the token \ptsupp{} in the input file is assigned to $\ktsupp$ and the
  function in eq.~(\ref{eq:suppfac}) is implemented in the user-defined
  subroutine {\ttfamily born\_suppression}.} The generated events now carry a
variable weight, equal to the inverse of the suppression factor, ${\cal
  S}\left(\kt\right)$, whose functional form may be changed by the user at
will, but keeping the same limiting behaviour as $\kt\to 0$ and $\kt \to
\infty$. The $\kt$ dependence of the suppression factor in
eq.~(\ref{eq:suppfac}) is such that the generation of low transverse momentum
events is relatively damped, while the whole transverse momentum region is
nearly uniformly populated up to momenta of the order of $\ktsupp$. As with
the generation cut, a more thorough description of these technicalities may
be found in ref.~\cite{Alioli:2010qp}.

To illustrate these technical features, in fig.~\ref{fig:GenCuts}
we show the jet
transverse-momentum spectrum obtained with different generation cuts, and
with a weighted sample.  The effect of the generation cut on the unweighted
sample is negligible above the $\pt$ value where the unweighted sample agrees
with the weighted one. Notice also that the weighted sample, in spite of
being smaller than the other two, comprising of just 1 million events,
populates the region of large transverse momenta well.

\section{Theoretical analysis and validation}
\label{sec:theory_bit}
In this section we present results obtained for dijet production in the
\POWHEGBOX, primarily to audit it and to verify its correctness. This is,
however, not quite a simple validation exercise but one very much
connected to the phenomenology of jet physics and jet production. In
particular we shall consider, in some detail, the interplay of the fixed
order component of the calculation and resummation effects.

Here, as with all results presented in this paper, we have used the
CTEQ6M~\cite{Pumplin:2002vw} parton distribution functions in generating our
predictions. Furthermore, unless otherwise stated we have used the seed-based
D0 midpoint cone algorithm, as implemented in the \FASTJET{}
package~\cite{Cacciari:2005hq,fastjetweb}, with a jet radius of $R=0.7$, an overlap
threshold $f=0.5$ and assuming the default values of the minimum jet $\Et$
and minimum jet $\Et$ ratio parameters (6~GeV and 0.5 respectively).
We have also made some studies using the \SISCONE{} and $\kt$
algorithms~\cite{Salam:2007xv,Ellis:1993tq,Catani:1993hr} which we shall
discuss in due course.

\subsection{NLO cross section}
\label{sec:nlo_xsec}
The full, regulated, NLO cross section is fundamental in the \POWHEG{}
algorithm in that it alone is used to generate the {underlying Born
configuration}, in this case a $2\rightarrow2$ QCD process, upon which the
whole event is founded.  The \POWHEGBOX{} framework has built into it the
facility to compute fixed-order NLO distributions for all simulations
based upon it. This feature is primarily intended as a diagnostic tool,
enabling users to check that the NLO cross section underlying the event
generation has been realised correctly within their code. We have made
use of this feature to check this delicate component of the simulation,
comparing predictions for a wide range of inclusive distributions against
the independent parton level program of ref.~\cite{Frixione:1997ks}.

In performing these cross checks, we chose to run both programs using a fixed
renormalization and factorization scale of 100~GeV. Furthermore, in our code,
we have used a tiny $\kt$ generation cut of 0.1~GeV and a Born suppression
parameter, $\ktsupp$, of 100~GeV. Taking the generation cut to such a small
value ensures that the results become insensitive to it,
while the Born suppression factor
compensates for the sharply falling $\kt$ spectrum such that values of the
transverse momentum up to a few hundred~GeV are uniformly sampled. The
program of ref.~\cite{Frixione:1997ks} requires a cut on the total transverse
energy of the final state which we have stated in the legend of each of the
plots. These choices are simply made to ensure a good yield of events in
the distributions under consideration, while respecting the jet $\Et$ cuts
applied in each case.

\begin{figure}
\begin{centering}
\epsfig{file=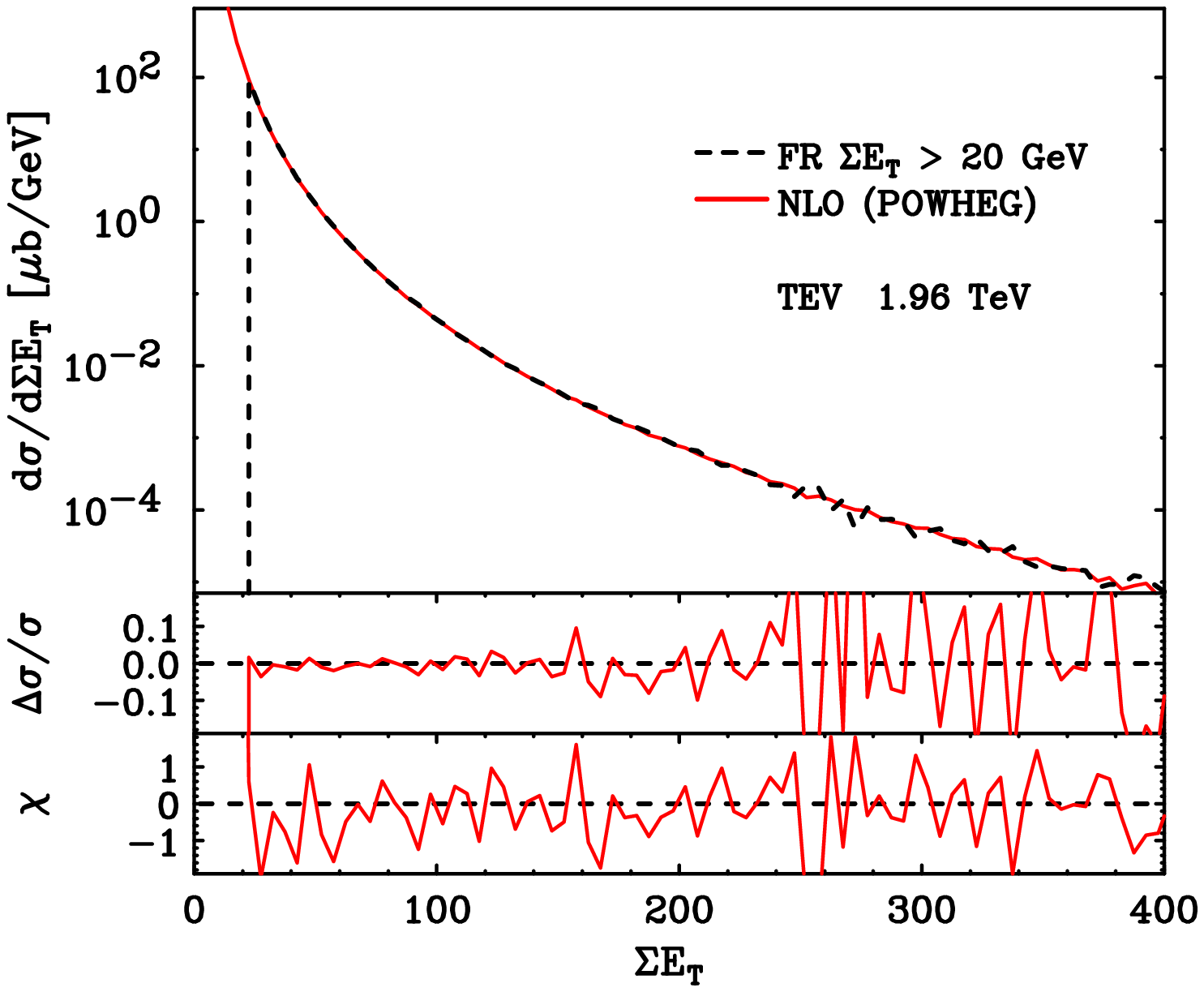,width=0.490\textwidth}
\hfill{}
\epsfig{file=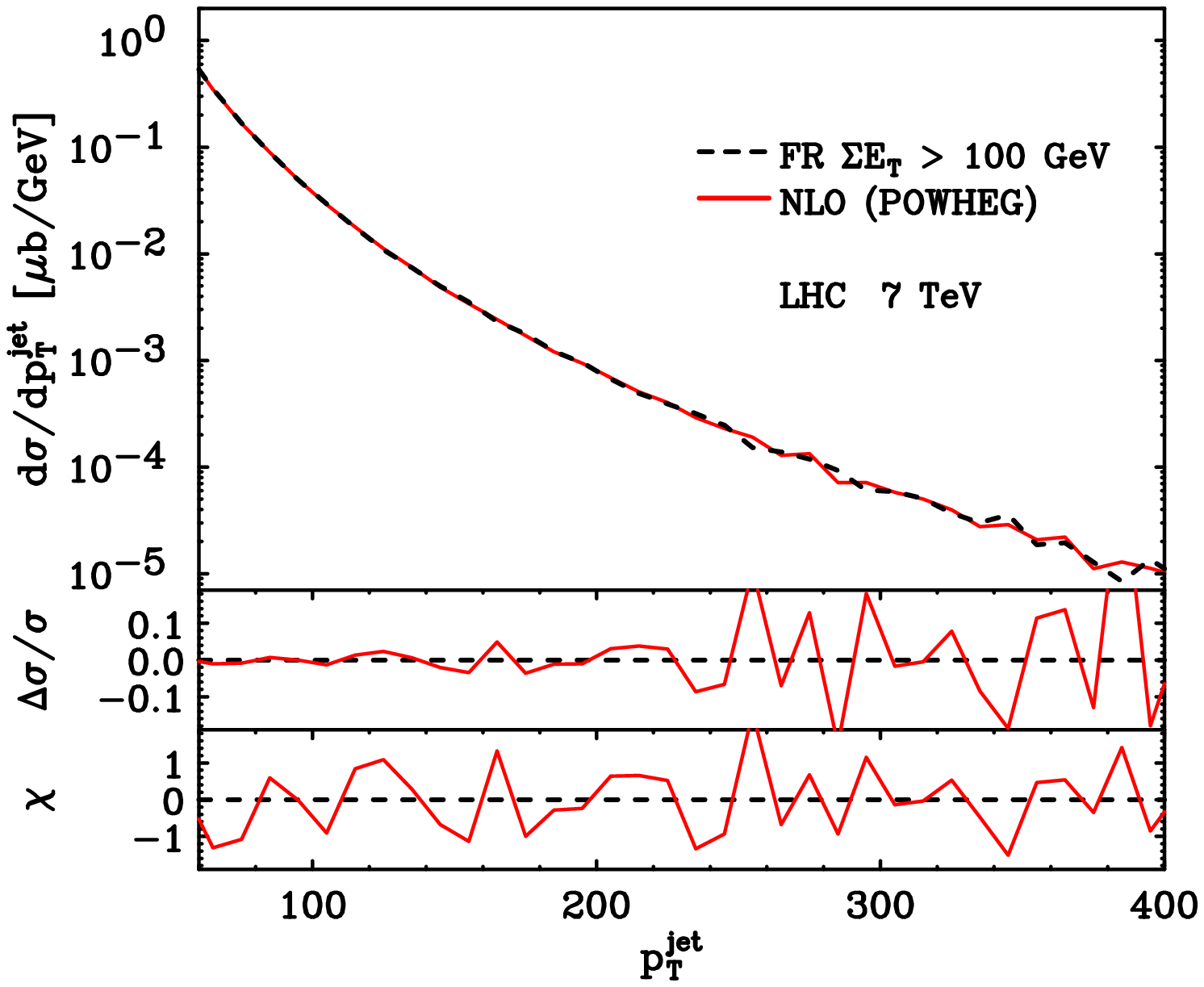,width=0.490\textwidth}
\par\end{centering}

\begin{centering}
\epsfig{file=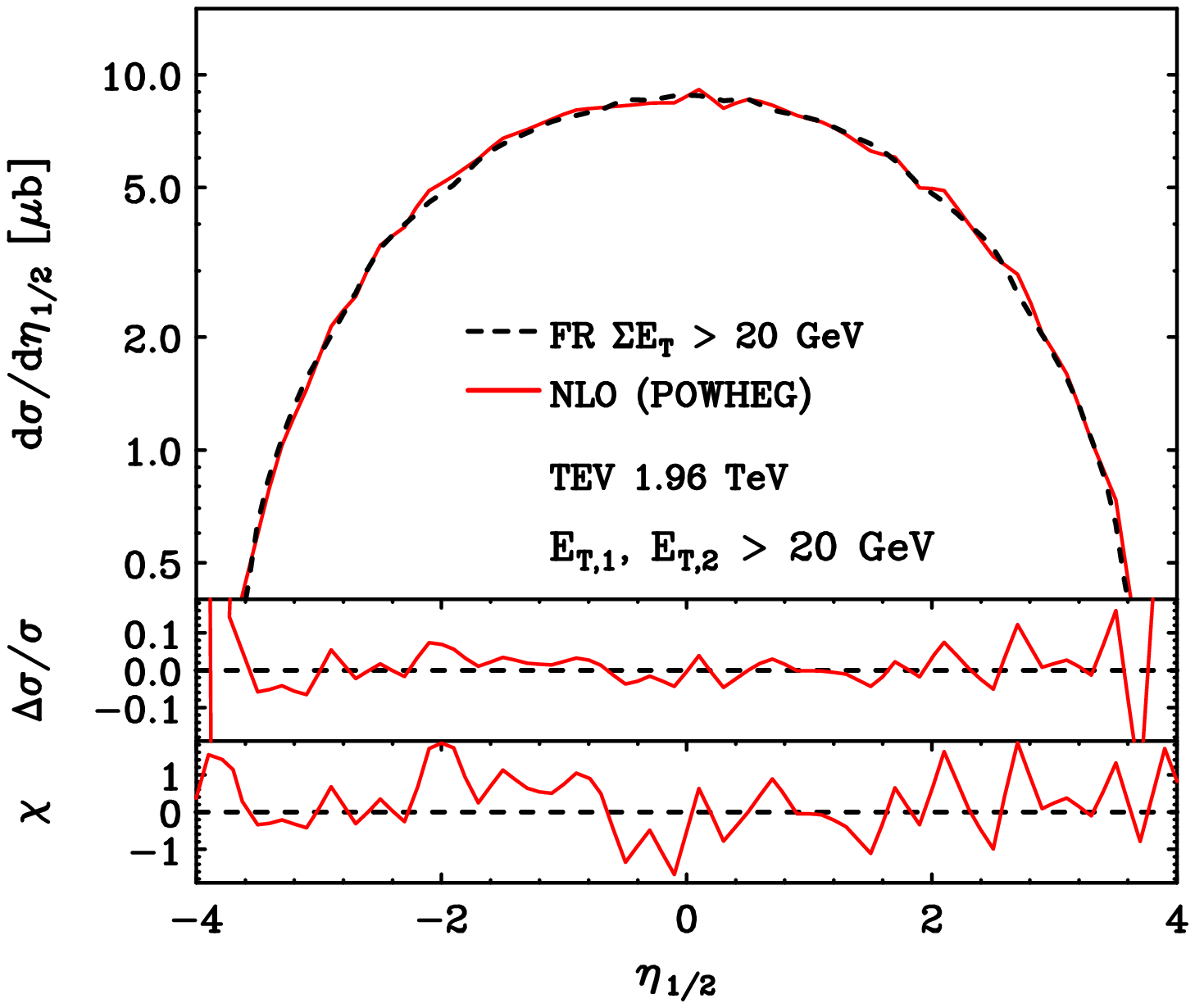,width=0.490\textwidth}
\hfill{}
\epsfig{file=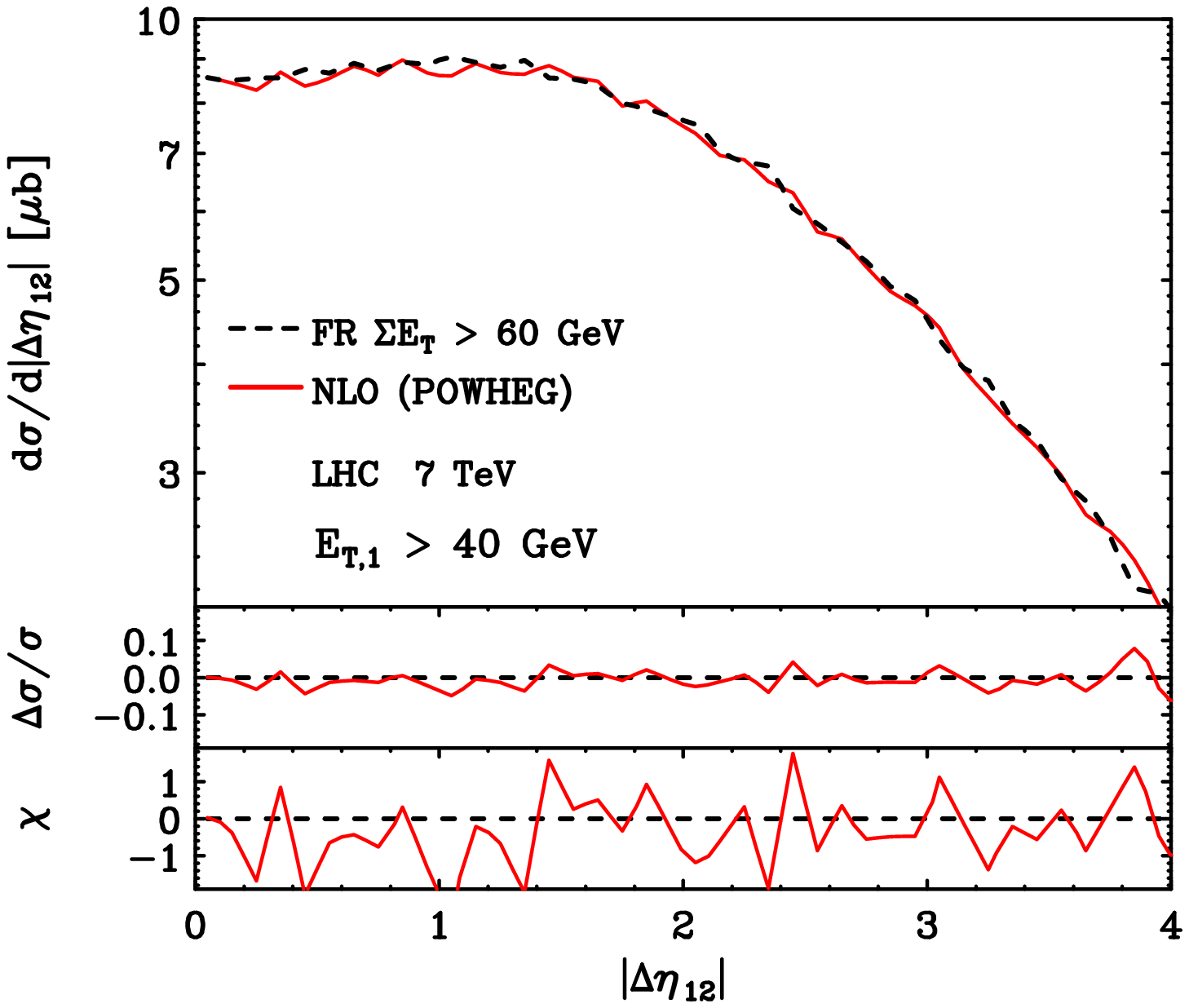,width=0.490\textwidth}
\par\end{centering}
\caption{\label{fig:NLO-FR} A sample of distributions
  demonstrating the validity of the underlying fixed-order component of the
  \POWHEG{} simulation (solid red), comparing it to the NLO calculation
  implemented in the program of ref.~\cite{Frixione:1997ks} (FR, black dashes).
  The left-hand column shows cross checks carried
  out for 1.96~TeV $p\bar{p}$ collisions, while the right-hand column
  concerns 7~TeV $pp$ collisions.  In the upper pair plots we show the total
  transverse energy spectrum (left) and the inclusive jet $\pt$ spectrum
  (right). Beneath these are the inclusive pseudorapidity distribution of the
  two highest transverse momentum jets and that of the pseudorapidity gap
  between them, for given cuts on their respective transverse energies.}
\end{figure}

In fig.~\ref{fig:NLO-FR} we display some results typical of these cross
checks. For each plot therein, the two lower plots show the relative
difference of the two calculations, $\Delta\sigma/\sigma$, and the difference
divided by the statistical error, $\chi$, so defined:
\beqn
\label{eq:deltasigma}
&&\frac{\Delta\sigma}{\sigma} = \frac{\sigma_1-\sigma_2}{\sigma_2}\,,\\
\label{eq:chi}
&&\chi = \frac{\sigma_1-\sigma_2}{\sqrt{\delta\sigma_1^2 +
    \delta\sigma_2^2}}\,. 
\eeqn
All of the distributions we have
studied from each code, like those shown in fig.~\ref{fig:NLO-FR}, have been
found to be fully consistent with one another.

\subsection{Hardest-emission cross section}
\label{sec:hardest_emission_xsec}
Having demonstrated the validity of the underlying NLO cross section, we turn
to examine the next phase of the event generation procedure, whereby the
hardest branching in the event is generated from the initial $2\rightarrow2$
underlying Born configuration. For such a configuration the {radiative
  variables}, determining the kinematics of this branching, are distributed
according to the product of a Sudakov form factor and the real emission cross
section divided by the Born cross section.  In this way all orders soft
resummation effects are included in the generation of this radiation in the
\POWHEG{} {hardest-emission cross section}~\cite{Nason:2004rx}. In the
following, we wish to assess the impact of such resummation on NLO accurate
distributions. To this end, we compare fixed-order NLO predictions, like
those in the previous subsection, against those obtained by analysing the
hardest-emission events in the Les Houches files {prior} to their showering
with \HERWIG{} or \PYTHIA{}.\footnote{By setting the {\tt testplots} token
  equal to {\tt 1} in the input file, two output files are generated that
  contain these NLO and \POWHEG{} hardest-emission distributions
  respectively.}

In the following analysis, we have used the default scales of
section~\ref{sec:scales_choice}. Note that using the $\pt$ of the radiation
as scale choice for the generation of the hardest emission has no bearing on
the distribution of the Born variables $\Phi_{B}$. In fact, given that the
same scale is adopted for $\bar{B}\left(\Phi_{B}\right)$ and the fixed-order
prediction, the generation of the underlying Born kinematics in the
\POWHEG{} simulation and the fixed-order calculation is identical, by
construction. This point should be borne in mind throughout this section, in
comparing the NLO prediction to those obtained with the \POWHEG{}
hardest-emission events.

\subsubsection{Parameters for the generation of the samples}
For the fixed-order computations, we have chosen a generation cut of 1~GeV on
the transverse momentum of the underlying Born configuration and a $\pt$
suppression factor parameter $\ktsupp$ of 50~GeV (see eq.~\ref{eq:suppfac}). In
producing Les Houches event files of hardest-emission events,
i.e.~events distributed according to the hardest-emission cross
section only, we have used a generation cut of 10~GeV and $\ktsupp=50$~GeV, for
980~GeV Tevatron beams, while, in simulating LHC events, we have used a 20~GeV
generation cut and no Born suppression factor. In the case of the hardest
emission events we neglect the effects of negative-weight events, whose
presence we have reduced to per mille levels by {folding} the radiative
phase space upon itself according to the technique described in
refs.~\cite{Nason:2007vt,Alioli:2010qp}.

\subsubsection{Inclusive distributions}
\begin{figure}
\begin{centering}
\epsfig{file=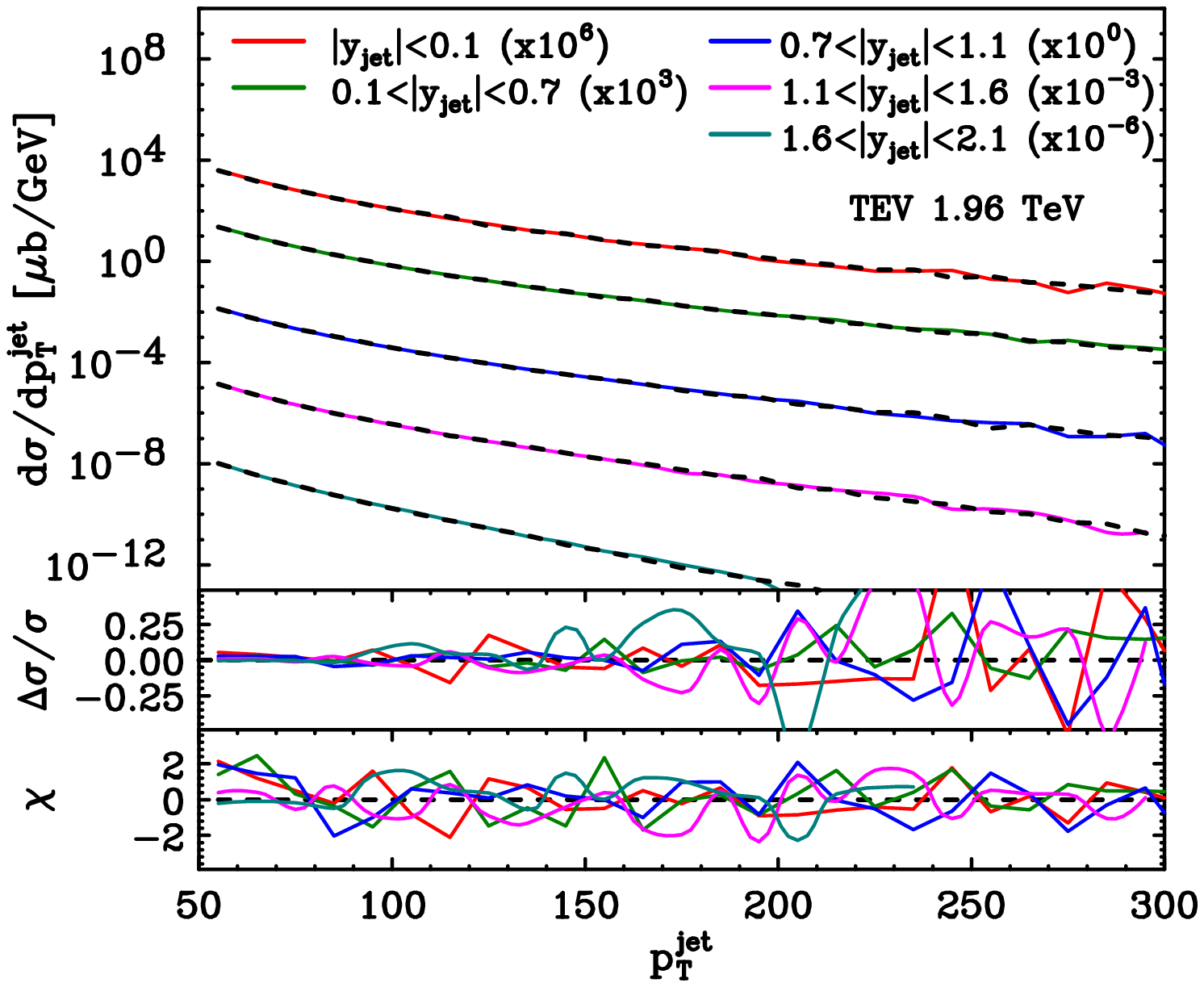,width=0.490\textwidth}
\hfill{}
\epsfig{file=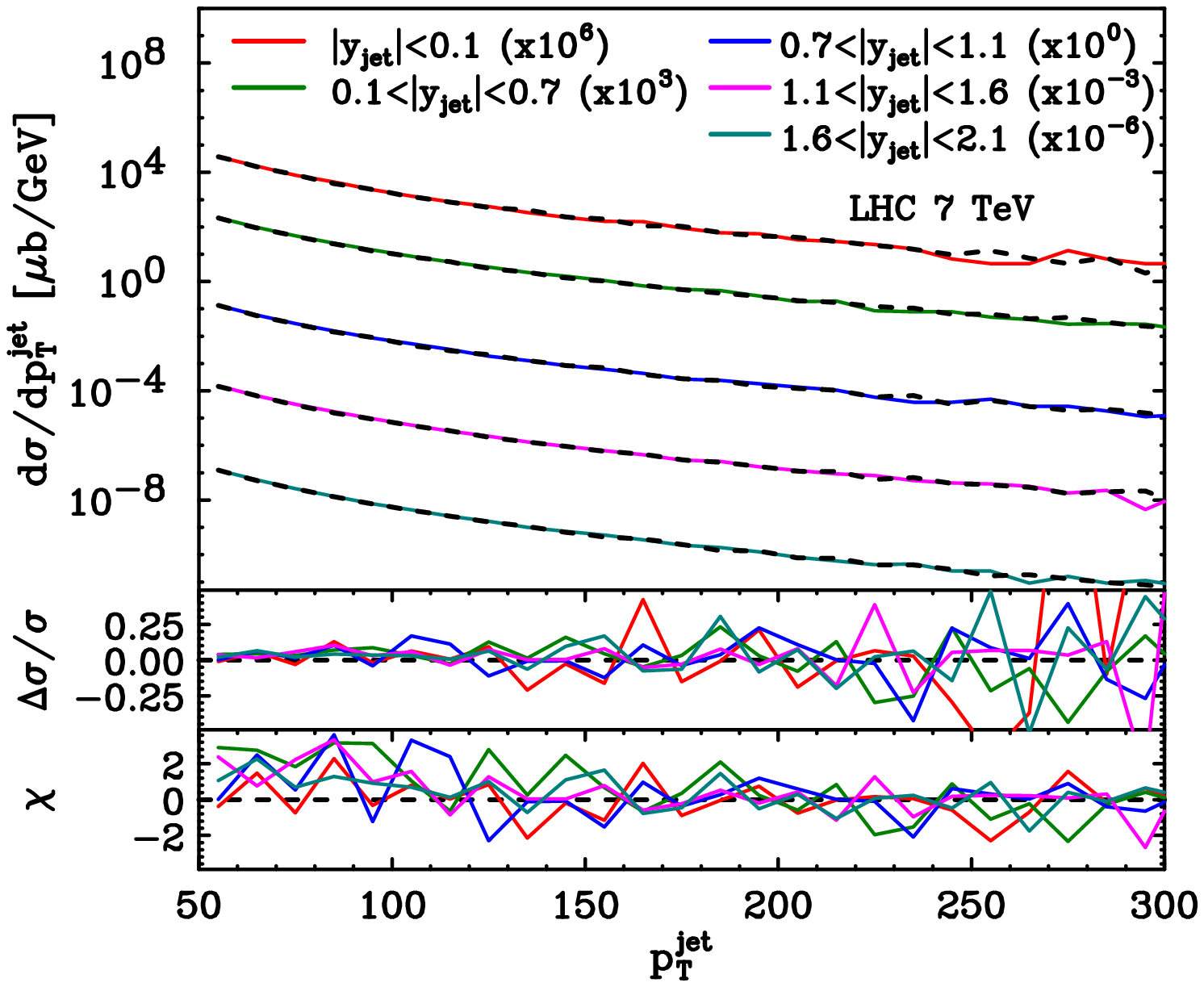,width=0.490\textwidth}
\par\end{centering}
\begin{centering}
\epsfig{file=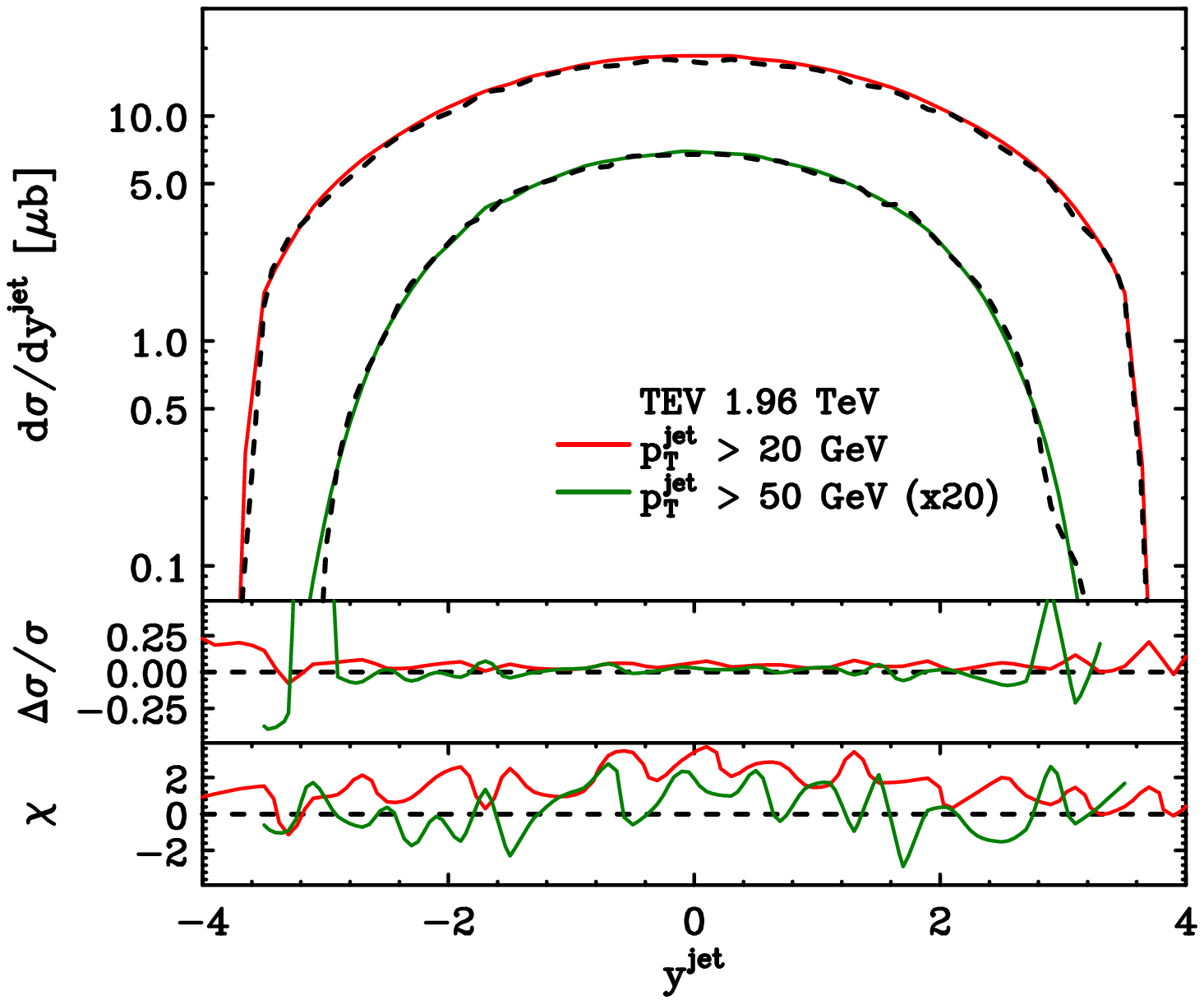,width=0.490\textwidth}
\hfill{}
\epsfig{file=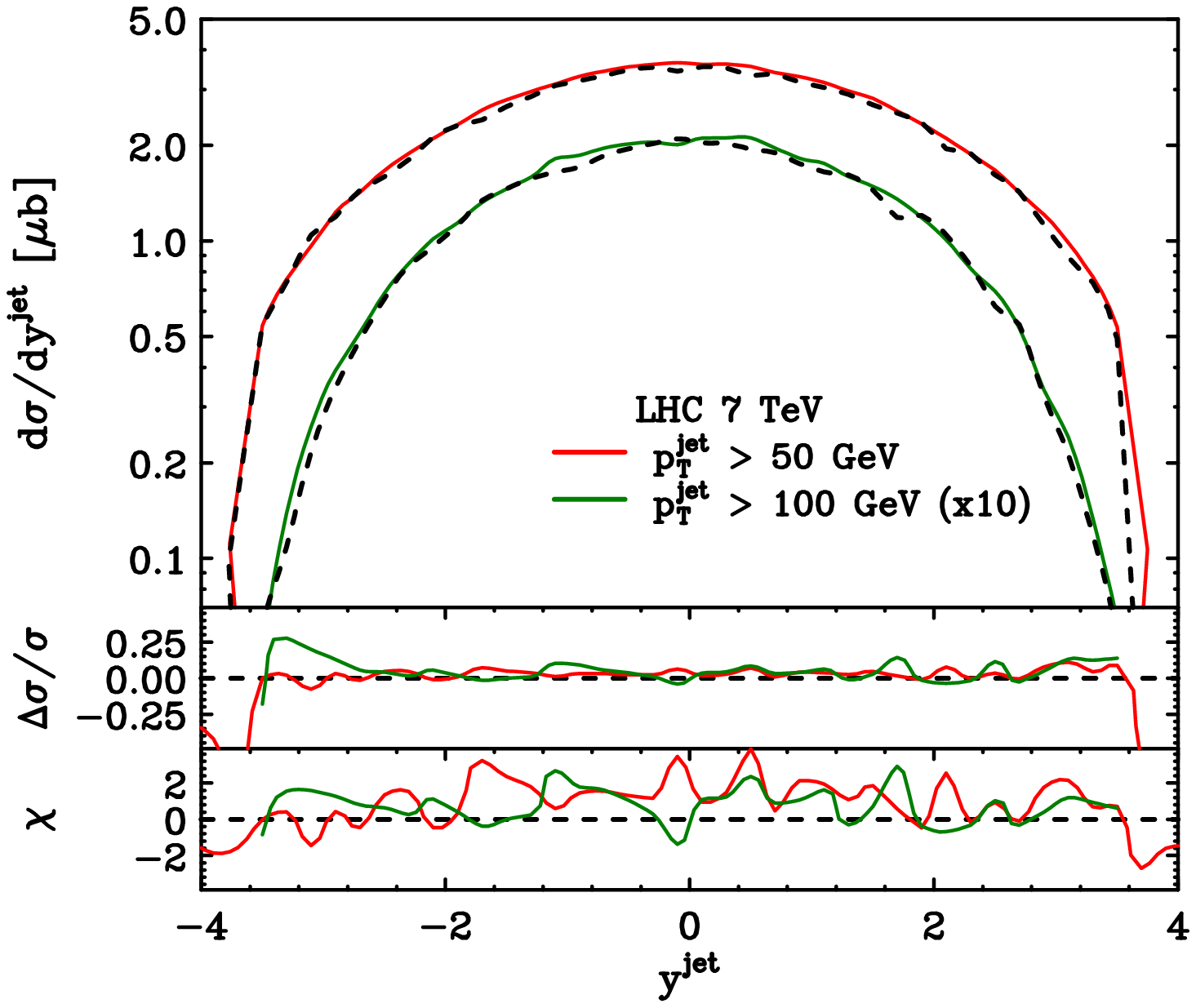,width=0.490\textwidth}
\par\end{centering}
\caption{\label{fig:LHEF-NLO-incl-pt-and-y} A comparison of the \POWHEG{}
  results, prior to showering (solid coloured lines), to the corresponding
  fixed-order NLO predictions (black dashes) of the inclusive jet
  transverse-momentum and rapidity distributions. Coloured curves in the
  upper two plots are drawn, from top to bottom, in order of increasing
  rapidity, while in the lower plots the result obtained with the
  greater of the two transverse-momentum cuts is lowermost.}
\end{figure}

Typically one expects that inclusive observables should exhibit a good level
of agreement between the NLO results and those of the \POWHEG{}
hardest-emission cross section. More specifically, for quantities that are
insensitive to Sudakov effects in the radiation of the third jet, the two
sets of results should exhibit deviations no greater than the corresponding
expected NNLO corrections. Precisely this behaviour is demonstrated in
fig.~\ref{fig:LHEF-NLO-incl-pt-and-y}, where the inclusive jet transverse
momentum and rapidity spectrum are shown, as given by the same analysis
procedure used by the CDF collaboration in ref.~\cite{Aaltonen:2008eq}.
Specifically, we cluster events according to the CDF midpoint cone algorithm,
with a jet radius parameter of $R=0.7$ and overlapping fraction $f=0.75$,
cutting events for which the ratio of the missing transverse energy
$E\!\!\!\!/_{\sss\rm T}$ to the total transverse energy, $\sum{\Et}$, fails
to satisfy $E\!\!\!\!/_{\sss\rm T}/\sqrt{\sum{\Et}}<\min\left(3+0.0125\times
\pt^{\max},6\right)$. By analogy with section~\ref{sec:nlo_xsec}, together
with the differential cross sections, we plot the relative difference of each
\POWHEG{} hardest-emission cross section (solid coloured lines) with respect
to the corresponding fixed-order NLO result (black dashes) and the
corresponding difference divided by the statistical error, $\chi$, as defined
in eqs.~(\ref{eq:deltasigma}) and~(\ref{eq:chi}).

\subsubsection{Jet cross sections with symmetric cuts}
\begin{figure}
\begin{centering}
\epsfig{file=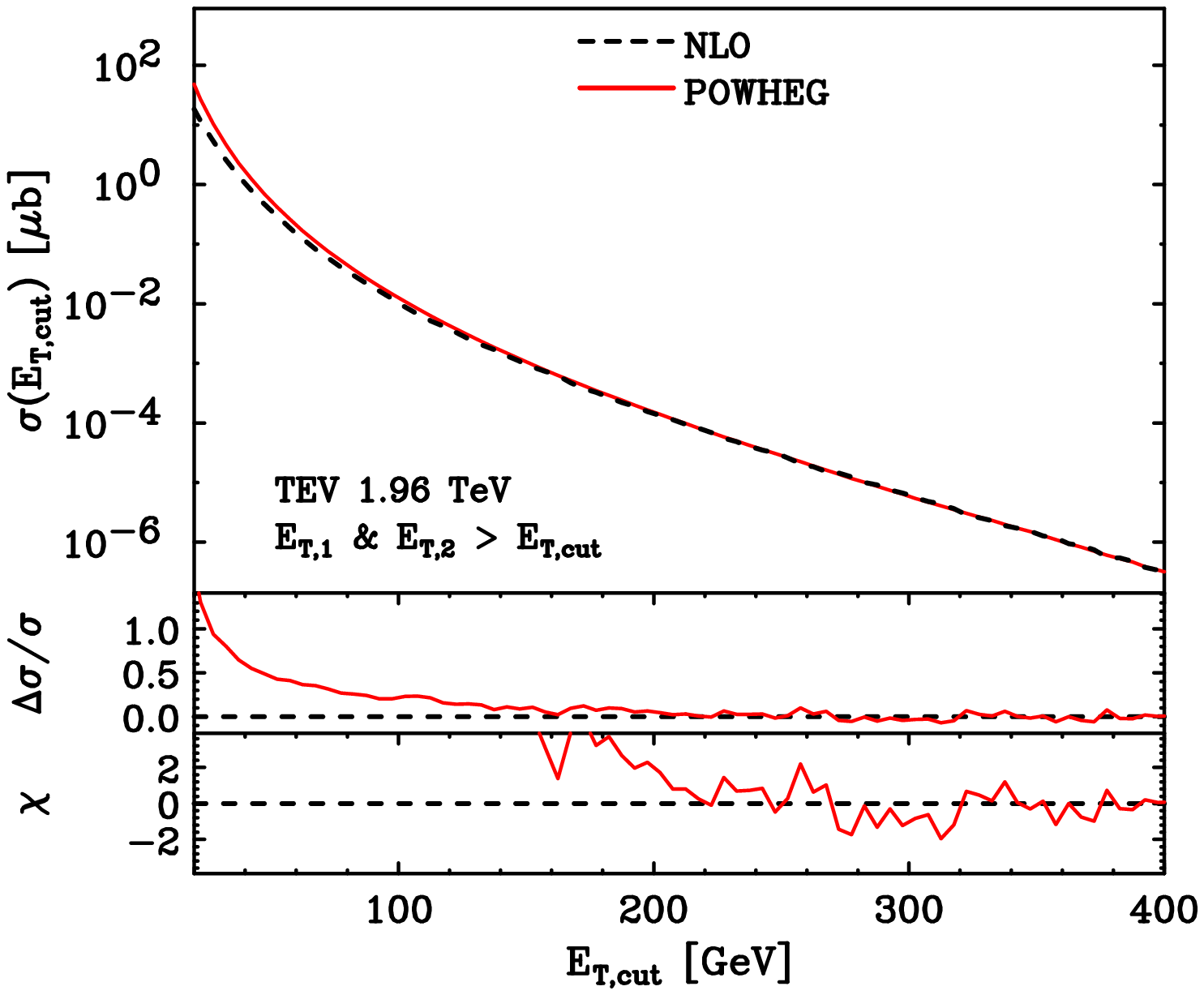,width=0.490\textwidth}
\hfill{}
\epsfig{file=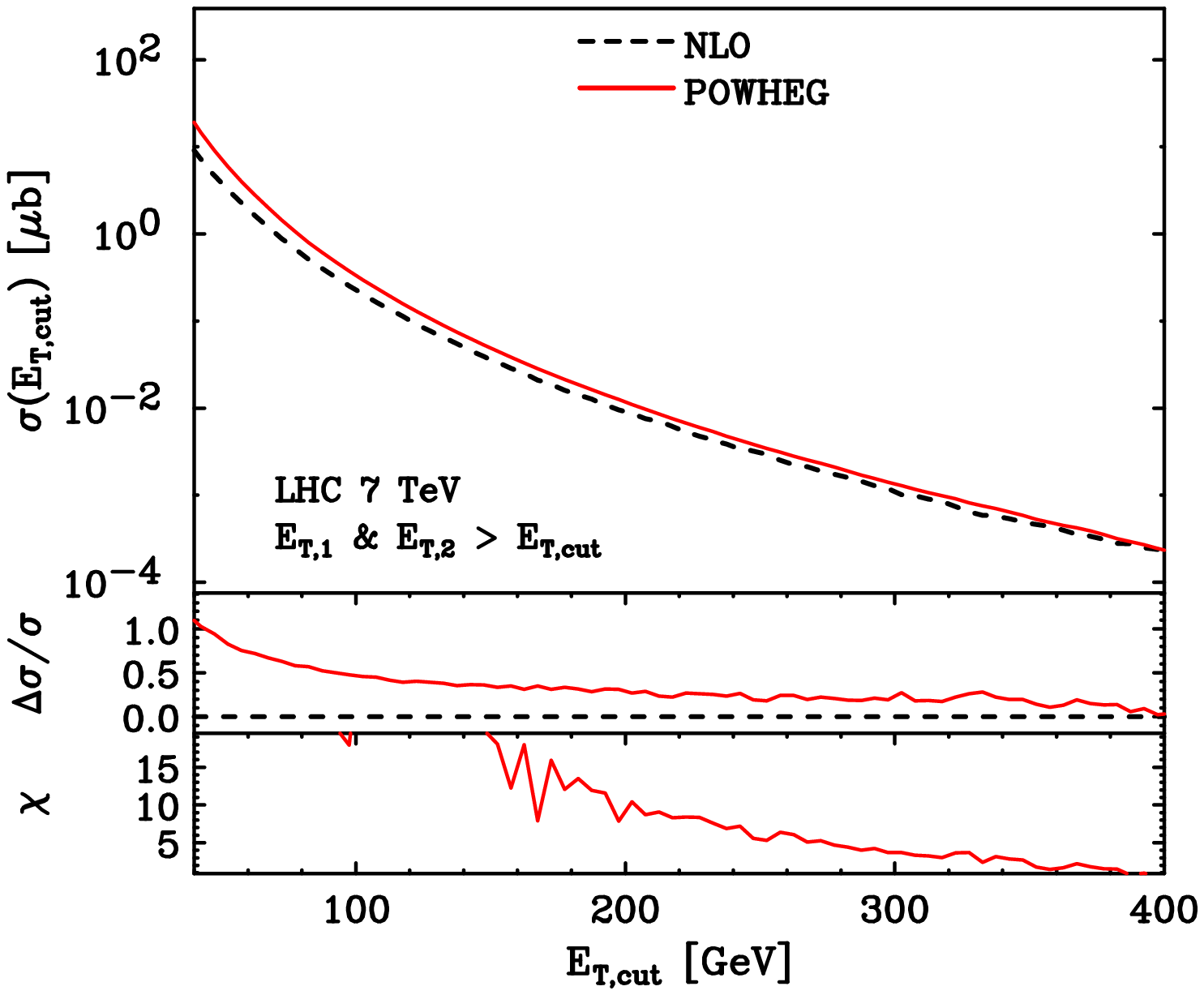,width=0.490\textwidth}
\par\end{centering}

\caption{\label{fig:LHEF-NLO-et1_et2_gt_d} Predictions for the fixed-order
  NLO cross sections to the analogous \POWHEG{} hardest-emission one, for
  symmetric cuts on the transverse energies of {both} the highest and second
  highest $\Et$ jets, at the Tevatron and LHC, in the left- and right-hand
  plots respectively.}
\end{figure}

We now examine the total cross section for jet production with symmetric cuts
on the transverse energy $\Et$ of the two leading jets. In
fig.~\ref{fig:LHEF-NLO-et1_et2_gt_d} we plot the total cross section as a
function of $E_{\sss\rm T,\mathrm{cut}}$, defined
to be the cut on the transverse energy of the two highest
transverse energy jets: $E_{\sss\rm T,1}>E_{\sss\rm T,\mathrm{cut}}$,
$E_{\sss\rm T,2}>E_{\sss\rm T,\mathrm{cut}}$.  For both plots, obtained at
Tevatron and LHC energies respectively, we show the fixed-order NLO
prediction as a dashed black line, with the corresponding \POWHEG{}
hardest-emission cross section, in solid red.
At first sight the disagreement between the fixed-order
and the \POWHEG{} results may not
seem so remarkable. However, a quick look at the distributions of
$\Delta\sigma/\sigma$ and $\chi$, in the lower panels, reveals that the
prediction of the resummed NLO prediction is around a factor of two higher
than that of the fixed-order NLO calculation, when $E_{\sss\rm
  T,\mathrm{cut}}$ tends to small values.
This large discrepancy is alarming, particularly given that there is
certainly nothing untoward about these cuts from the point of view of
infrared safety.  However, instabilities of NLO jet production cross sections
in the presence of symmetric cuts on jet transverse energies have been noted
and studied in the past, in
lepton-hadron~\cite{Klasen:1995xe,Frixione:1997ks} and hadron-hadron
collisions~\cite{Banfi:2003jj}.

\begin{figure}
\begin{centering}
\epsfig{file=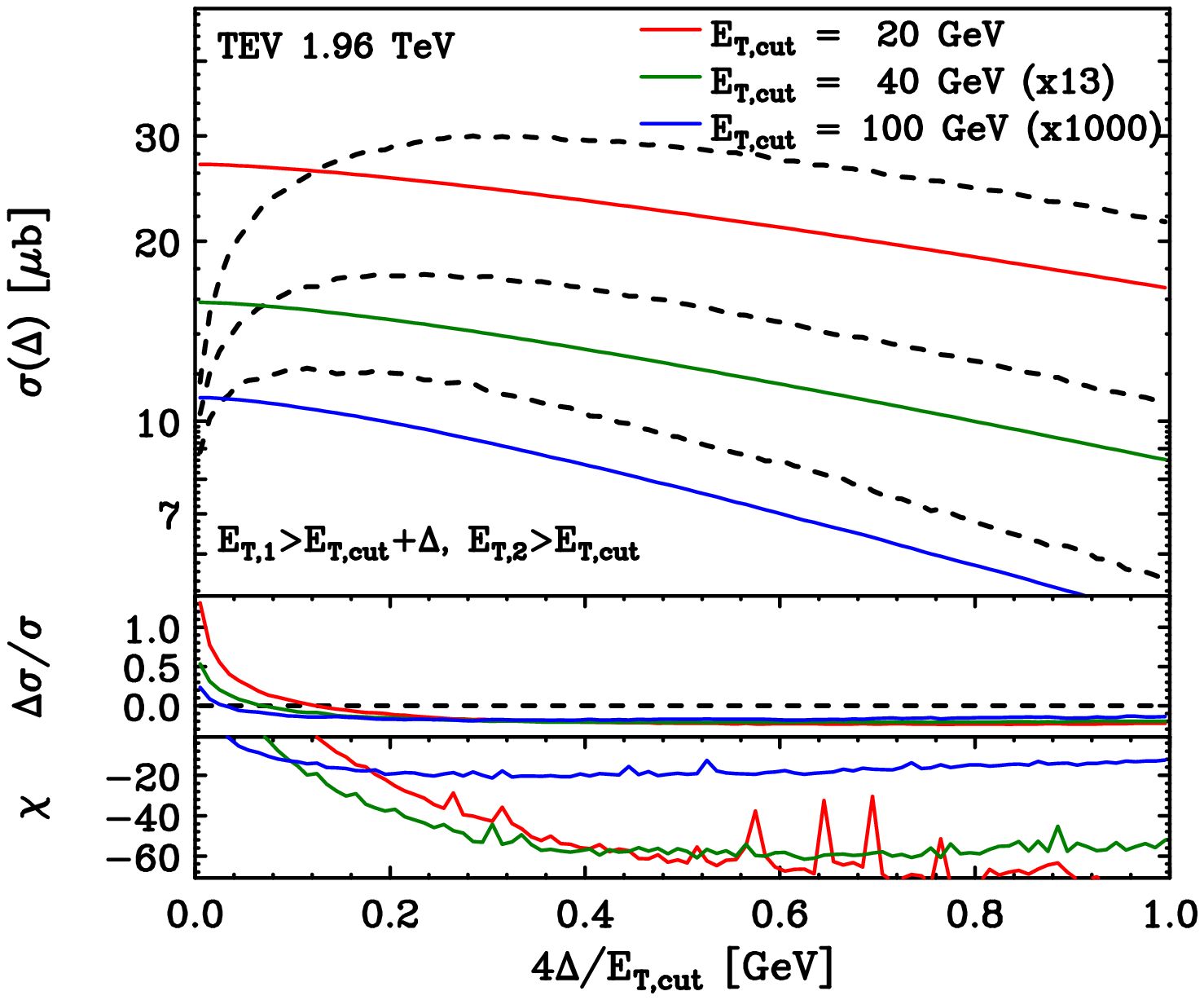,width=0.490\textwidth}
\hfill{}
\epsfig{file=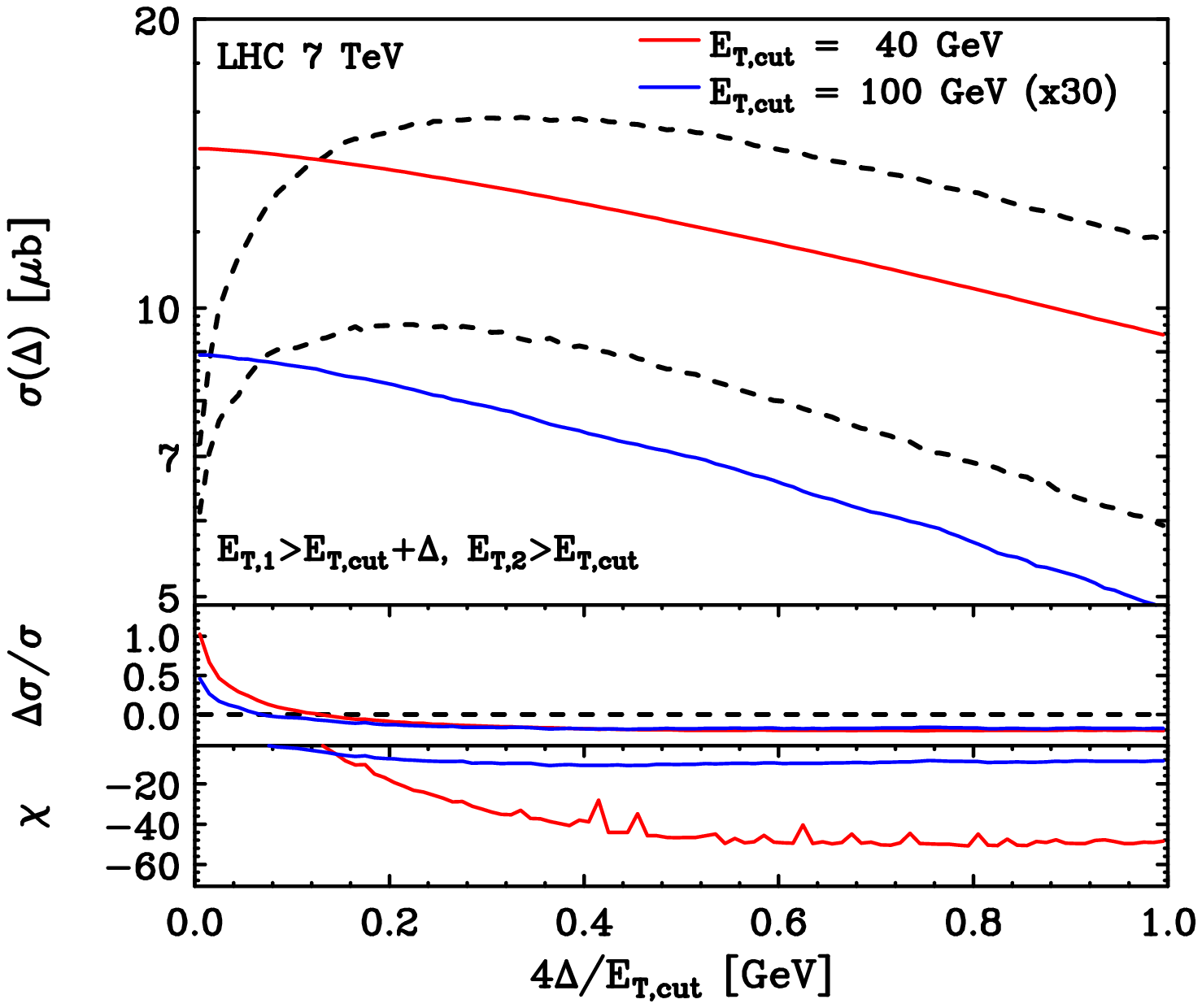,width=0.490\textwidth}
\par\end{centering}
\caption{\label{fig:LHEF-NLO-fr_3} Predictions for
  fixed-order NLO cross sections (black dashes) compared with those of the
  \POWHEG{} hardest-emission cross section
  (coloured solid lines), as a function of the transverse energy cut
  parameters $\Delta$ and $E_{\sss\rm T,\mathrm{cut}}$, defined by the
  relations: $E_{\sss\rm T,1} > E_{\sss\rm T,\mathrm{cut}} + \Delta$,
  $E_{\sss\rm T,2} > E_{\sss\rm T,\mathrm{cut}}$, with $E_{\sss\rm T,1}$ and
  $E_{\sss\rm T,2}$ being the transverse energies of the leading and
  next-to-leading jets, ordered according to transverse momentum. Results for
  the Tevatron on the left, and for the LHC on the right. Coloured lines are
  ordered from top to bottom with increasing $E_{\sss\rm T,\mathrm{cut}}$.}
\end{figure}

In order to reconcile the predictions of the NLO and \POWHEG{} hardest
emission cross sections we have carried out a similar analysis to that
performed in ref.~\cite{Frixione:1997ks} in the context of two-jet
photoproduction in lepton-hadron collisions, the results of which are
displayed in fig.~\ref{fig:LHEF-NLO-fr_3}. Here we have considered the total
cross section as a function of $\Delta$ which parametrises the degree to
which the cuts on the leading- and next-to-leading-$\Et$ jets are asymmetric:
$E_{\sss\rm T,1}>\Delta+E_{\sss\rm T,\mathrm{cut}}$, $E_{\sss\rm
  T,2}>E_{\sss\rm T,\mathrm{cut}}$, the limit $\Delta\rightarrow0$ therefore
corresponding to the case of symmetric cuts. In the case of the fixed-order
predictions, for LHC and Tevatron collider configurations, and for all
studied values of $E_{\sss\rm T,\mathrm{cut}}$, we observe the same behaviour
as in ref.~\cite{Frixione:1997ks}. In particular, counter to one's
expectations based on simple phase-space considerations, we find that the
cross section is not monotonically increasing with decreasing $\Delta$ but
rather it rises gradually to a peak as $\Delta$ tends to small values, before
falling sharply as $\Delta\rightarrow0$. This is in contrast to the
\POWHEG{} prediction which simply continues to rise.

In the case of two-jet photon production the nature of this peculiar turn
over in the NLO distribution was heuristically and assuredly explained by the
authors of ref.~\cite{Frixione:1997ks} as being due to the emergence of
large, dominant, logarithmic terms of the form $-\Delta\log\Delta$ in the
real part of the NLO cross section. This functional dependence on $\Delta$ is
plainly manifest around the $\Delta\rightarrow0$ region in our fixed-order
predictions. Although our study concerns dijet hadroproduction, the
explanation advocated in ref.~\cite{Frixione:1997ks} readily applies here too
without modification, since exchanging the initial-state photon for an
initial-state parton does not qualitatively affect the leading collinear
singular behaviour of the real cross section.

It is also stated in ref.~\cite{Frixione:1997ks} that the fall in the fixed
order predictions generated by the $-\Delta\log\Delta$ term is symptomatic of
the truncation of perturbative series at NLO, and that the resummation of
higher order soft-virtual corrections will oppose this effect. Such
resummation is implicit in the hardest-emission cross section, via the
\POWHEG{} Sudakov form factor, which acts precisely in this way, inhibiting
the same soft and collinear emissions which give rise to the
$-\Delta\log\Delta$ terms in the model of ref.~\cite{Frixione:1997ks}.
Having suppressed these spurious strong dynamical contributions
appropriately, the \POWHEG{} predictions, shown as coloured lines in
fig.~\ref{fig:LHEF-NLO-fr_3}, do not follow the same trends set by the
fixed-order predictions, but simply decrease along with the available phase
space.

\subsubsection*{The r\^ole of the underlying Born}
\begin{figure}
\begin{centering}
\epsfig{file=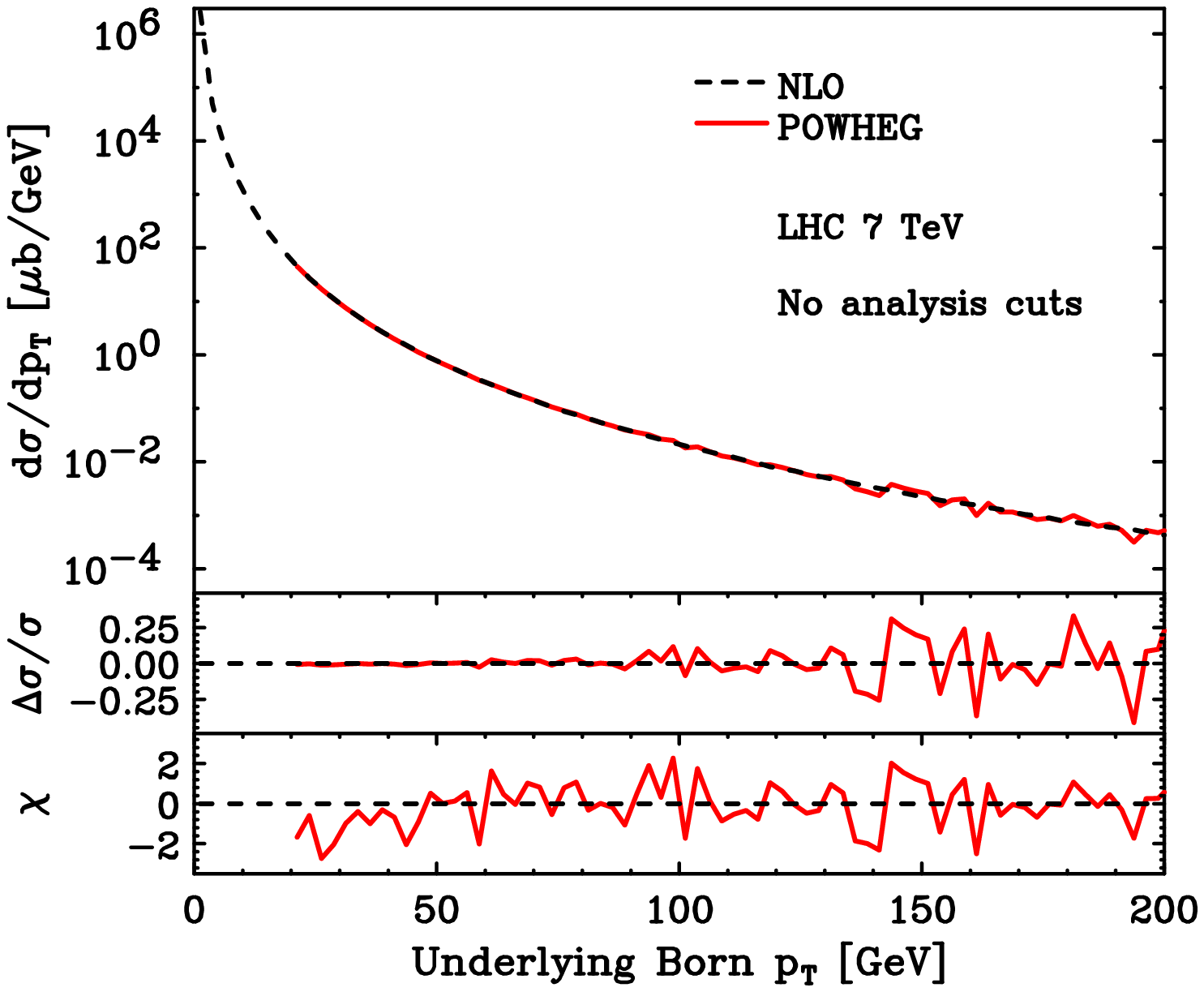,width=0.490\textwidth}
\hfill{}
\epsfig{file=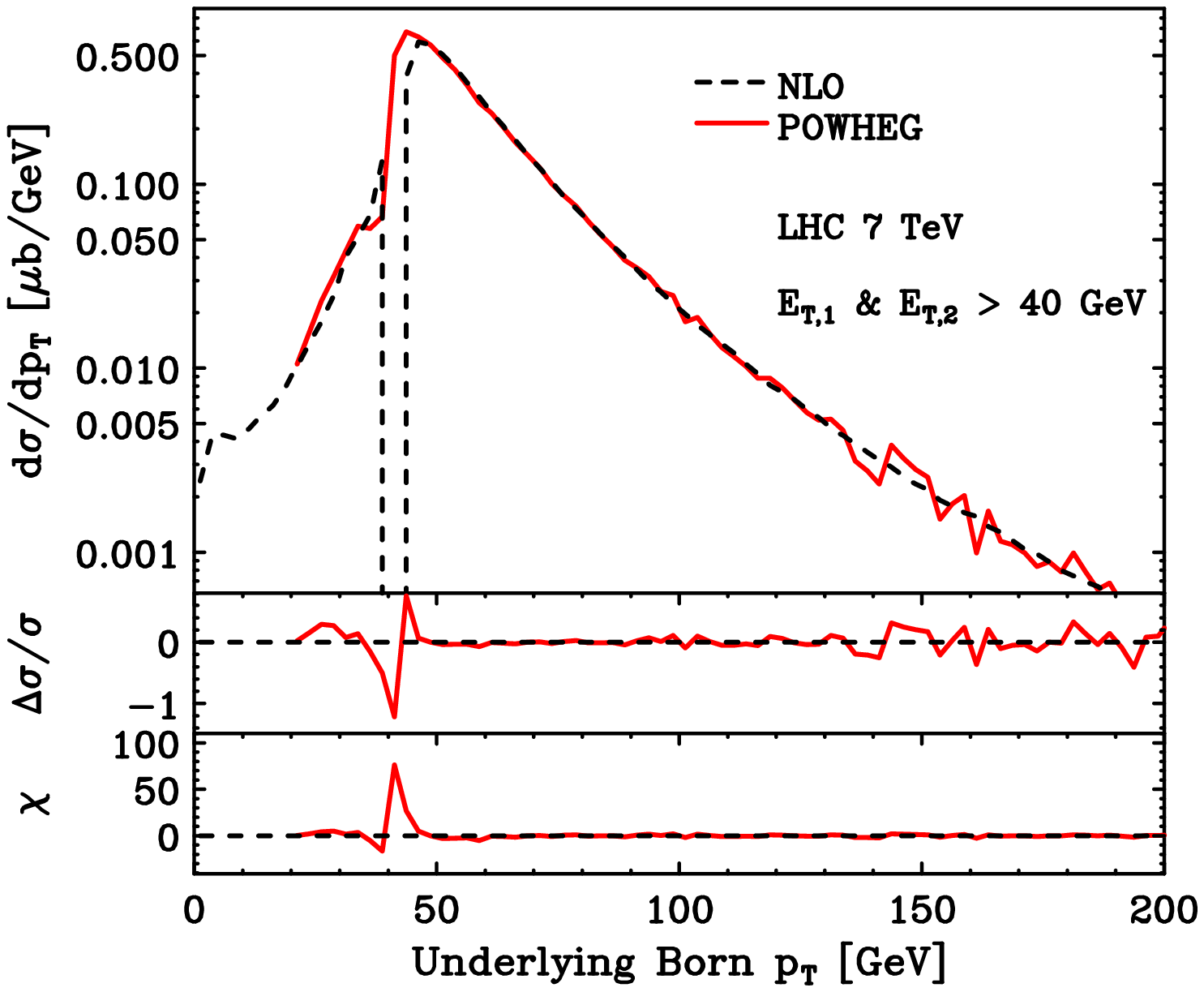,width=0.490\textwidth}
\par\end{centering}
\begin{centering}
\epsfig{file=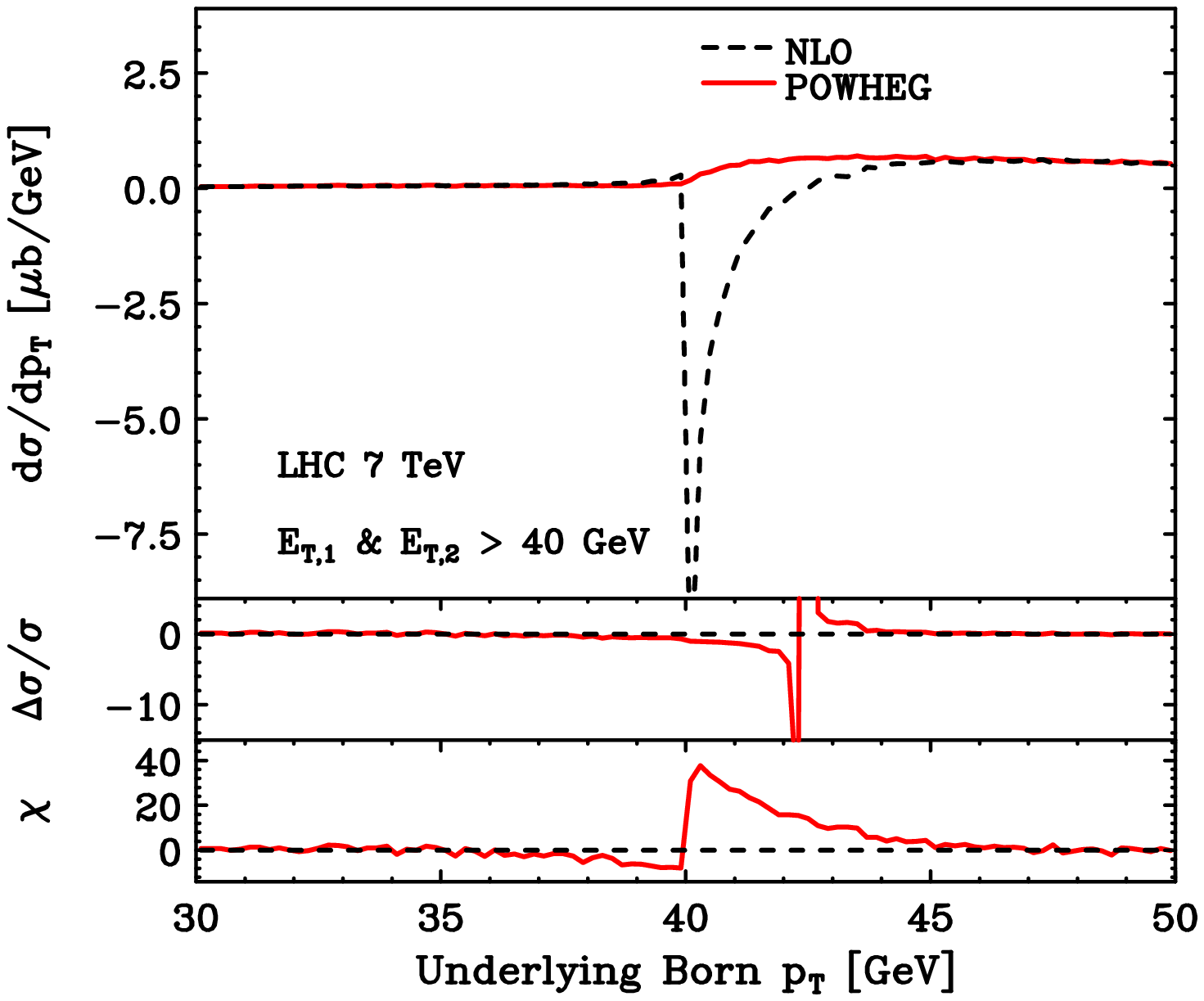,width=0.490\textwidth}
\hfill{}
\epsfig{file=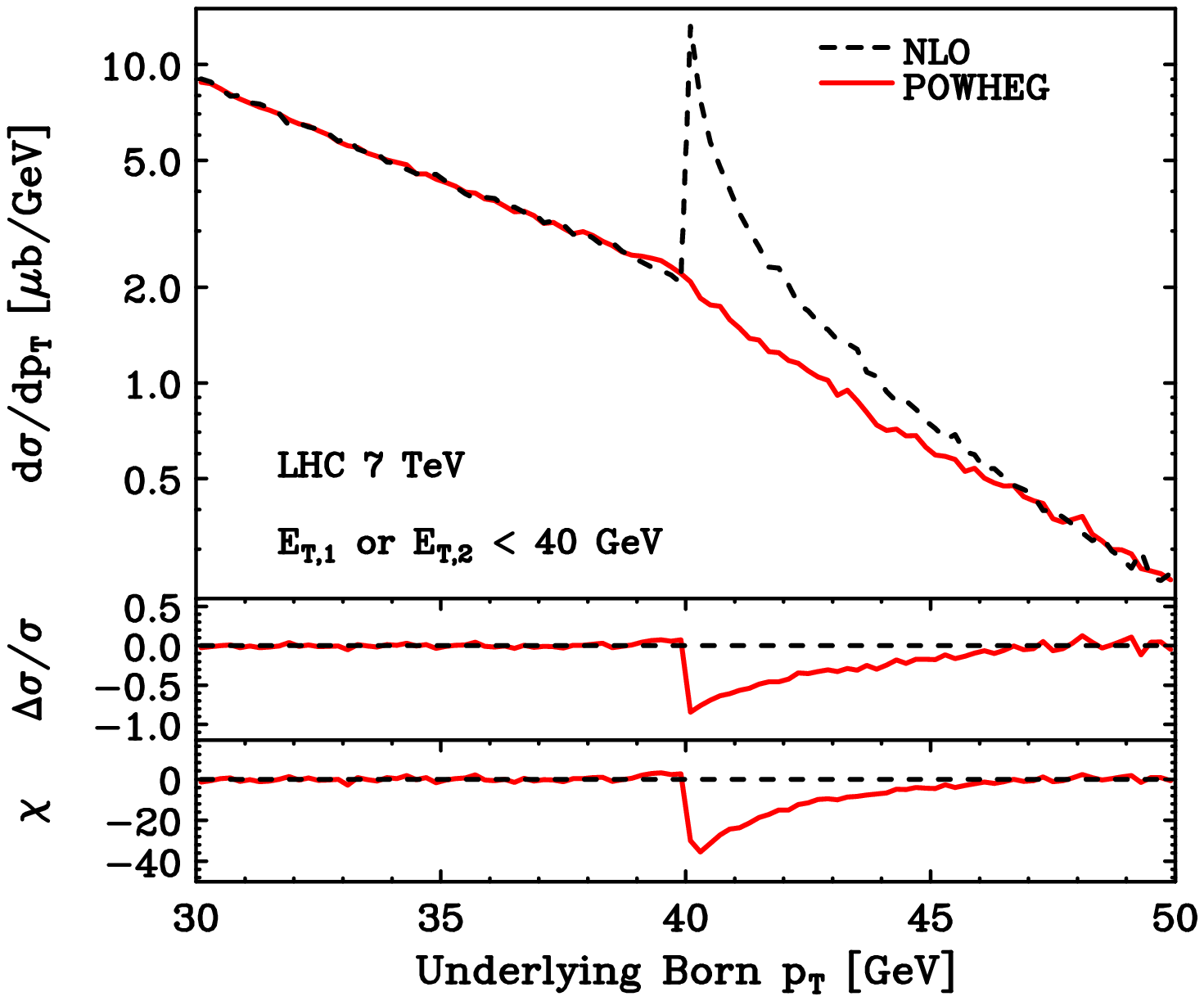,width=0.490\textwidth}
\par\end{centering}

\caption{\label{fig:LHEF-NLO-et1_et2_gt_d_UB_plots} The transverse-momentum
  spectra for the underlying Born configuration having applied symmetric cuts
  on the transverse energies of the jets in 2- and 3-parton events and, in
  the case of the fixed-order NLO results, to the counter-events too. The
  results from \POWHEG{} hardest-emission events are shown in solid red,
  while the corresponding fixed-order NLO predictions are drawn as dashed
  black lines.}
\end{figure}

In \POWHEG{}, the division of the real-emission phase space into
that of the underlying Born configuration and that of the hardest branching
lends itself to an alternative, more mechanical, understanding of
these effects. In particular, it is instructive to consider
{a posteriori} how events passing and/or failing the symmetric
cuts have originated, i.e.~the associated $2\rightarrow 2$ underlying
Born kinematics. For the case at hand, since transverse energy cuts
are applied to the two leading jets, we are especially interested
to know if and how events {migrate} across the cut depending
on the transverse momentum of their underlying Born. We stress
that this transverse momentum, as with all quantities deriving from
the underlying Born kinematics, $\Phi_{B}$, is infrared safe; this
follows directly from the definition of the mappings used to relate
underlying Born kinematics to those of real emission configurations.

Accordingly, in fig.~\ref{fig:LHEF-NLO-et1_et2_gt_d_UB_plots} we show four
distributions of the underlying Born transverse momentum. In the first plot
on the upper left-hand side we show predictions for this quantity as given by
NLO QCD (black dashes) and the hardest-emission cross section (red solid)
with no analysis cuts applied. As stated earlier, these two distributions are
the same by construction, the only difference being the use of a higher
generation cut on the $\pt$ of the underlying Born kinematics in the latter
case (10~GeV as opposed to 1~GeV).  In the upper right-hand corner we plot
the same quantity but now only for events which pass the symmetric cuts of
40~GeV in $\pt$ on the final-state jets. Above the cut, the
distribution falls rapidly, as in the first plot where no cuts were applied.
Conversely, between 0 and 40~GeV the distribution increases.  Below
$\pt=40$~GeV the distribution is populated exclusively by events initially
comprised of two partons with $\pt<40$~GeV, which radiate to yield events
containing two jets with $\Et>40$~GeV: the rise in the distribution towards
40~GeV simply reflects the fact that less radiative phase space is required
to produce an event passing the cut when the underlying Born configuration is
closer to passing the cut itself.

It is quite remarkable that one can see very good agreement between these NLO
and hardest-emission cross sections, at the level of about 10\%, for almost
all of the range in $\pt$, yet from fig.~\ref{fig:LHEF-NLO-et1_et2_gt_d} we
know that the NLO cross section for these cuts is around half that of the
\POWHEG{} hardest-emission cross section. One may think that the extra phase
space available in the case of the fixed-order calculation, due to having
used a 1~GeV instead of a 10~GeV generation cut in the underlying Born $\pt$
may cause the discrepancy. However, this could only explain an excess of the
NLO result, whereas it is the hardest-emission cross section which is the
greater of the two. Moreover, by closer inspection of the results, one can
see that the additional region populated by the fixed-order calculation
contributes only 1.7\% of the total cross section.

In fact, the factor of two deficit in the NLO calculation with respect to that
of the hardest-emission cross section can be wholly attributed to the
apparently empty bin between 40 and 42.5~GeV. A close up of this $\pt$ region
can be seen in the lower left-hand plot and the same distribution, in the
same region, can be seen for events {failing} the symmetric cuts in the lower
right-hand plot. The unstable nature of the fixed-order calculation is very
clear as a discontinuity in the first derivative of the $\pt$ distribution of
the underlying Born kinematics.  Moreover, we can see that in the spurious
bin in the NLO distribution, $40.0<\pt<42.5$~GeV, the negative-weight
counter-events, having $2\rightarrow2$ kinematics specified by $\Phi_{B}$,
pass the cut by construction, while the corresponding three-body real
emission events, constructed from {exactly the same} $\Phi_{B}$ and
additional radiative variables $\Phi_{R}$, migrate below the cut (bottom
right plot). The fact that the NLO cross section is seen to be negative in
the region just above 40~GeV, together with the correspondence in the size of
the excess (deficit) in the NLO events failing (passing) the cut, is an
unmistakable sign that this mechanism is in effect. By contrast, in the
\POWHEG{} case, one has uniquely positive-weighted events as opposed to
events and counter-events, moreover, the generation of the soft and collinear
emissions, which cause the real radiation events to fail the jet $\Et$ cut by
only a small amount, but in large numbers, is Sudakov suppressed. Hence, the
\POWHEG{} prediction exhibits no such anomalous behaviour, instead it rises
to a smooth, rounded, peak which falls away to the left, just before the
40~GeV mark is reached.

In summary, the NLO dijet cross section, for the case of symmetric cuts on
the leading jets, while being formally infrared safe, exhibits
pathological behaviour e.g.~the fall in the total cross section encountered
as $\Delta\rightarrow0$ in fig.~\ref{fig:LHEF-NLO-fr_3} and the discontinuous
distributions which it predicts for the transverse momentum of the underlying
Born configurations in fig.~\ref{fig:LHEF-NLO-et1_et2_gt_d_UB_plots}.  These
eccentricities are the result of an acute sensitivity of the cross section to
soft emission effects, which are not properly handled in the fixed-order
computation. The mechanism behind this sensitivity can be clearly understood
in terms of soft, three-body, real-radiation events migrating below the cuts
while their corresponding two-body counter-events remain above it, in the NLO
case (fig.~\ref{fig:LHEF-NLO-et1_et2_gt_d_UB_plots}). The inclusion of the
proper Sudakov suppression for this soft radiation in \POWHEG{} largely fixes
this abnormal behaviour, leading to more physical predictions.

In parting we wish to make two further qualifying remarks regarding these
conclusions. Firstly, as one might expect, developing the \POWHEG{} events
more fully, by including the subsequent parton showering and hadronization,
depletes the jet transverse energies through emission of out-of-cone
radiation, lessening the level of disagreement. However, for the distribution
shown in fig.~\ref{fig:LHEF-NLO-et1_et2_gt_d}, the showered and unshowered
predictions always agree to within 15\%. Secondly, we note that it may be
tempting to think that the anomalous behaviour shown here by the fixed-order
results may somehow arise through a deficiency, e.g.~a lack infrared safety,
in the jet algorithm; yet we have repeated this exercise using the \SISCONE{}
($R=0.7$, $f=0.5$) and inclusive $\kt$ jet algorithms ($R=0.7$) finding
qualitatively the same features: a large deficit of the NLO prediction with
respect to the hardest-emission cross section, by at least a factor of two in
both cases, and a solitary, highly negative bin entry at $40.0<\pt<42.5$~GeV
in the transverse momentum spectrum of the underlying Born configuration.

\subsubsection{The invariant-mass distribution}
\begin{figure}
\begin{centering}
\epsfig{file=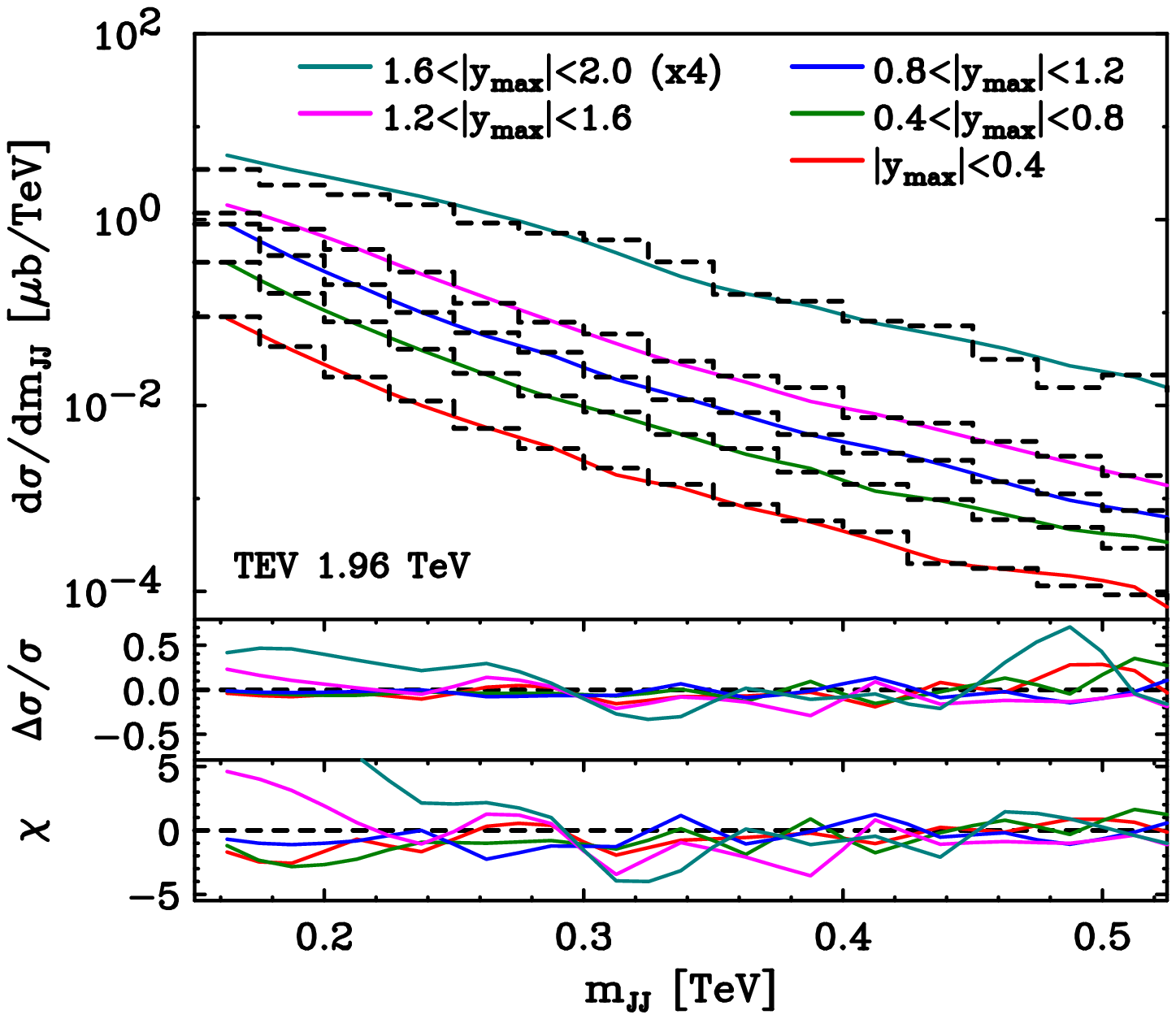,width=0.490\textwidth}
\hfill{}
\epsfig{file=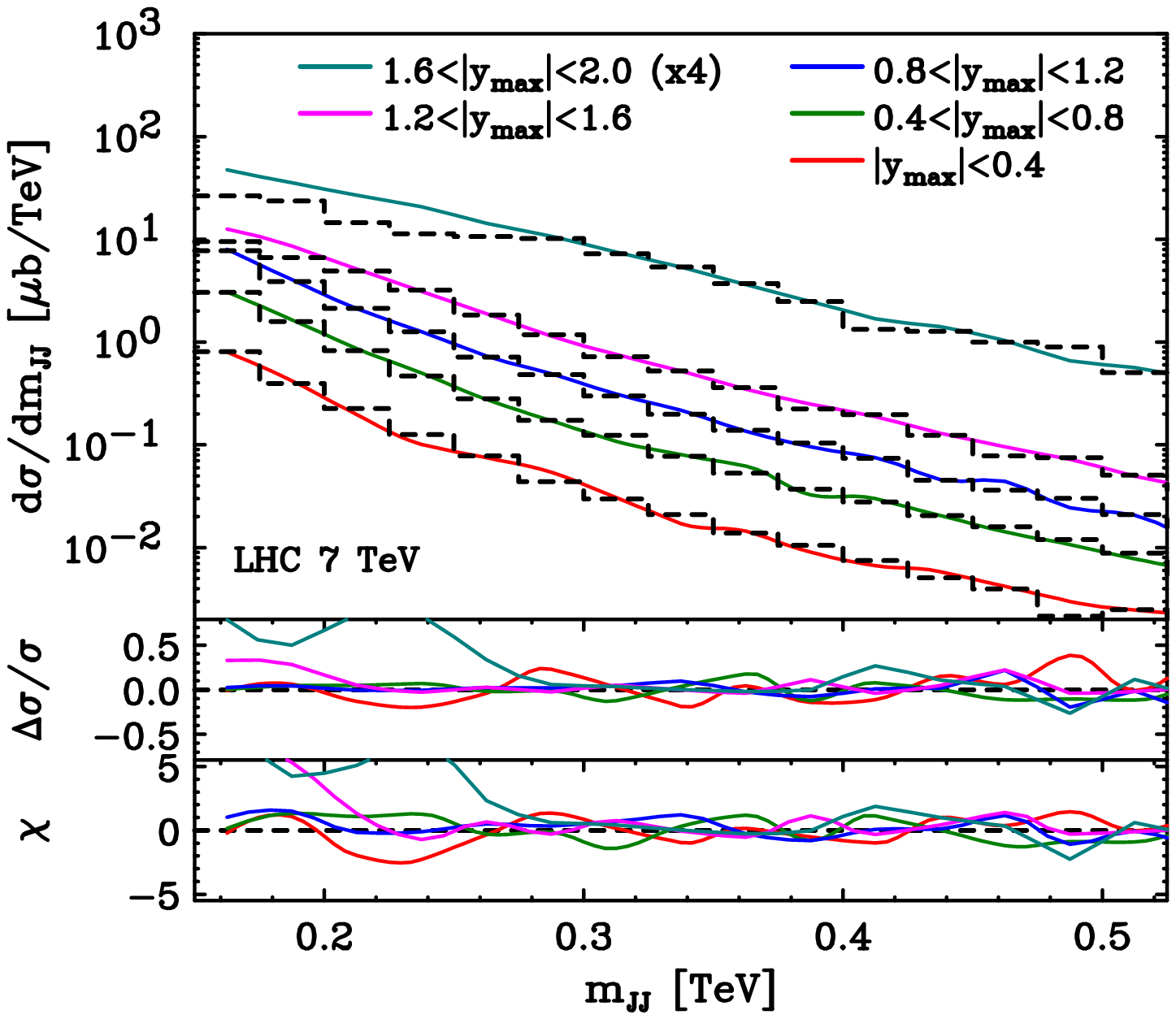,width=0.490\textwidth}
\par\end{centering}
\caption{\label{fig:LHEF-NLO-1} In coloured lines the \POWHEG{}
  hardest-emission distributions for the dijet invariant mass (prior to
  parton showering) with their analogous fixed-order NLO counterparts
  overlaid (black dashes). Coloured lines are ordered from bottom to top in
  increasing rapidity intervals.}
\end{figure}
As an interesting example of the effect of symmetric cuts on
experimentally-measurable distributions, we present in
fig.~\ref{fig:LHEF-NLO-1} the dijet invariant-mass distributions, resulting
from analysis of the \POWHEG{} hardest-emission events (coloured lines),
superimposed on the corresponding fixed-order predictions (black dashes), for
both Tevatron and LHC collider configurations.  In keeping with the analysis
of ref.~\cite{Abazov:2010fr}, we impose a symmetric 40~GeV
transverse-momentum cut on the two leading $\pt$ jets and we constrain the
rapidity of the most forward of these, $|y|=\max(|y_1|,|y_2|)$, to lie within
one of several bins (shown inset).  It is clear that these selection criteria
favour events in which the rapidities of the two highest $\pt$ jets have
opposite sign, their absolute value lying in the same bin.  The rapid fall in
the cross section as the invariant mass increases is, of course, expected
from simple phase space considerations, while the growth with increasing
$|y|$ reflects the fact that, at fixed invariant mass, higher jet rapidities
correspond to smaller angle, lower transverse-momentum scatterings, which
yield larger cross sections.

As with the case of the inclusive jet transverse-momentum spectrum, here we
find that the level of agreement between the \POWHEG{} and NLO predictions is
generally very good, with the exception of the highest rapidity bins in the
low invariant mass region, where the former is found to exhibit a 50\%{}
excess.  In light of our previous discussion on the effects of symmetric jet
$\Et$ cuts this excess is readily explainable. Note that the invariant mass
spectrum shown here begins at 160~GeV, corresponding to the production of
pairs of jets in the central region with transverse energies of around
80~GeV.  However, should the leading jets be produced at high rapidities, the
cross section will be dominated by pairs of back-to-back jets with smaller
$\pt$ and a larger rapidity separation. It is easy to check that for
rapidities around 1.6 the transverse momentum of the jets in the back-to-back
configuration approaches 40~GeV. The effects of the symmetric $\pt$ cuts then
play an increasingly significant role and so the previously discussed
\POWHEG{} excess over fixed-order predictions for such cuts becomes visible.

\subsubsection{Jet cross sections with a cut on a single jet}
\begin{figure}
\begin{centering}
\epsfig{file=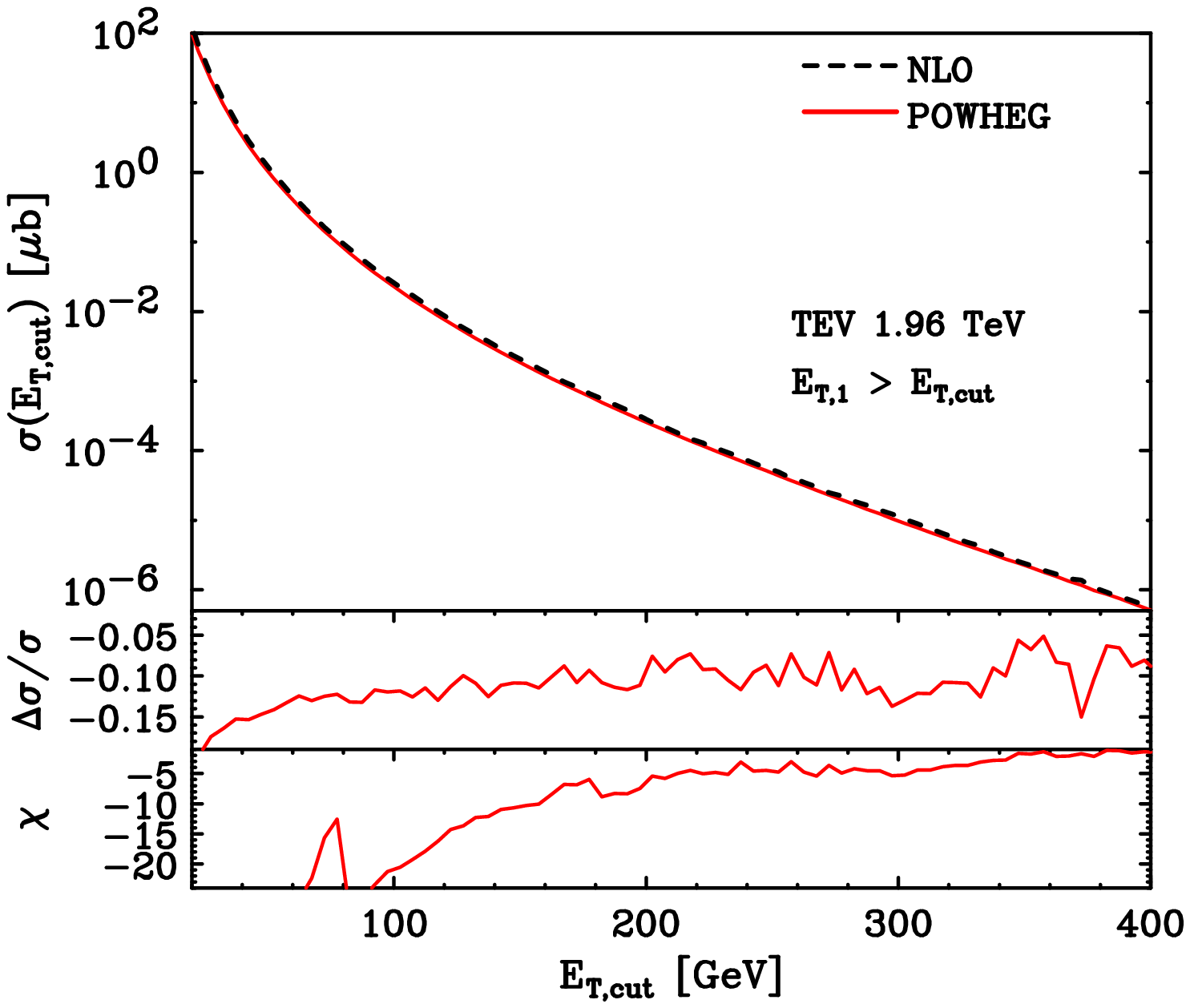,width=0.490\textwidth}
\hfill{}
\epsfig{file=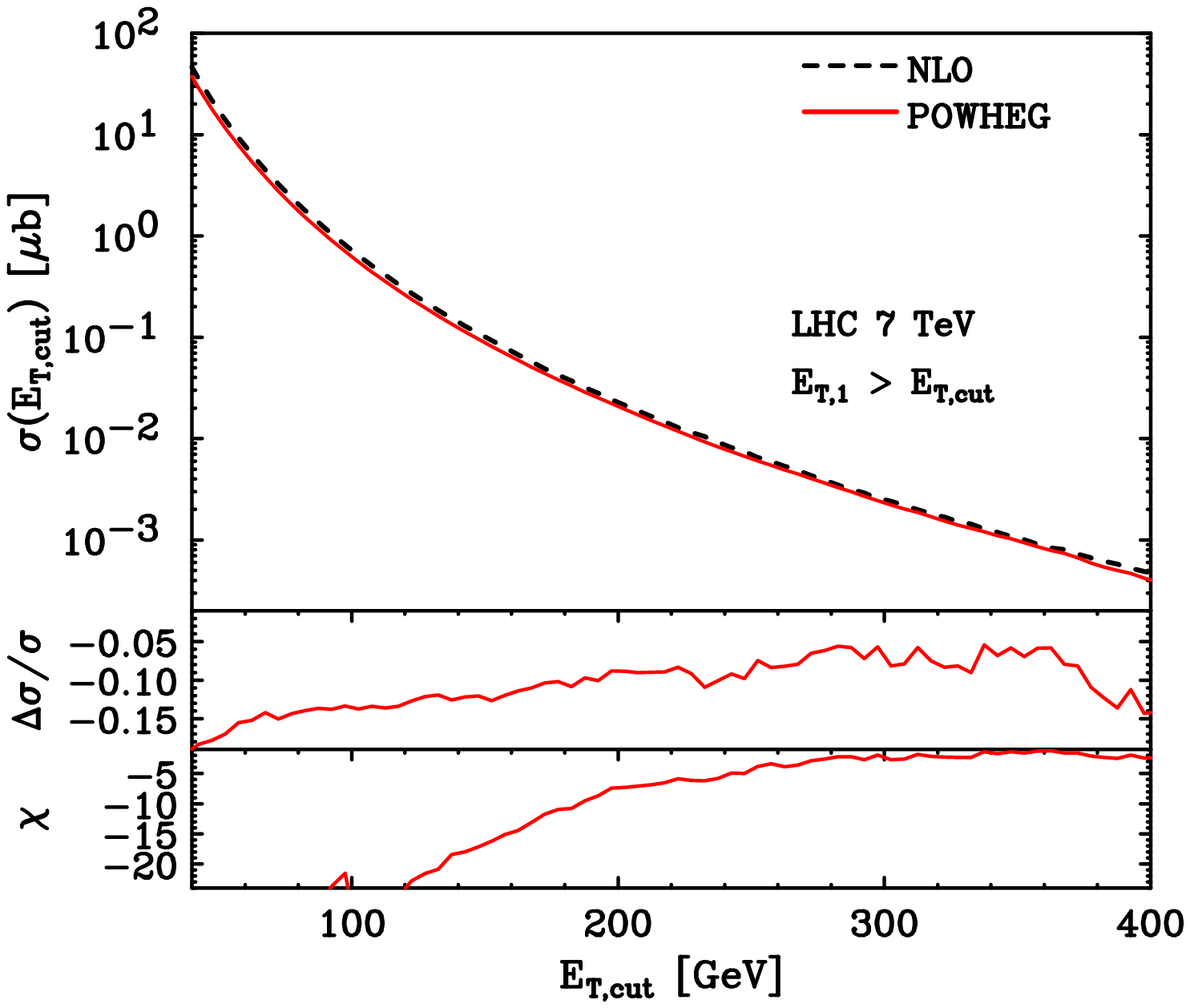,width=0.490\textwidth}
\par\end{centering}

\caption{\label{fig:LHEF-NLO-et1_gt_d} As in
  fig.~\ref{fig:LHEF-NLO-et1_et2_gt_d}, here we compare fixed-order NLO
  cross sections (black dashes) to the corresponding \POWHEG{} hardest-emission
  cross section (solid red), however, in this case we have applied the jet
  transverse-energy cut to the highest-$\Et$ jet alone.}
\end{figure}

As stated previously, the frailties of NLO calculations for dijet production
with symmetric cuts were first noted quite some time ago.  This has led to
the general consensus that asymmetric cuts on jet transverse energies should
be used in theoretical studies and in confronting data with next-to-leading
order QCD predictions. This being so, we then compare the \POWHEG{}
hardest-emission cross section against fixed-order predictions also for
asymmetric cuts. We reiterate that, for inclusive quantities, insensitive to
Sudakov effects in the radiation of the third jet, the two sets of results
should exhibit deviations from one another not greater than those expected
relative to the corresponding NNLO predictions.

In fig.~\ref{fig:LHEF-NLO-et1_gt_d} we plot the total cross section as a
function of $E_{\sss\rm T,\mathrm{cut}}$, defined to be the cut on the
transverse energy of the highest transverse-energy jet: $E_{\sss\rm
  T,1}>E_{\sss\rm T,\mathrm{cut}}$. This is a more inclusive quantity with
respect to the case of symmetric cuts (but still less inclusive with respect
to the jet inclusive cross section). As with
fig.~\ref{fig:LHEF-NLO-et1_et2_gt_d_UB_plots}, the fixed-order results are
given by the dashed black lines while those from the \POWHEG{}
hardest-emission cross section are shown in red.  In this case, the
$\Delta\sigma/\sigma$ and $\chi$ distributions, in the lower panels, reveal a
statistically significant excess of the fixed-order prediction over that of
the resummed calculation, in the region of 10\% to 20\% across most of the
range in $E_{\sss\rm T,\mathrm{cut}}$.  This level of disagreement may not be
particularly bothersome in the context of a differential distribution in some
infrared safe quantity, and one can expect that parton showering and
hadronization may combine to alter this picture by a similar amount.
However, these differences occur at the level of the total inclusive cross
section subject to a single cut, $E_{\sss\rm T,1}>E_{\sss\rm
  T,\mathrm{cut}}$, and therefore demand some degree of scrutiny.

\subsubsection*{The r\^ole of the underlying Born}
\begin{figure}
\begin{centering}
\epsfig{file=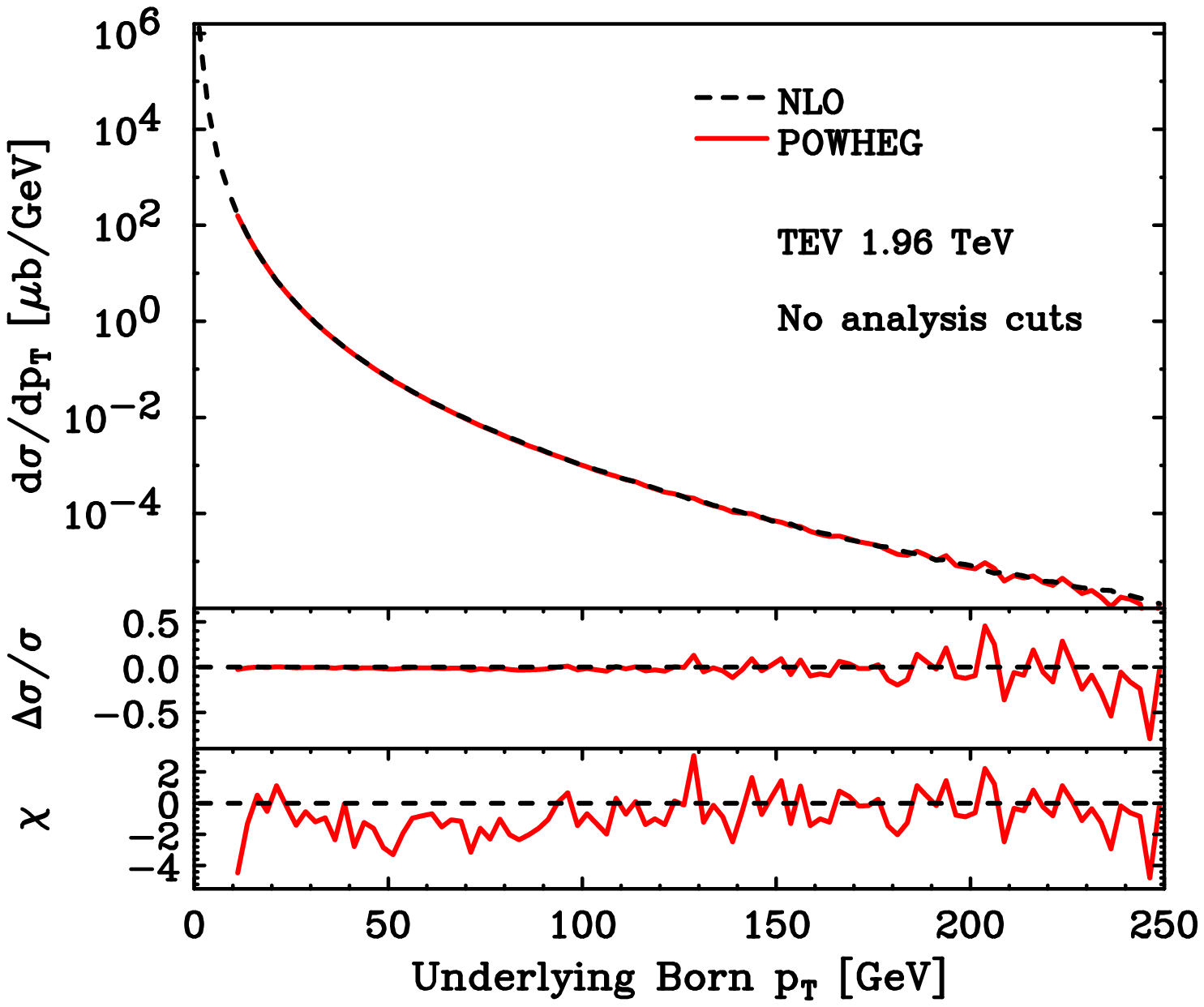,width=0.490\textwidth}
\hfill{}
\epsfig{file=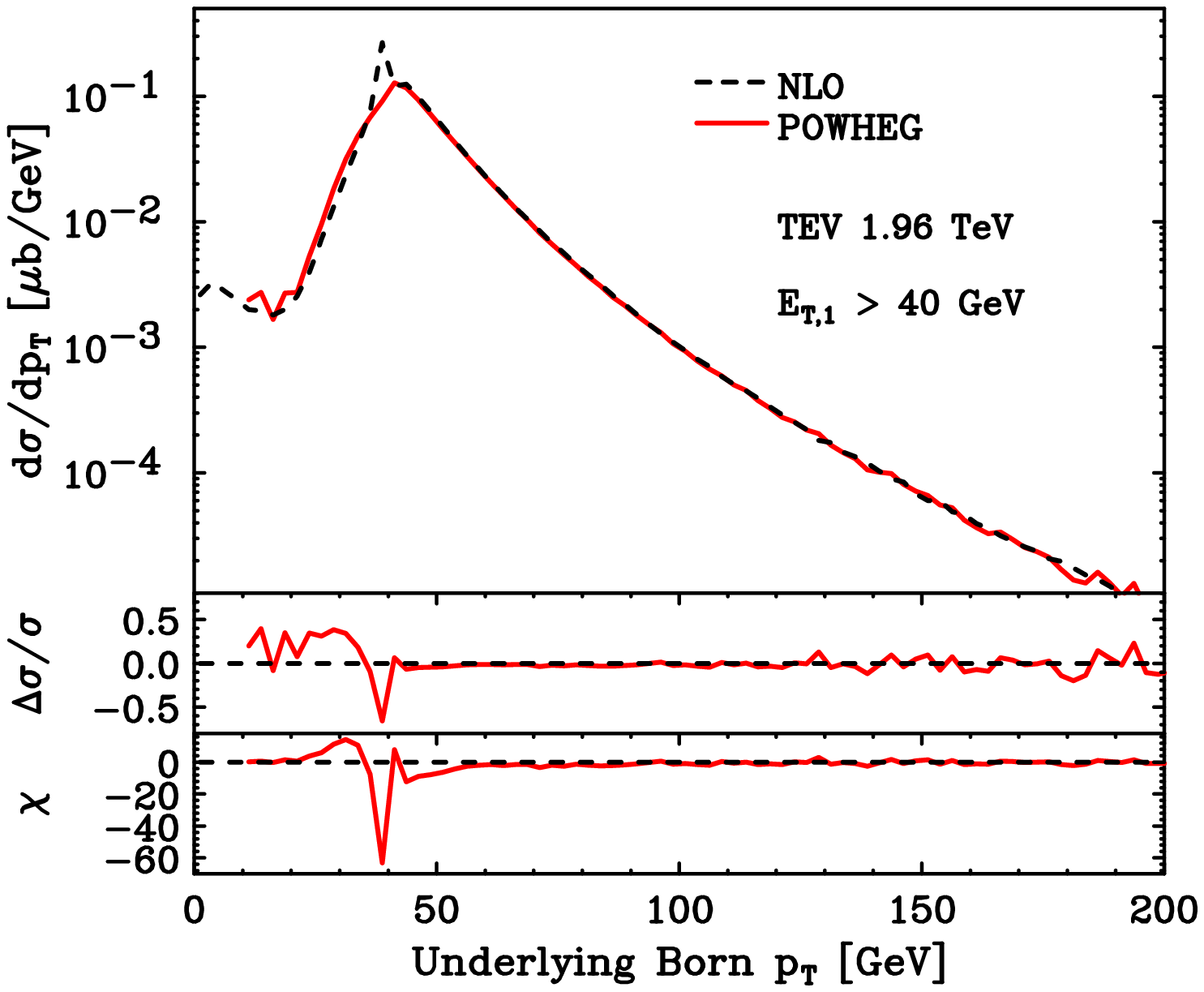,width=0.490\textwidth}
\par\end{centering}
\begin{centering}
\epsfig{file=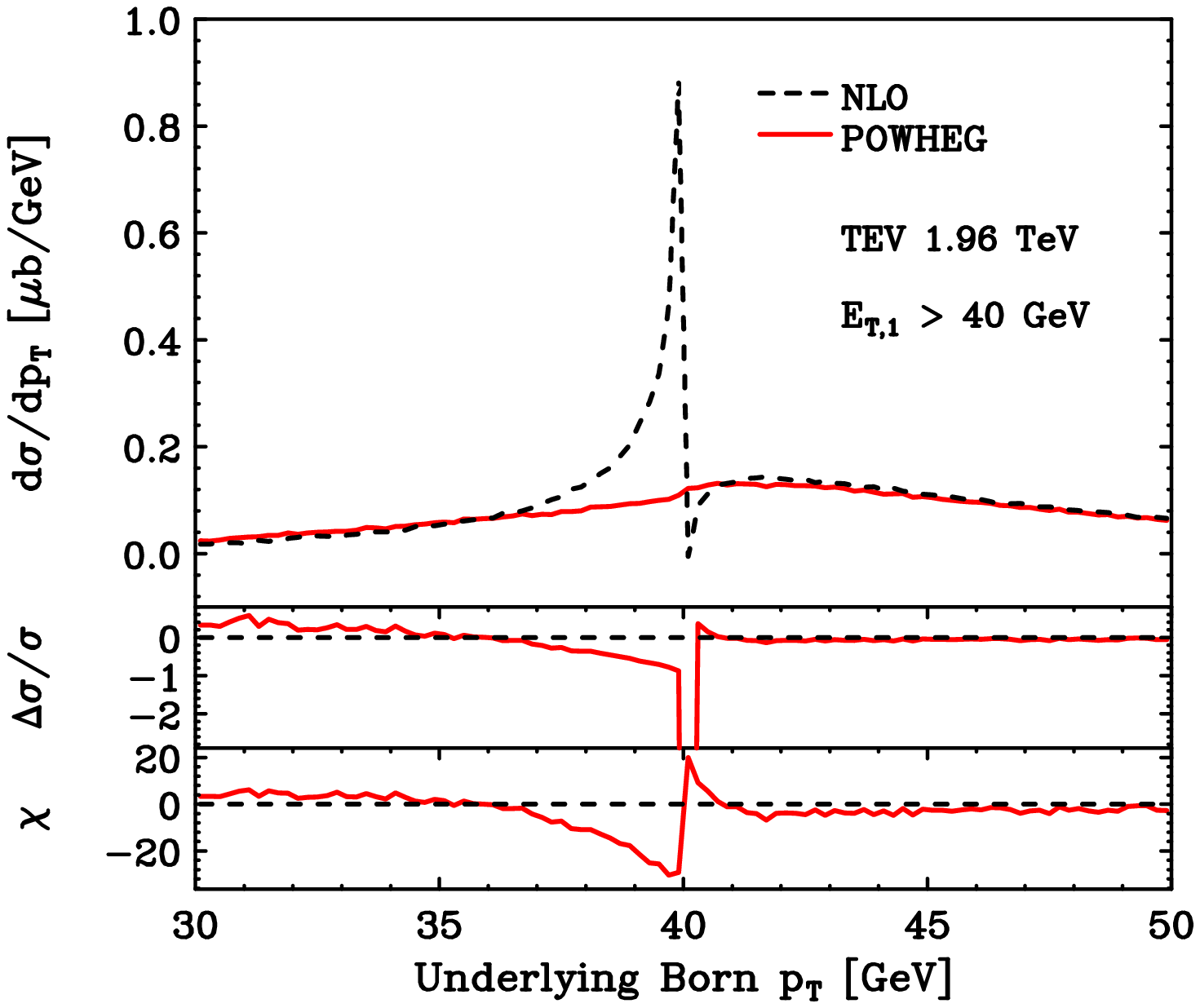,width=0.490\textwidth}
\hfill{}
\epsfig{file=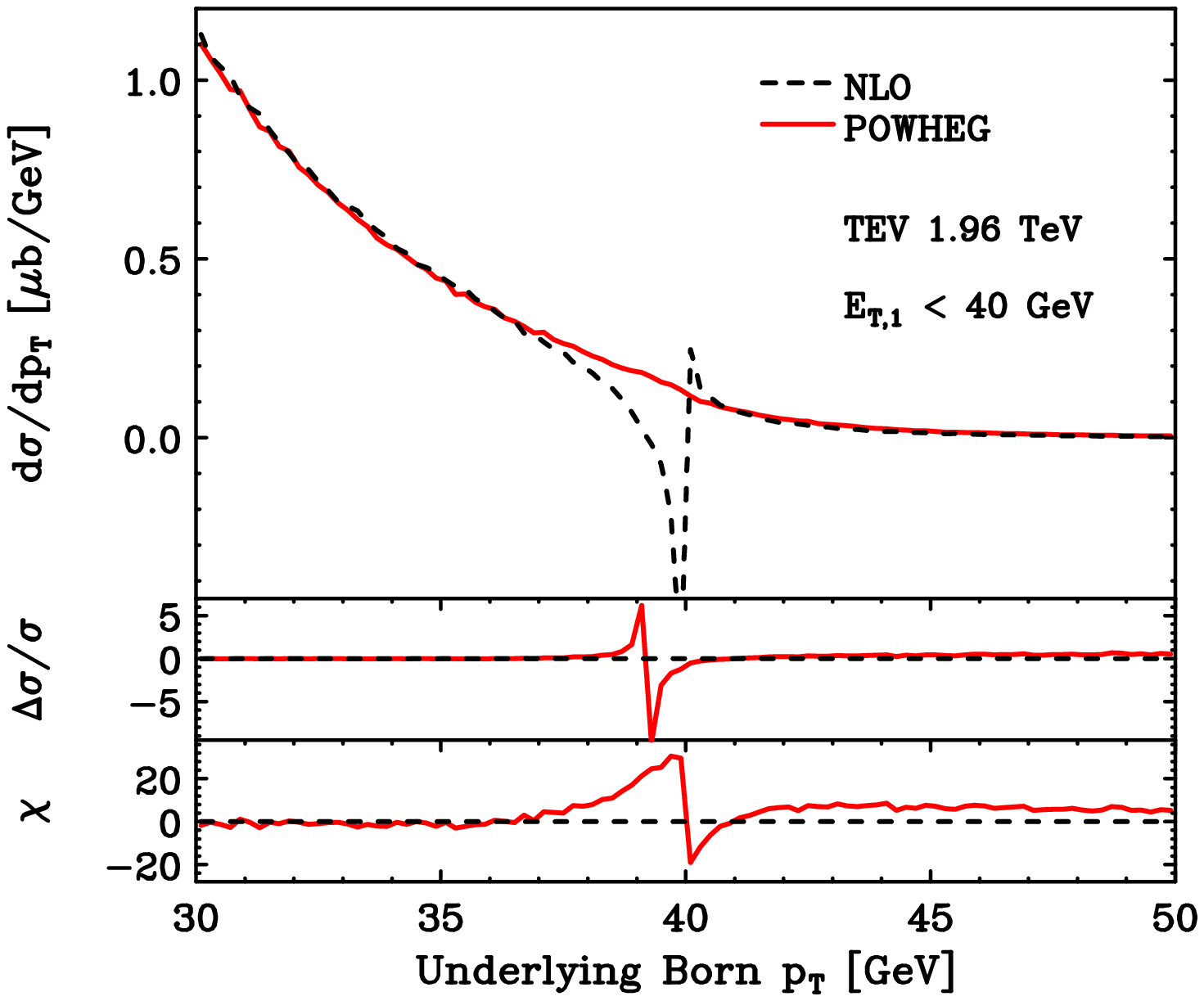,width=0.490\textwidth}
\par\end{centering}
\caption{\label{fig:LHEF-NLO-et1_gt_d_UB_plots} The transverse-momentum
  spectra for the underlying Born configuration having applied a cut only on
  the highest $\Et$ jet. The results from \POWHEG{} hardest-emission events
  are shown in solid red, while the corresponding fixed-order NLO predictions
  are drawn as dashed black lines.}
\end{figure}

As we have remarked before, since the underlying Born variables, $\Phi_{B}$,
are generated by the exact same mechanism in the fixed order and \POWHEG{}
predictions, the source of any disparity between the two must
follow from the different distributions of the radiative variables,
$\Phi_{R}$, governing the kinematics of real radiation, for a given
$\Phi_{B}$. Hence, here again, we choose to interrogate the events passing and
or failing the jet $\Et$ cut regarding the transverse momentum of their
underlying Born kinematics.

In fig.~\ref{fig:LHEF-NLO-et1_gt_d_UB_plots} we plot the same set of
distributions as in fig.~\ref{fig:LHEF-NLO-et1_et2_gt_d_UB_plots} for the
case of a cut on the $\Et$ of the leading jet alone, assuming a
$\sqrt{s}=1.96$~TeV proton-antiproton collider.  The
resulting picture reveals many of the same features seen in the analogous
symmetric $\Et$ jet cut analysis. The integrated bin contents confirms that
the prediction from the hardest-emission cross section is 15\% below that of
the fixed-order result. Moreover, we find that the difference due to the fact
that the region below 10~GeV is unpopulated by \POWHEG{} -- due to the use of
a higher generation cut -- accounts for only 1\% of this. It turns out that
the remaining 14\% of the excess is attributed to a lone spike in the fixed
order distribution, first seen in the bin at $37.5<\pt<40.0$~GeV,
in the plot shown in the upper right-hand corner of
fig.~\ref{fig:LHEF-NLO-et1_gt_d_UB_plots}. This region of the distribution is
shown magnified in the plot in the bottom left corner of the same figure,
with the complementary distribution for events failing the
$\Et>40$~GeV cut shown alongside it.

Remarkably we see, once again, that the fixed-order distribution is prone to
pathological behaviour, being discontinuous in its first derivative at
$\pt=40$~GeV, while the corresponding \POWHEG{} result is smooth and
physical. The fixed-order distributions show how events, in which the $\pt$
of the underlying Born kinematics is below, yet close to, the cut, migrate
above it, through the emission of radiation. The related, negative weight,
two-body counter-events have the kinematics of the underlying Born
configuration, hence, these remain below the cut where they give rise to a
negative cross section for events {failing} it in the region near
40~GeV. The fact that the distribution diverges as the 40~GeV cut is
approached clearly indicates that the majority of the upward migration is due
to soft radiation. Of course, the rate of these soft emissions in the fixed
order calculation is highly erroneous, containing no Sudakov suppression
factor, unlike the \POWHEG{} hardest-emission cross section. This explains
why a similar diverging and discontinuous distribution is not present in its
associated predictions.

Once again, we may conclude here that the predictions of the hardest-emission
cross section offer a considerably improved description with respect to their
fixed-order counter parts. Nevertheless, in this case, unlike in the case of
symmetric jet $\Et$ cuts, this point should not be overemphasised, since it
must be remembered that the differences 
were only at the level of 10-15\%. Parton showering and hadronization, in the
case of \POWHEG{}, and non-perturbative correction factors, in the case of
fixed-order computations, contribute to shift these predictions by similar
degrees.

Before continuing we wish to allay any concerns which the reader may have as
to the physical nature of the analyses depicted in
figs.~\ref{fig:LHEF-NLO-et1_et2_gt_d_UB_plots}
and~\ref{fig:LHEF-NLO-et1_gt_d_UB_plots}.  In particular, one might well
wonder to what extent the transverse momentum of the underlying Born
configuration is a {physical} quantity and so query the validity of our
explanations.  Taking into account the parton showering, hadronization and
underlying event effects that occur in reality, it is certainly the case that
this quantity is not experimentally measurable. However, from the point of
view of the three-body events originating from the NLO and \POWHEG{}
hardest-emission cross sections, one can construct effectively a jet
algorithm, based on the \POWHEG{} phase-space factorisation and mappings,
which clusters them back to a two-body underlying Born configuration.

In any case, to quell any doubts that may have arisen, we point out that we
have repeated the analysis surrounding these figures plotting, instead of the
underlying Born $\pt$, the average transverse momentum of a pair of jets,
$\left<\pt\right>$, obtained by applying the {exclusive} $\kt$ jet algorithm,
demanding it returns always just two jets, using the $\Et$ recombination
scheme (rather than the default $E$ scheme). This effectively projects the
three-body real-emission kinematics to a massless two-body configuration, as
in the underlying Born $\Phi_{B}$. Clearly this quantity is very closely
related to the underlying Born transverse momentum, converging to it in the
limit of soft and collinear emissions.  It should then come as no surprise
that, in so doing, we see essentially the same distributions, with the same
structure and features, as in
figs.~\ref{fig:LHEF-NLO-et1_et2_gt_d}--\ref{fig:LHEF-NLO-et1_gt_d_UB_plots}.
In particular we note the continuing presence of the discontinuities in the
fixed-order
predictions. At a quantitative level, the fact that we used the $\kt$ jet
clustering algorithm to define all jets in the analysis changes things but
not in such a way as to alter our conclusions. Whereas in the symmetric $\Et$
cut case, with the D0 midpoint cone algorithm, the cross section for
$E_{\sss\rm T,1}>40$~GeV and $E_{\sss\rm T,2}>40$~GeV revealed a deficit of
105\% with respect to the \POWHEG{} prediction, falling to 3\% on neglecting
the large negative bin at $40.0<\pt<42.5$~GeV in the underlying Born $\pt$
distribution, when using the $\kt$ algorithm, with $R=0.7$, the deficit was
instead 250\% reducing to 22\% on omitting the same bin in the
$\left<\pt\right>$ distribution. In the case of the lone $E_{\sss\rm T,1}$
cut scenario, the fixed-order prediction with the D0 midpoint cone algorithm
exhibited a 16\% excess which dropped to 1\% when omitting the
$37.5<\pt<40.0$~GeV bin in the underlying Born $\pt$ distribution, while
using the $\kt$ algorithm we find the excess is 16\%, reducing to less than
4\% when the same bin is omitted from the $\left<\pt\right>$
distribution. Thus our conclusions based on analysing the underlying Born
transverse momentum distribution can certainly be understood as being
unequivocally physical.

Before leaving this discussion, we would like to point out that,
in a recent publication~\cite{Rubin:2010xp}, the problem of large, unphysical
NLO corrections in dijet production has also been considered from a
different perspective. It will be interesting to compare our findings
with those of that work.

\subsubsection{Features of the hardest emission}
\begin{figure}
\begin{centering}
\epsfig{file=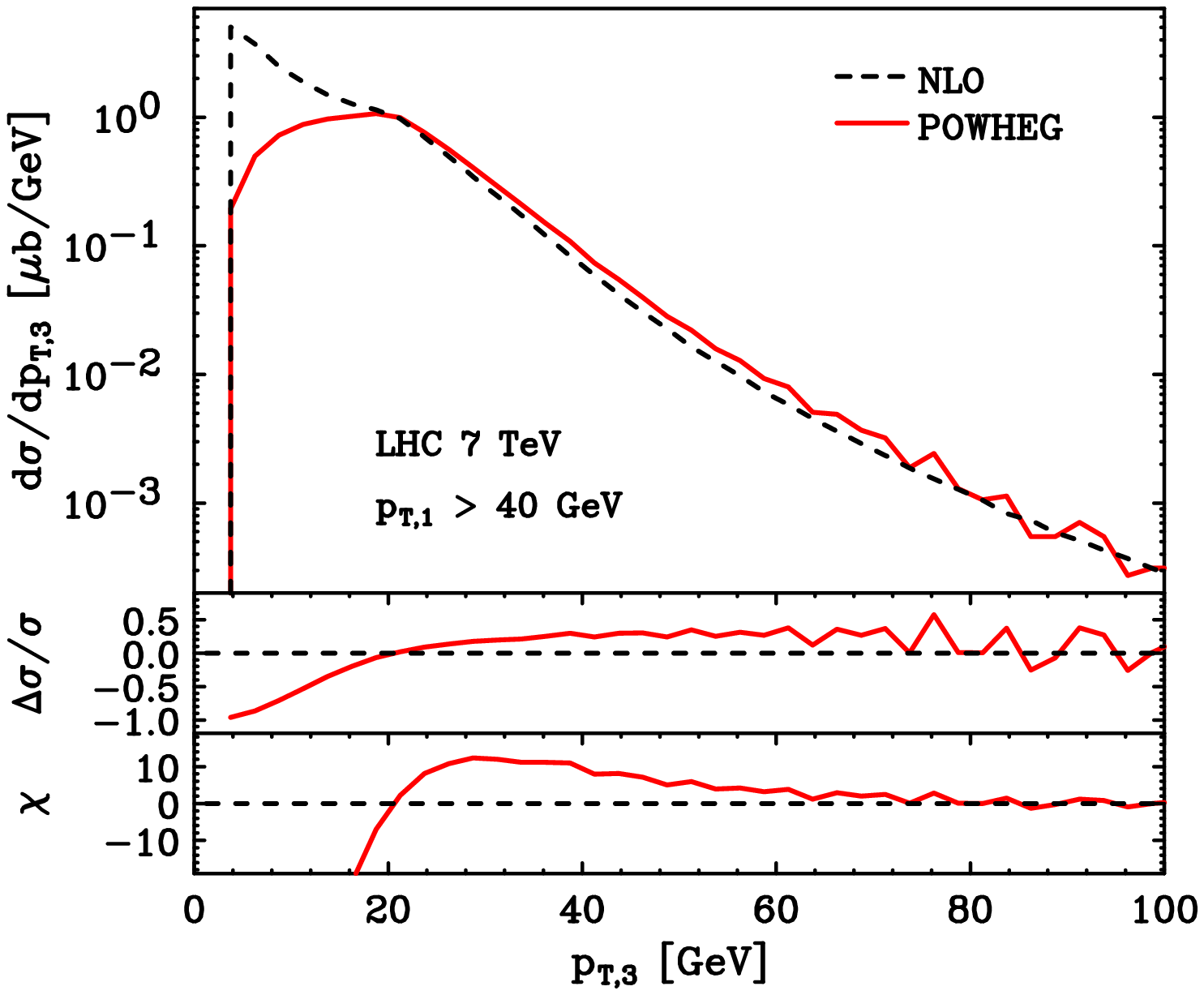,width=0.490\textwidth}
\epsfig{file=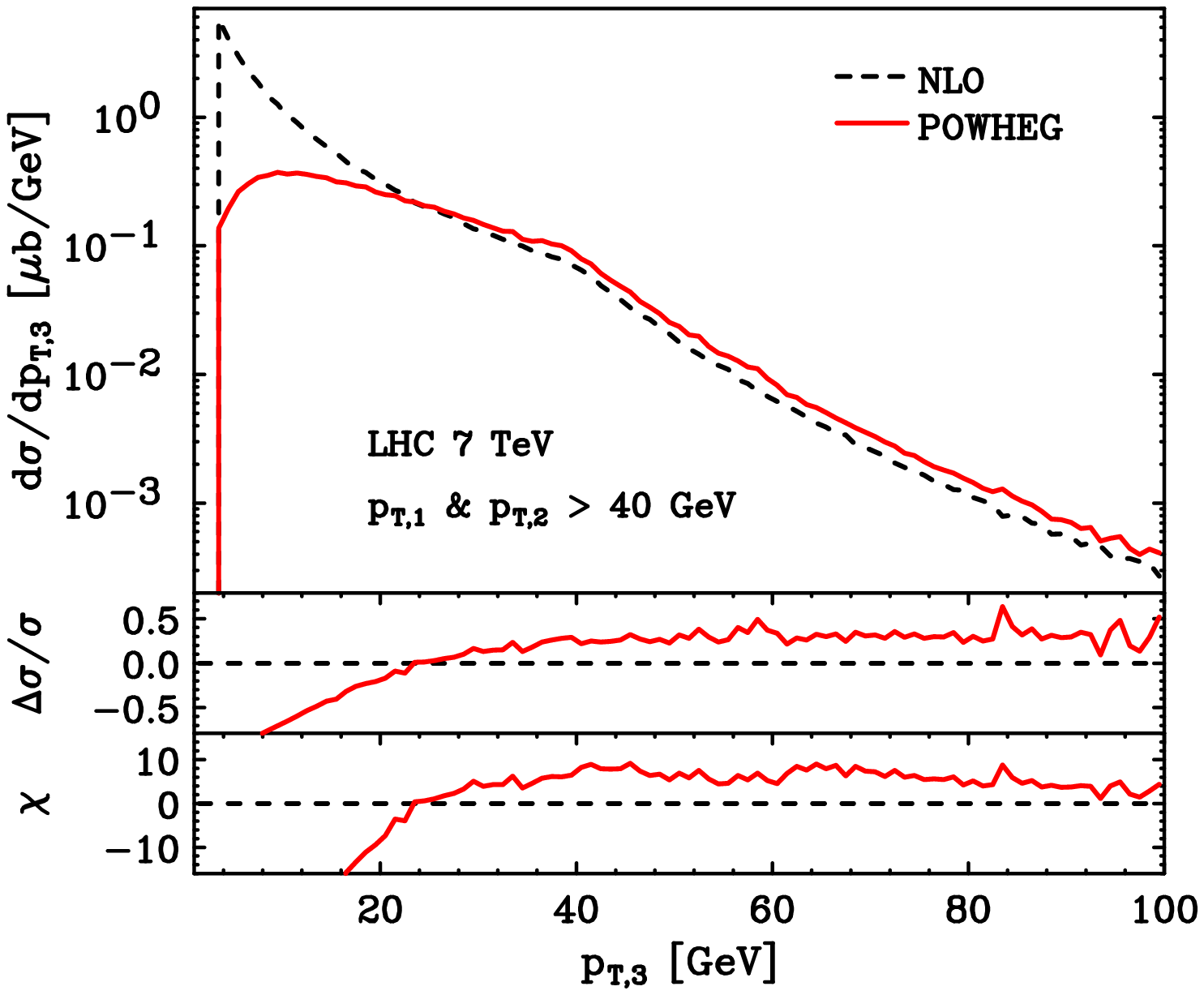,width=0.490\textwidth}
\par\end{centering}
\caption{\label{fig:lhef-nlo-pt3} The transverse-momentum spectrum of the
  third hardest jet as given by the NLO cross section (black dashes) and
  \POWHEG{} without parton showering (solid red). In the left-hand plot we have
  applied a transverse-momentum cut on the hardest (highest $\pt$) jet
  alone, while on the right-hand side a symmetric cut on the two hardest
  jets has been used.}
\end{figure}

In dijet production, the radiation generated by \POWHEG{} is
generally softer than the other two jets. In order to investigate its
radiation pattern, in the left plot of fig.~\ref{fig:lhef-nlo-pt3} we
display the transverse momentum of the third highest $p_{T}$ jet at the LHC,
in events where the leading jet has $\pt$ greater than 40~GeV, while on the
right we show the same spectrum, with the second jet also subject to
the same cut. A slight kink is visible at around 20~GeV in the former case
and at 40~GeV in the latter.  In the first instance, this feature can be
ascribed to the fact that the majority of selected events will be such that
the hardest jet is near the cut, with the balancing second and third jets
having a relatively small angular separation, thus bounding the $\pt$ of the
third jet to be less than 20~GeV. In the second case, the fact that the cross
section falls rapidly as the transverse momentum of the two leading jets
increases, favours them having $\pt$ close to 40~GeV in events passing the
cut, hence, by definition, the $\pt$ of the third jet will tend to be limited
to this value. These arguments are easier to understand by considering the
fixed-order predictions, since resummation effects in the \POWHEG{} case add
another layer of subtlety in the low $\pt$ region.  Lastly, we point out that
the vanishing of both the NLO and \POWHEG{} distributions below 3~GeV arises
from our use of the D0 jet algorithm, which discards jets with transverse
energy below 3~GeV.

With the origins of the kinks understood we can safely say that, for both
distributions shown, the differences of the NLO predictions with respect to
those of \POWHEG{} display the usual features. In the low-$\pt$ region we see
that the NLO results tend to diverge while those of \POWHEG{} are affected by
Sudakov damping. At larger transverse momenta, the Sudakov suppression
effects disappear and the \POWHEG{} result tends to the fixed-order one,
multiplied by a factor of
$\bar{B}\left(\Phi_{B}\right)/B\left(\Phi_{B}\right)$ from the hardest
emission cross section \cite{Nason:2004rx,Frixione:2007vw}. Note that, from
the point of NLO accuracy, the presence or absence of such a factor is
formally irrelevant, since, in the high-$\pt$ regime, the distribution of
events is governed by the real emission cross section, about which the
$\bar{B}\left(\Phi_{B}\right)/B\left(\Phi_{B}\right)$ factor produces
modifications of NNLO significance only. For a more detailed explanation of
this point, in the context of Higgs boson production via gluon fusion, we
refer the reader to refs.~\cite{Alioli:2008tz,Hamilton:2009za}.

\subsubsection[Jet structure: the $\pt^{\rm rel}$ distribution]
{Jet structure: the $\boldsymbol{\pt^{\rm rel}}$ distribution}
\begin{figure}
\begin{centering}
\epsfig{file=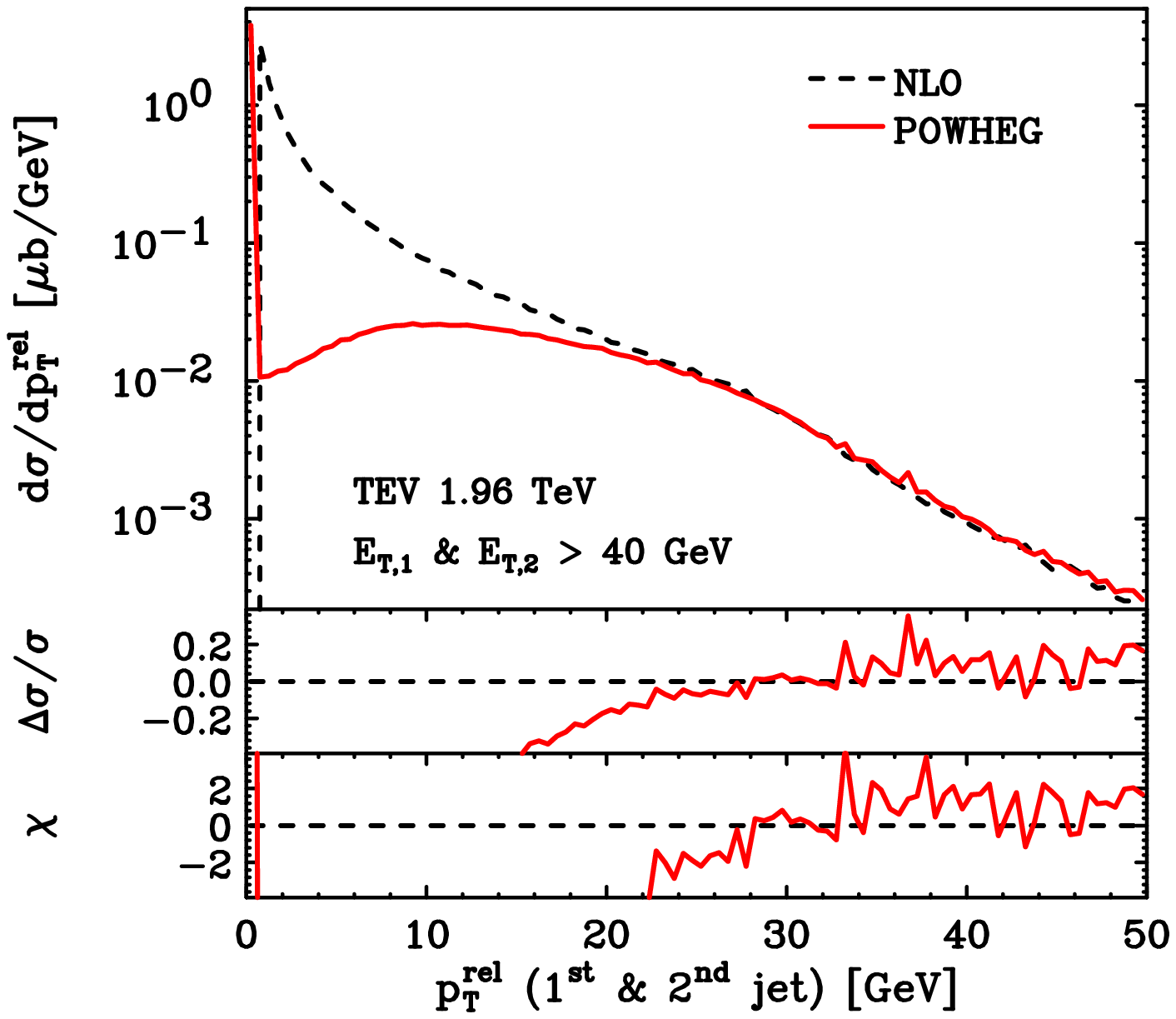,width=0.490\textwidth}
\hfill{}
\epsfig{file=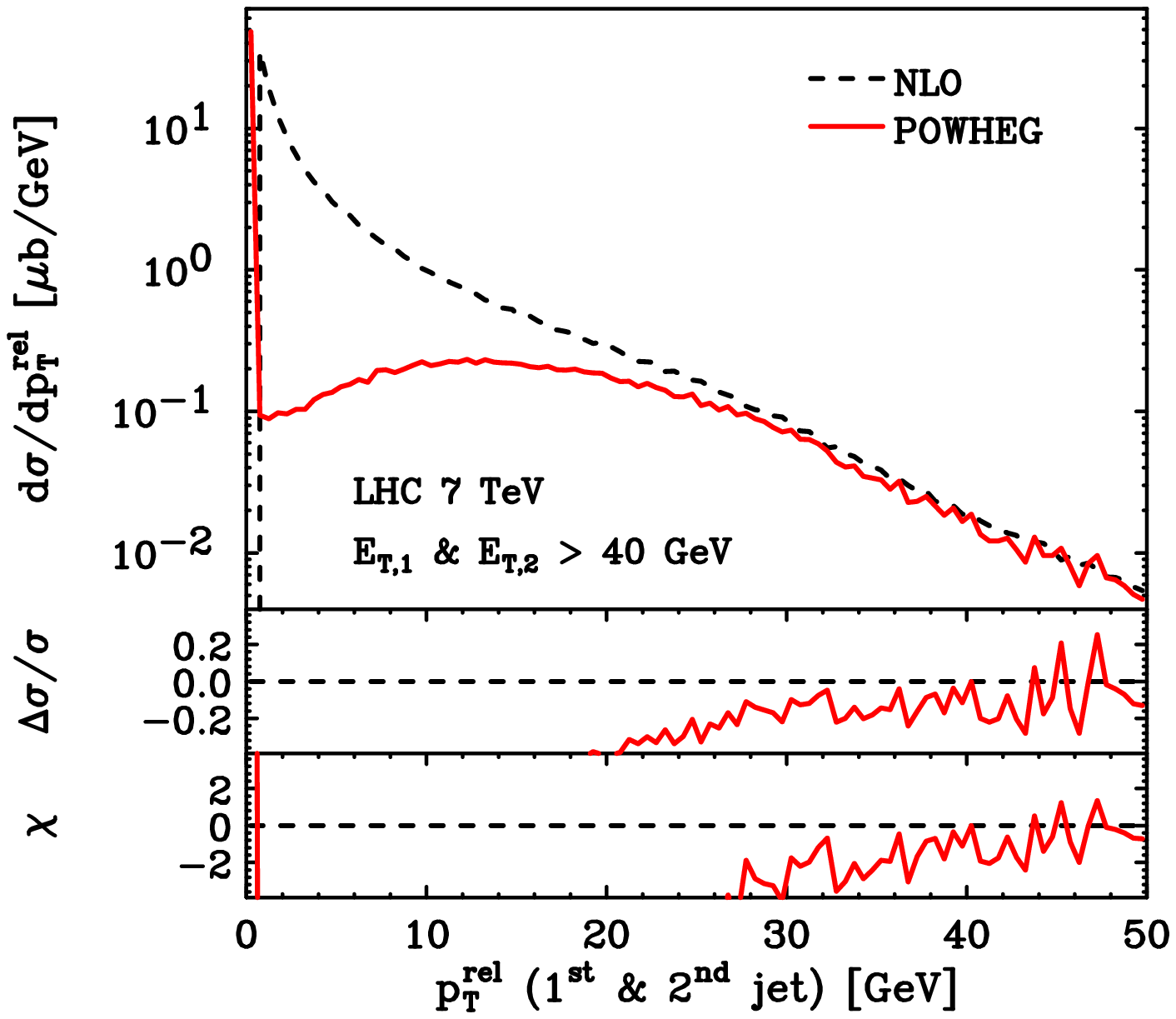,width=0.490\textwidth}
\par\end{centering}
\caption{\label{fig:lhef-nlo-ptrel} On the left- and right-hand side we show
  the distribution of the $\pt^{\rm rel}$ distribution for the two highest
  transverse-momentum jets, as given by \POWHEG{} (solid red) and NLO QCD
  (black dashes), at the Tevatron and LHC.}
\end{figure}
In order to investigate the jet structure, in fig.~\ref{fig:lhef-nlo-ptrel}
we show the scalar sum of the relative transverse momenta of the particles in
the $j^{\rm th}$ jet, $\pt^{{\rm rel},j}$ ($j=1,2$),
defined with respect to the jet axis, in the frame
where the jet has zero rapidity:
\begin{equation}
\pt^{{\rm rel},j}=\sum_{i\in j} \frac{| \vec{k}_i \times
  \vec{p}_{j}|}{|\vec{p}_{j}|}\,,
\end{equation}
where $k_i$ denotes the momentum of the $i^{\rm th}$ particle, and 
$p_j$ is the momentum of the $j^{\rm th}$ jet.
As throughout this section, we
have compared the predictions of the \POWHEG{} hardest-emission cross section
(solid red) against the corresponding fixed-order predictions (black dashes).
Each of the two hardest jets in the event gives rise to an entry in this
histogram, however, in dealing with the \POWHEG{} hardest-emission events and
their fixed-order counterparts, the final-states only consist of two or three
partons. This being so, in the first bin we see an accumulation of events in
the \POWHEG{} case and a negative result in the fixed-order prediction due to
two-parton counter events. All other bins are filled by three-parton events,
which are clustered into two jets.

In both plots we see that the fixed-order result exhibits a mild
(logarithmic) divergence for small values of $\pt^{\rm rel}$. On the other
hand, the \POWHEG{} prediction displays strong Sudakov damping and a sharp
positive peak for $\pt^{\rm rel}=0$.  Although the \POWHEG{} predictions for
particularly exclusive observables (like the one we are considering here) may
appear better behaved than the pure NLO one, they are also plainly
unphysical, as evidenced by the first bin of these histograms. We reiterate
that, in terms of radiation, these bare \POWHEG{} events contain only the
hardest emission. Besides the erroneous peak, the Sudakov suppression here is
also somewhat spurious, since it is not simply due to the inhibition of
radiation around the leading jet axis but also to the fact that no harder
radiation is allowed to come from the other initial-state or final-state
partons. Only after the \POWHEG{} output is interfaced to a parton shower can
the shape of the Sudakov region be correctly modeled, with the peak at
$\pt^{\rm rel}=0$ also disappearing.

These $\pt^{\rm rel}$ distributions remind us the limitations of the
\POWHEG{} hardest-emission cross section alone: despite offering, in general,
a greatly improved description with respect to fixed-order methods, it will
naturally fail to describe exclusive observables sensitive to the emission of
more than one parton.

\subsection{Parton showering and hadronization}
\label{sec:showered_events}
In this section we investigate the effects of showering and hadronization on
the \POWHEG{} results discussed previously. When showering the hardest emission
event with \PYTHIA~6.4.21~\cite{Sjostrand:2006za} and
\HERWIG{}~6.510~\cite{Corcella:2000bw,Corcella:2002jc}  we
have used their default settings, with no underlying event and multiple-parton
interactions. Jet reconstruction was performed using jet algorithms and
parameters specified at the beginning of section~\ref{sec:theory_bit}.

\subsubsection{Inclusive distributions}
\begin{figure}
\begin{centering}
\epsfig{file=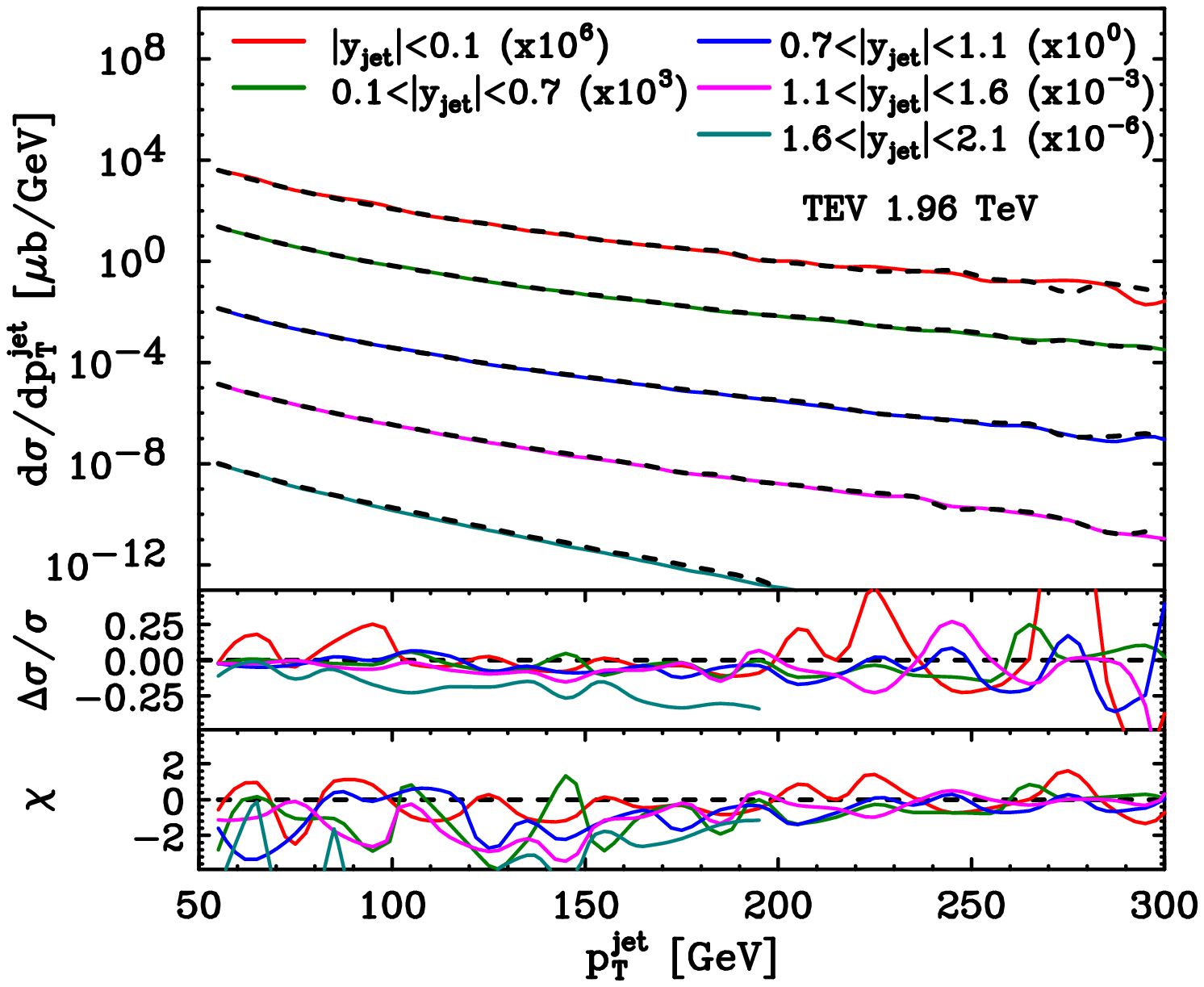,width=0.490\textwidth}
\hfill{}
\epsfig{file=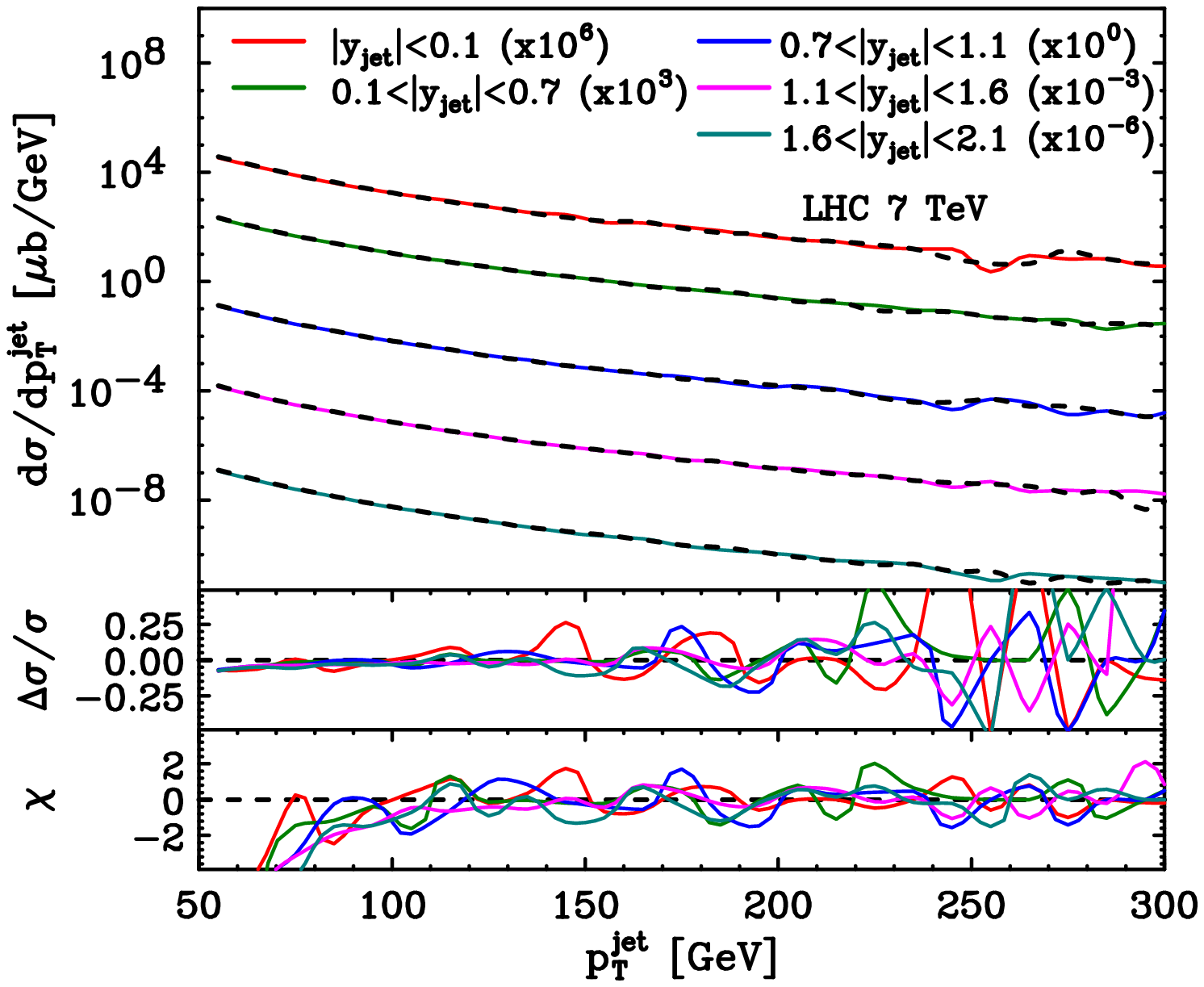,width=0.490\textwidth}
\par\end{centering}
\begin{centering}
\epsfig{file=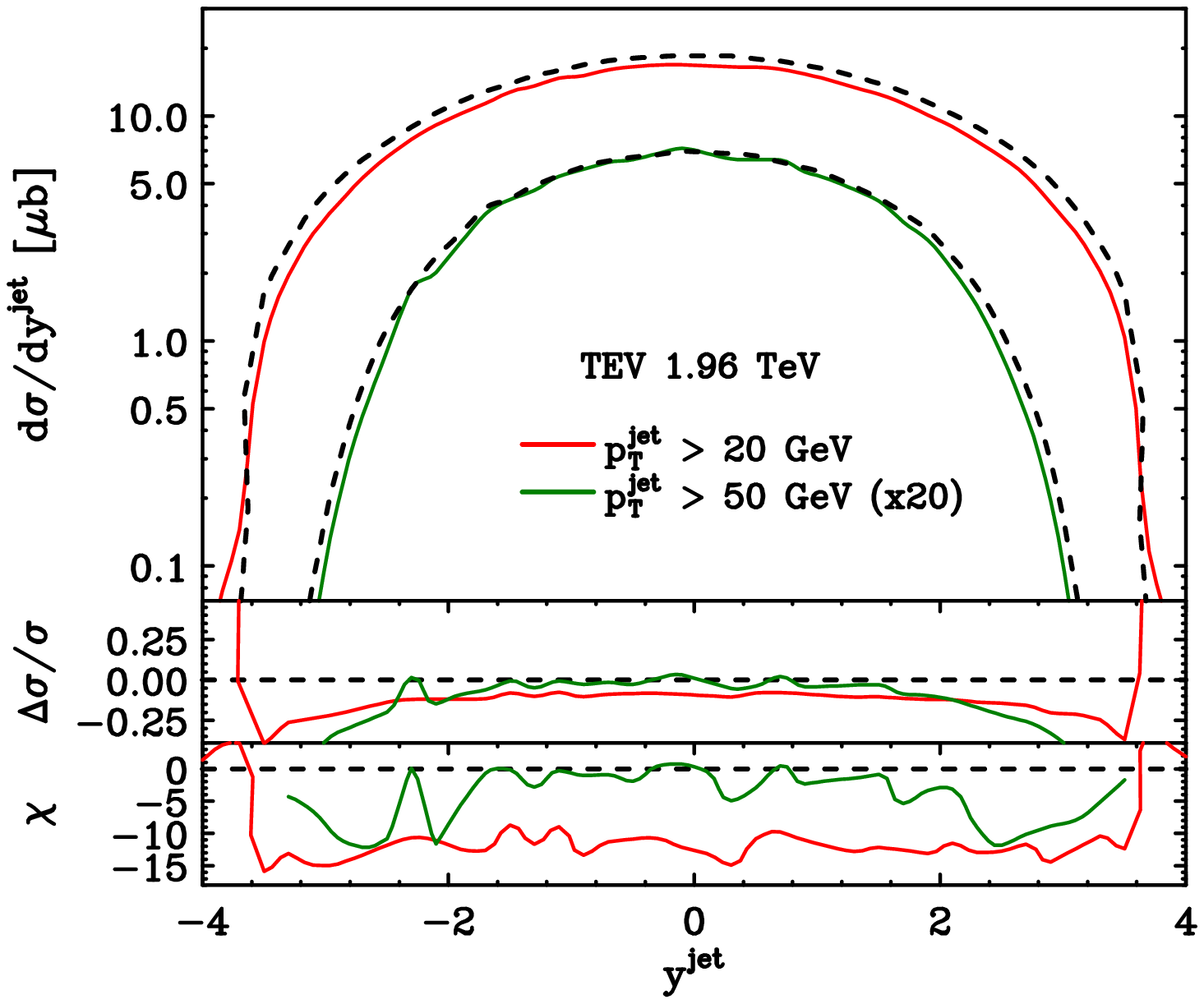,width=0.490\textwidth}
\hfill{}
\epsfig{file=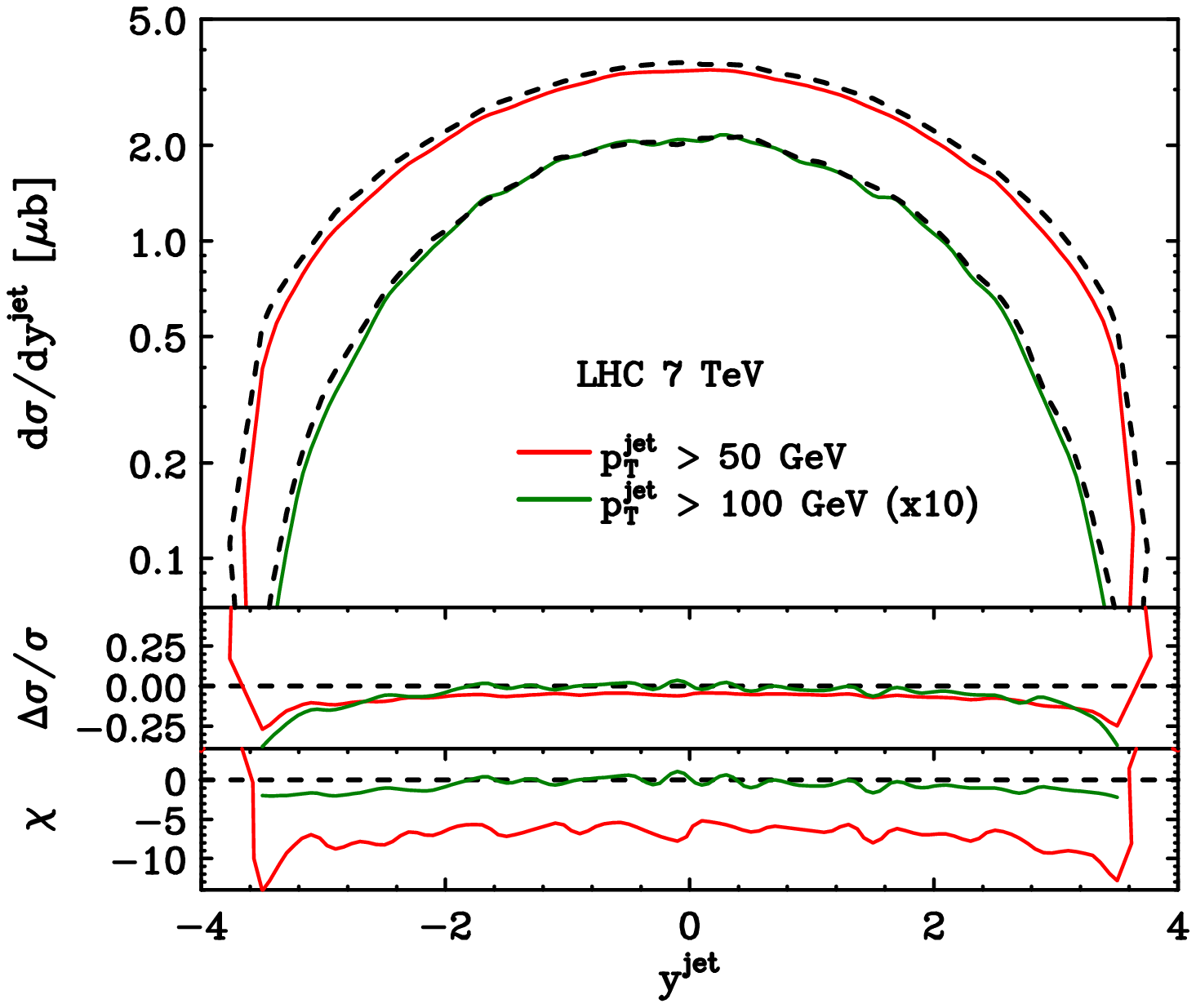,width=0.490\textwidth}
\par\end{centering}
\caption{\label{fig:LHEF-PY-incl-pt-and-y} \POWHEG{} hardest-emission
  cross sections for the inclusive jet transverse-momentum and rapidity spectra,
  i.e.~prior to showering (black dashes), overlaid on those obtained when
  the corresponding hardest-emission events have been showered with \PYTHIA{}
  (solid coloured lines). The transverse-momentum spectra have been binned
  according to the rapidities of the jets, the results being ordered from top
  to bottom with increasing jet rapidity, while in the case of the rapidity
  spectra their ordering is as in the legend.}
\end{figure}
In fig.~\ref{fig:LHEF-PY-incl-pt-and-y} we show, again, the inclusive jet
transverse-momentum and rapidity spectra, where this time the black dashed
lines pertain to the \POWHEG{} hardest-emission events, and their coloured
counterparts correspond to those obtained from the analysis of the aforesaid
events when evolved to the hadron level by \PYTHIA{}.
In general the two sets of results are seen to agree well, as expected given
that these are very inclusive quantities.  One can also see that the
redistribution of the momenta of the hard partons, among those generated by
the shower, slightly depletes the transverse momentum of the jets, leading to
a slight excess of the parton level, hardest-emission cross section with
respect to the hadron level results.

\subsubsection[The $R$ dependence of the jet cross section]
{The $\boldsymbol{R}$ dependence of the jet cross section}
\label{sec:R_dep}
\begin{figure}
\begin{centering}
\epsfig{file=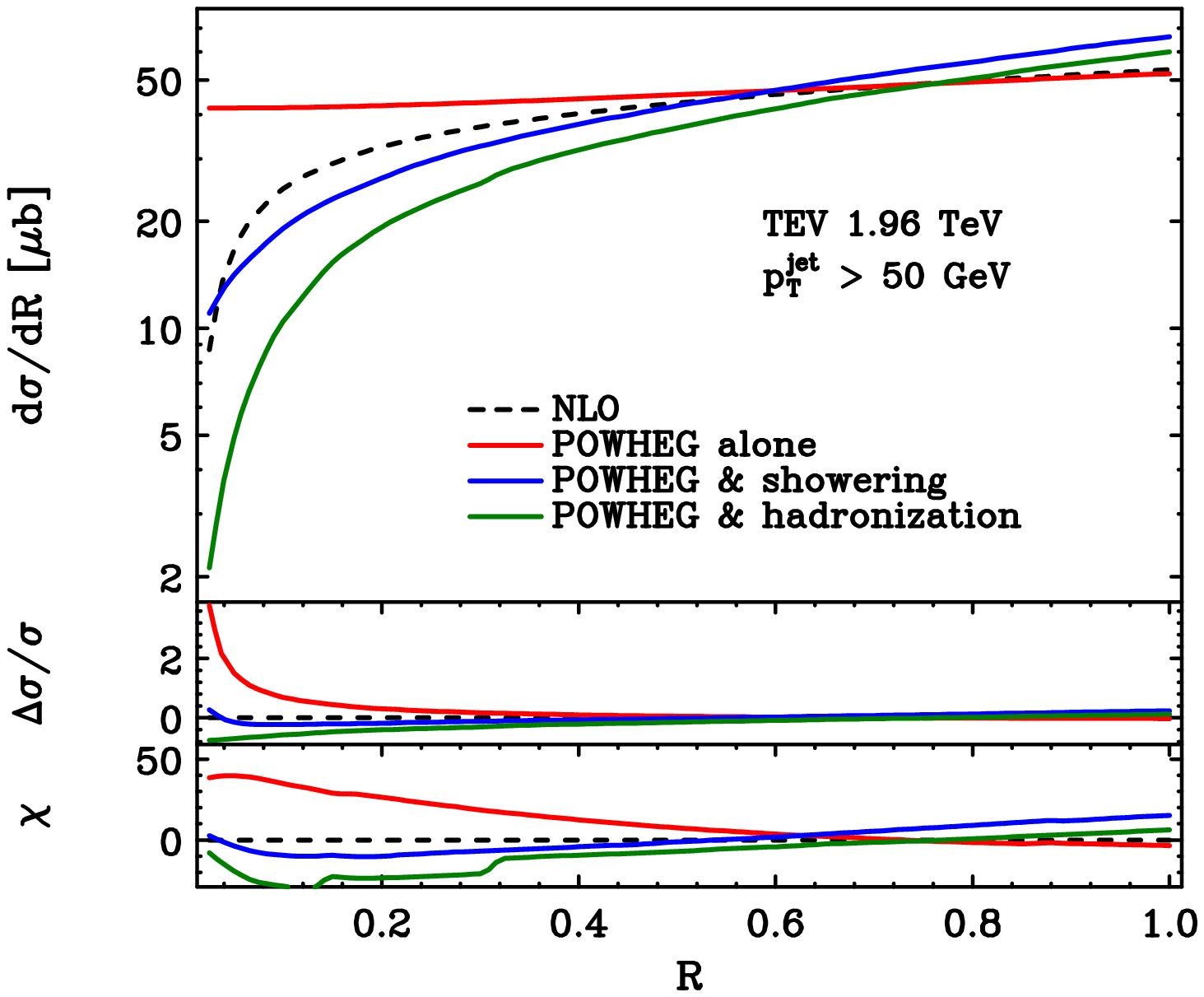,width=0.490\textwidth}
\hfill{}
\epsfig{file=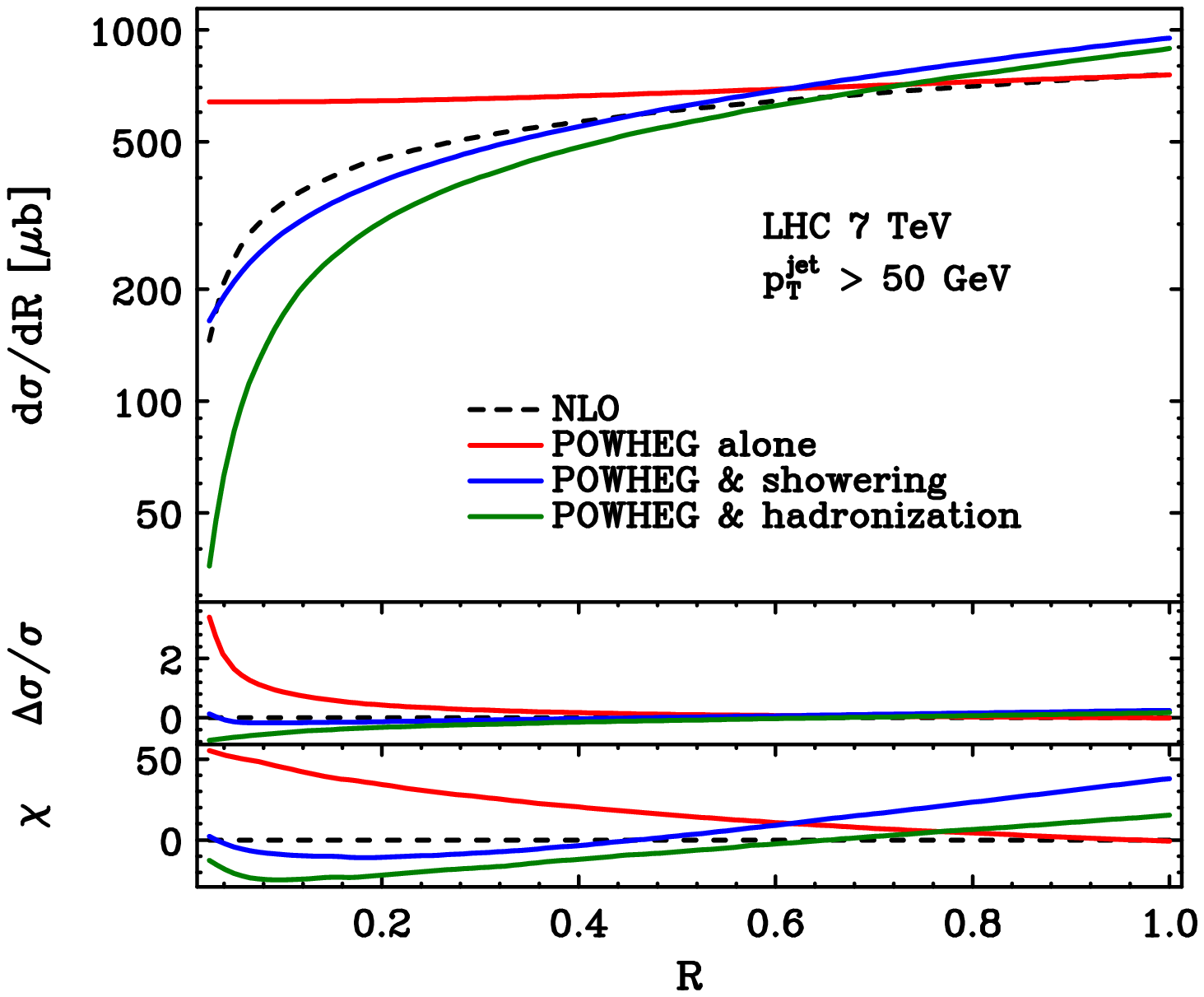,width=0.490\textwidth}
\par\end{centering}
\caption{\label{fig:NLO-LHEF-PY-R} The inclusive jet cross section for jets
  with transverse momenta in excess of 50~GeV, as a function of the jet
  radius parameter $R$. The black dashed line represents the fixed-order NLO
  prediction, while the coloured lines correspond to those obtained with
  bare \POWHEG{} events (red), showered \POWHEG{} events
  (blue) and showered and hadronized \POWHEG{} events (green).}
\end{figure}
As stated in the previous section, in our studies thus far, we have always
taken the value of the jet radius parameter, $R$, to be 0.7. This relatively
large choice ensures good agreement between partonic and showered jets.  It
is time now, however, to assess the $R$ dependence of the jet cross section,
especially in view of the fact that at the LHC smaller values are often used.
In fig.~\ref{fig:NLO-LHEF-PY-R} we display the $R$ dependence of the
inclusive jet cross section, for jets with transverse momenta greater than
50~GeV, at various levels of the simulation chain: NLO, hardest emission,
showered, and hadron-level events. The same configuration of \PYTHIA{} as
above has been used for the last two steps.  Unsurprisingly we can see that
for $R=0.7$ and above there is generally good agreement between the different
predictions, while the opposite is not true.

We notice, in particular, a marked difference between the pure NLO and bare
\POWHEG{} results, which deserves some explanation. First of all, we remind
the reader that the $R$ dependence, as an observable, is similar to the
transverse momentum of the third jet, or to the $\pt^{\rm rel}$, in that it
is not influenced by virtual corrections, but only by the real radiation. We
thus expect it to display, at the NLO level, an unphysical behaviour, in the
form of a logarithmic divergence at small $R$, as observed. We also expect
that the (unshowered) \POWHEG{} result will smear this divergence with a
Sudakov form factor.  This is indeed seen to be the case but the effect of
the smearing is so strong that the \POWHEG{} results displays a very mild $R$
dependence, up to the point where the NLO and the \POWHEG{} predictions
merge.  On the other hand, the showered results display a stronger (and more
physical) $R$ dependence at small $R$.

Although strange, it is easy to convince ourselves that this behaviour is
correct. In fact, as in the case of the $\pt^{\rm rel}$ distribution, since
the bare \POWHEG{} events comprise of a single emission, the $R$ dependence
induced by further emissions from the initial-state and final-state
recoiling partons is absent. The fact that only one of the four partons in
the underlying Born configuration emits suggests that roughly 3/4 of the
cross section should exhibit no $R$ dependence.
One should also consider that the \POWHEG{} Sudakov form factor will inhibit
small angle radiation, thus making the $R$ dependence even smaller.
Furthermore, in the hardest-emission events, the Sudakov suppression is much
stronger than the one that applies to a single radiating parton, since it
also includes contributions corresponding to the non-emission of the other
partons.  Thus, the bare \POWHEG{} prediction of the $R$ dependence is not
realistic; the correct behaviour is only obtained here after the subsequent
shower is turned on.

This is fully analogous to the case of the $\pt^{\rm rel}$ distribution in
fig.~\ref{fig:lhef-nlo-ptrel}, where the \POWHEG{} distribution displays a
(delta function) peak at $\pt^{\rm rel}=0$ and a broad Sudakov
shoulder. Further showering naturally transforms this peak into other Sudakov
shoulders, one for each jet in the event, overlapping with the original one
and making it narrower.  In the present case, the subsequent shower turns the
flat $R$ dependence due to the non-emitting partons into a positive slope,
since 
more energy is dissipated outside the jet cone, as $R$ decreases.

We notice that hadronization effects tend to further increase the slope of
the $R$ dependence. This is a known effect, since hadron formation will
further randomize the particles' momenta,
driving even more energy out of
the cone. It is also
known that the underlying event (not included here)
counteracts the effect of hadronization,
since it generates soft hadrons, that
bring more energy into the jet cone, with a probability proportional to its
area.

\subsubsection{More exclusive distributions}
\begin{figure}
\begin{centering}
\epsfig{file=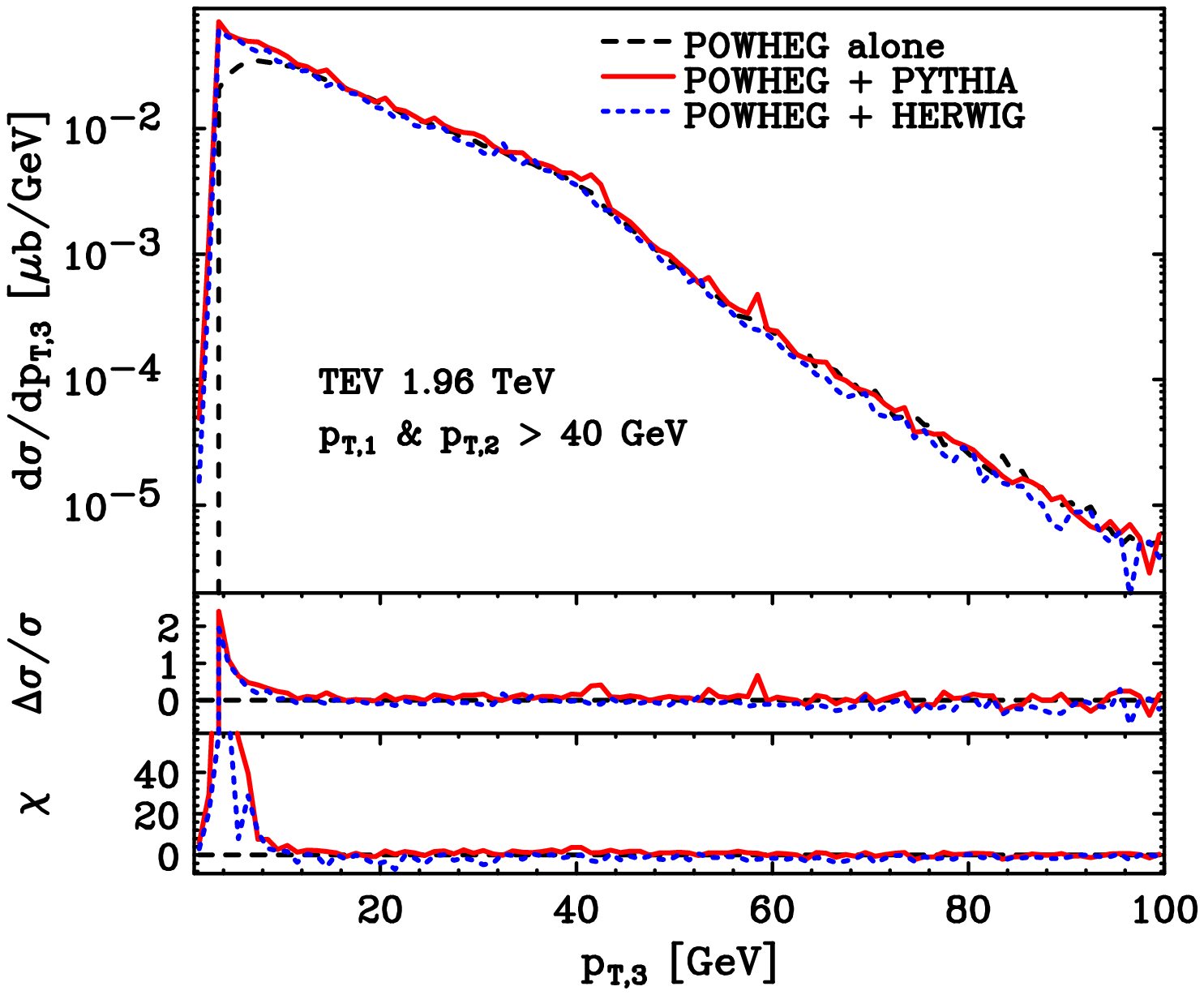,width=0.490\textwidth}
\hfill{}
\epsfig{file=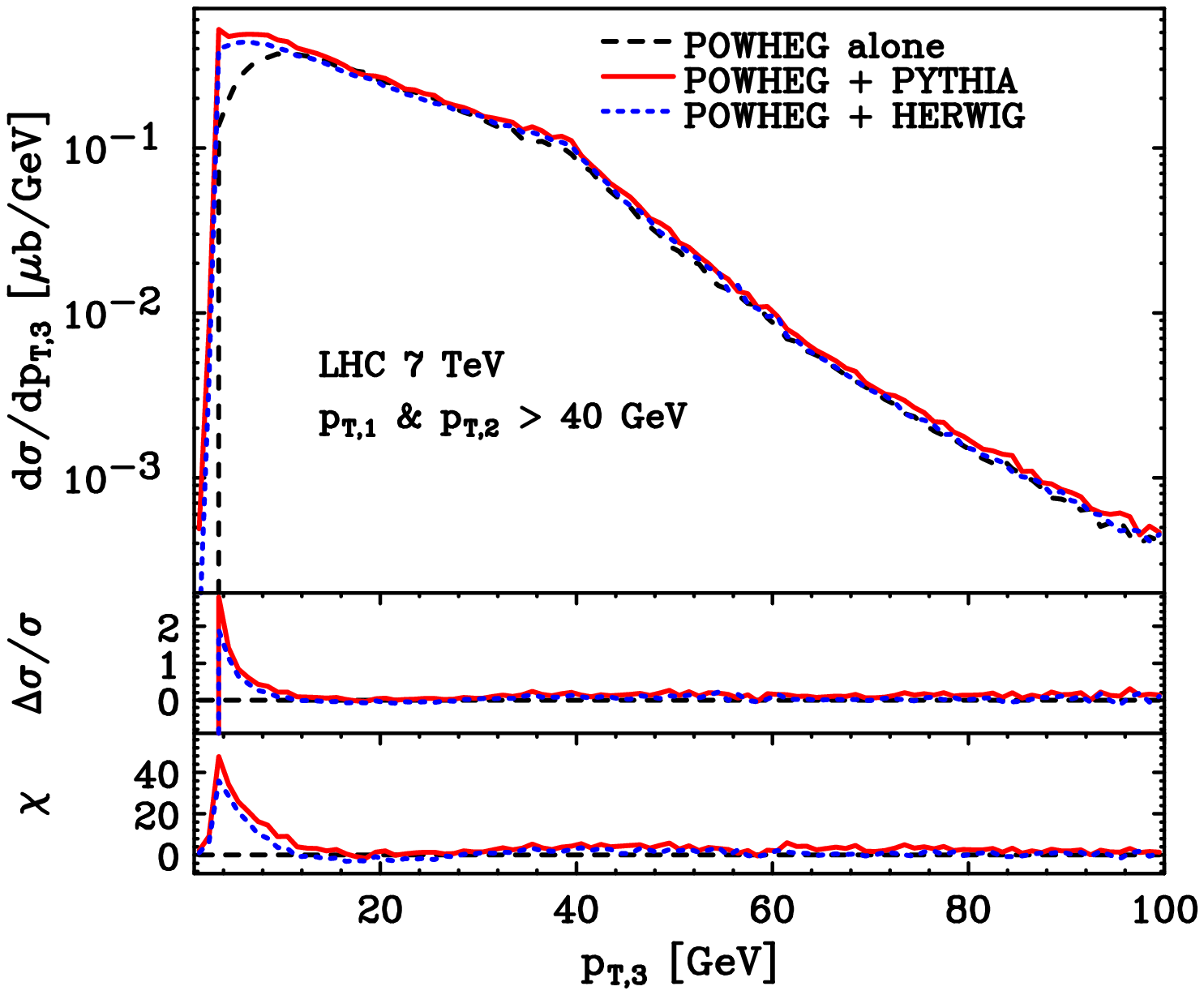,width=0.490\textwidth}
\par\end{centering}
\caption{\label{fig:LHEF-PY-pt3} The transverse-momentum spectrum of the
  third hardest jet in dijet production. The dashed black lines correspond to
  the predictions of the \POWHEG{} hardest-emission cross section, while the
  solid red and dotted blue lines correspond to showering the associated events
  using \PYTHIA{} and \HERWIG{}, omitting underlying-event activity.}
\end{figure}
We now consider the effects of showering and hadronization on more exclusive
observables.  We use again \PYTHIA{} with the aforementioned default setting,
and \HERWIG{}~\cite{Corcella:2000bw,Corcella:2002jc} with default values of
the parameters. In both cases, no multi-particle and underlying-event effects
are considered.

In fig.~\ref{fig:LHEF-PY-pt3}, we re-examine the $\pt$ spectrum of the third
jet.  Here we see that the action of showering and hadronization is
especially manifest for low $p_{\sss\rm T,3}$ values. The behaviour shown here
supports the analysis of the $R$ dependence: the effect of further
showering undoes the strong Sudakov suppression of the low $p_{\sss\rm T,3}$
region, imposed by the \POWHEG{} hardest-emission cross section, yielding a
result that is more peaked there.  In
comparing to the right-hand plot in fig.~\ref{fig:lhef-nlo-pt3} we can see,
however, that the increase due to further showering is still much below that
of the NLO result. In other words, some Sudakov suppression correctly
remains, although not as much as the bare \POWHEG{} result displays. Lastly
we remark that, in contrast to the bare \POWHEG{} predictions, the showered
events populate the region of $\pt$ below 3~GeV, since the cut in the D0 jet
algorithm applies to the transverse energy of the jet cones, which can be
smaller than the $\pt$ for the showered events.

\begin{figure}
\begin{centering}
\epsfig{file=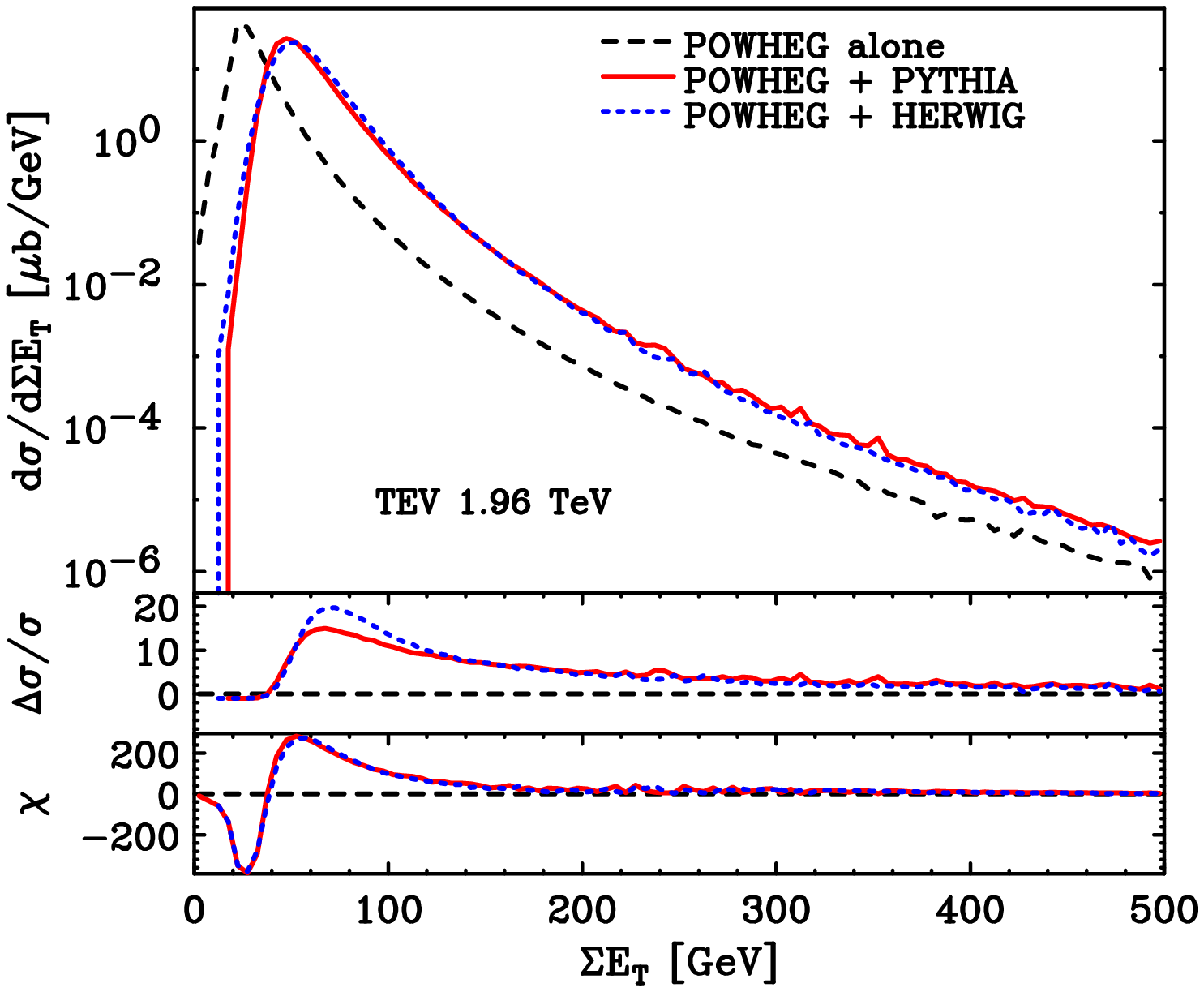,width=0.490\textwidth}
\hfill{}
\epsfig{file=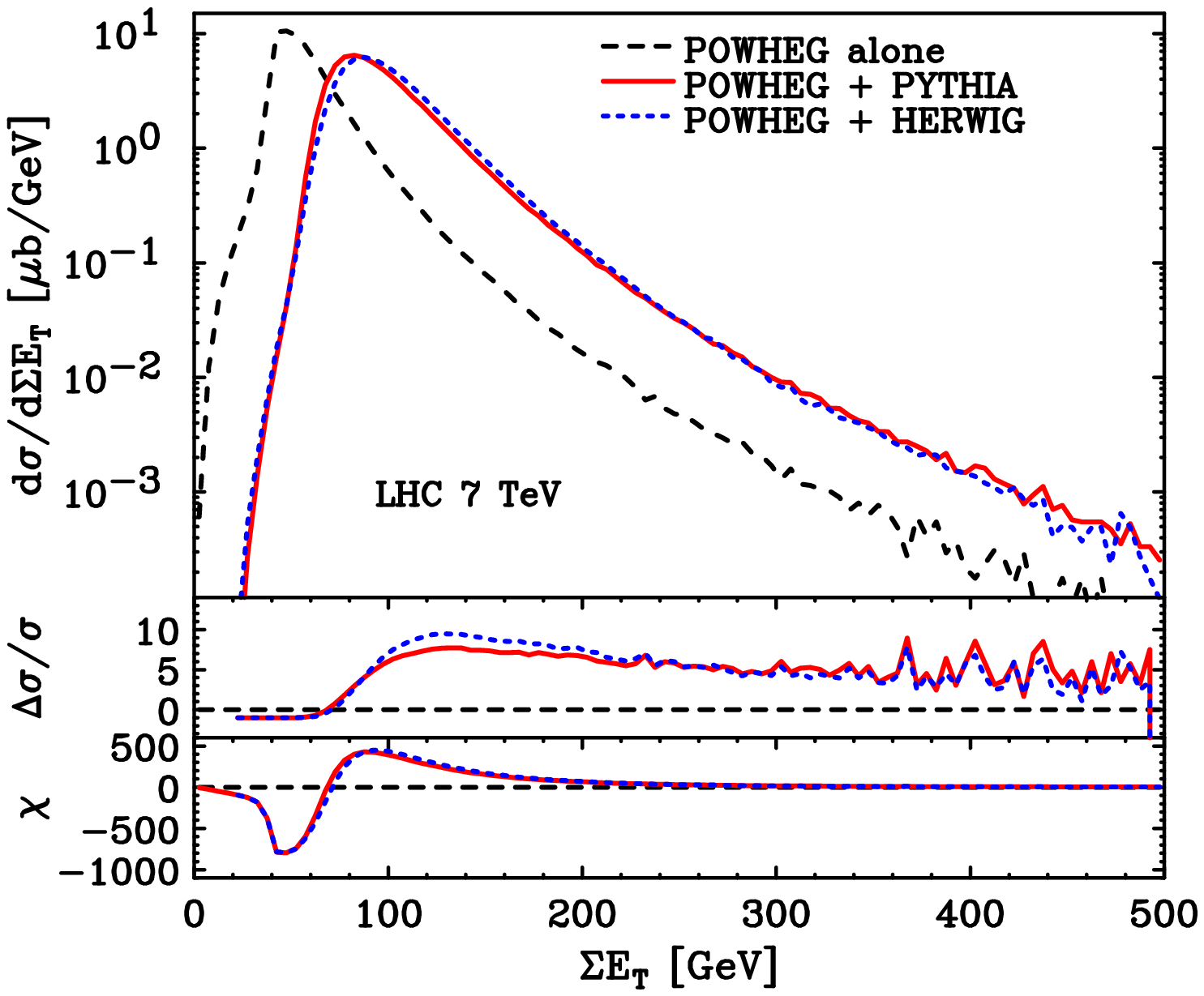,width=0.490\textwidth}
\par\end{centering}
\caption{\label{fig:LHEF-PY-totET} The total transverse-energy spectrum,
  namely, the scalar sum of the transverse energies of all {particles} in
  each event, at Tevatron (left) and LHC (right) energies. As in
  fig.~\ref{fig:LHEF-PY-pt3}, the black dashed lines represent the
  predictions of the \POWHEG{} hardest-emission cross section alone, while
  the solid red and dotted blue lines correspond to showering the hardest
  emission events using \PYTHIA{} and \HERWIG{}, neglecting underlying-event
  effects.}
\end{figure}
The effect of parton showering and hadronization on the scalar sum of the
transverse energy of all particles is shown in fig.~\ref{fig:LHEF-PY-totET}.
In the results obtained with both \PYTHIA{} and \HERWIG{} we find a dramatic
shift in the distribution to higher energies, of approximately 50~GeV.  This
result should, however, not be considered alarming. The activity accompanying
the showered events always leads to large multiplicities and, for example,
even the production of 100 soft particles, with momenta of the order of
500~MeV, is capable of raising the total transverse energy by 50~GeV. We
should thus consider this large discrepancy between the unshowered and the
showered result as fully understandable.

\section{Phenomenology}
\label{sec:phenomenology}
In this section, we compare the predictions for dijet production obtained
with the \POWHEGBOX, fully showered by
\PYTHIA~6.4.21~\cite{Sjostrand:2006za}, including hadronization,
underlying-event and multi-particle effects as defined by the Perugia~0 tune,
with some Tevatron and LHC published measurements. For ease of notation, in
the following, we will refer to the theoretical results simply as the
`\POWHEG{} results'. All the aforementioned effects from \PYTHIA{} are
always included.

In order to efficiently populate the high-$\pt$ regions of phase space, we
have always used weighted-event samples in this section. On rare occasions we
find that, when using \PYTHIA{} for showering, large spikes with large errors
appear in the distributions. We have not observed any such behaviour in the
bare \POWHEG{} output. We have concluded that this unpleasent feature is due
to the fact that, in exceptional cases, low transverse-momentum
hardest-emission events, with a large weight, can be promoted to high
transverse-momentum events by showering. 
In order to circumvent the problem, on the few occasions on which it has
occurred here, we have merged histograms obtained by showering the same
hardest-emission events using \PYTHIA{} runs with different random seeds,
replacing these anomalous bins with spikes by the corresponding result
obtained in other \PYTHIA{} runs, with smaller error. Unfortunately, the only
way guaranteed to avoid this problem is to use unweighted samples with
different generation cuts to cover the whole transverse-momentum spectrum.

The purpose of this section is, to some extent, to validate our code with
real data rather than performing exhaustive tests on jet physics. We thus
limit ourselves to a single PDF set, CTEQ6M, and we use, as renormalization
and factorization scale in the evaluation of $\bar{B}\left(\Phi_{B}\right)$,
the transverse momentum of the underlying Born configuration, as described in
section~\ref{sec:scales_choice}.

\subsection{Multiple-parton interactions in \PYTHIA{} and the \POWHEG{} 
jet generator} 
The recent versions of \PYTHIA{} include a model of multiple-parton
interactions~(MPI)~\cite{Sjostrand:2004pf} that improves the
description of the underlying event accompanying the hard scattering
process. A multiple interaction is essentially a $2\to 2$ parton scattering
arising from the collision of the remnants of the incoming nucleons. In the
\PYTHIA{} model, the scale that limits the transverse momentum 
of these processes is set to be quite high, much above the
scale of the hard process under examination. The logic behind this choice is
well documented in the \PYTHIA{} manual: when considering a process like, for
example, $Z$ production, there is a finite probability that a pair of jets
with transverse momenta larger than half the invariant mass of the $Z$ are
produced, and the \PYTHIA{} authors want these events to be effectively
generated.  However, if the process in question is jet production, this
approach may lead to overcounting. Therefore, \PYTHIA{} inhibits this
behaviour when generating jets, limiting the MPI scale to that of the jets in
the primary interaction.  When interfacing \PYTHIA{} to the \POWHEG{} jet
program (and, for that matter, to any matrix-element generator for jets)
using the user-process interface, there is no way for \PYTHIA{} to know that
it is showering a jet process. The user should therefore force \PYTHIA{} to
limit the scale of the MPI to the hardness of the primary process.\footnote{In
the dijet case this scale corresponds to the transverse momentum of the jets.}
\PYTHIA{} provides a method to do this, which is to set the parameter
{\tt MSTP(86)=1}. We have collected the sequence of \PYTHIA{} calls we have
used in appendix~\ref{sec:PY_settings}.

\subsection{Tevatron results}
In this section we illustrate the comparison between the results obtained
with the \POWHEGBOX{} and data from the CDF and D0 Collaborations at the
Tevatron, running at a center-of-mass energy of 1.96~TeV.

We have generated a sample of roughly $5$ millions weighted events, with a
cut on the Born transverse momentum of 1~GeV and with $\ktsupp=600$~GeV, in
order for the events to cover a region in transverse momentum up to 1~TeV.
We have used the folding 1-5-1,\footnote{That is to say, setting
{\tt ifoldcsi}$=1$, {\tt ifoldy}$=5$ and
{\tt ifoldphi}$=1$ in the {\tt powheg.input} file. See
refs.~\cite{Nason:2007vt,Alioli:2010qp} for a detailed explanation.}
getting a fraction of negative-weight events below
5\permil, that we have disregarded.

\begin{figure}
\begin{centering}
\epsfig{file=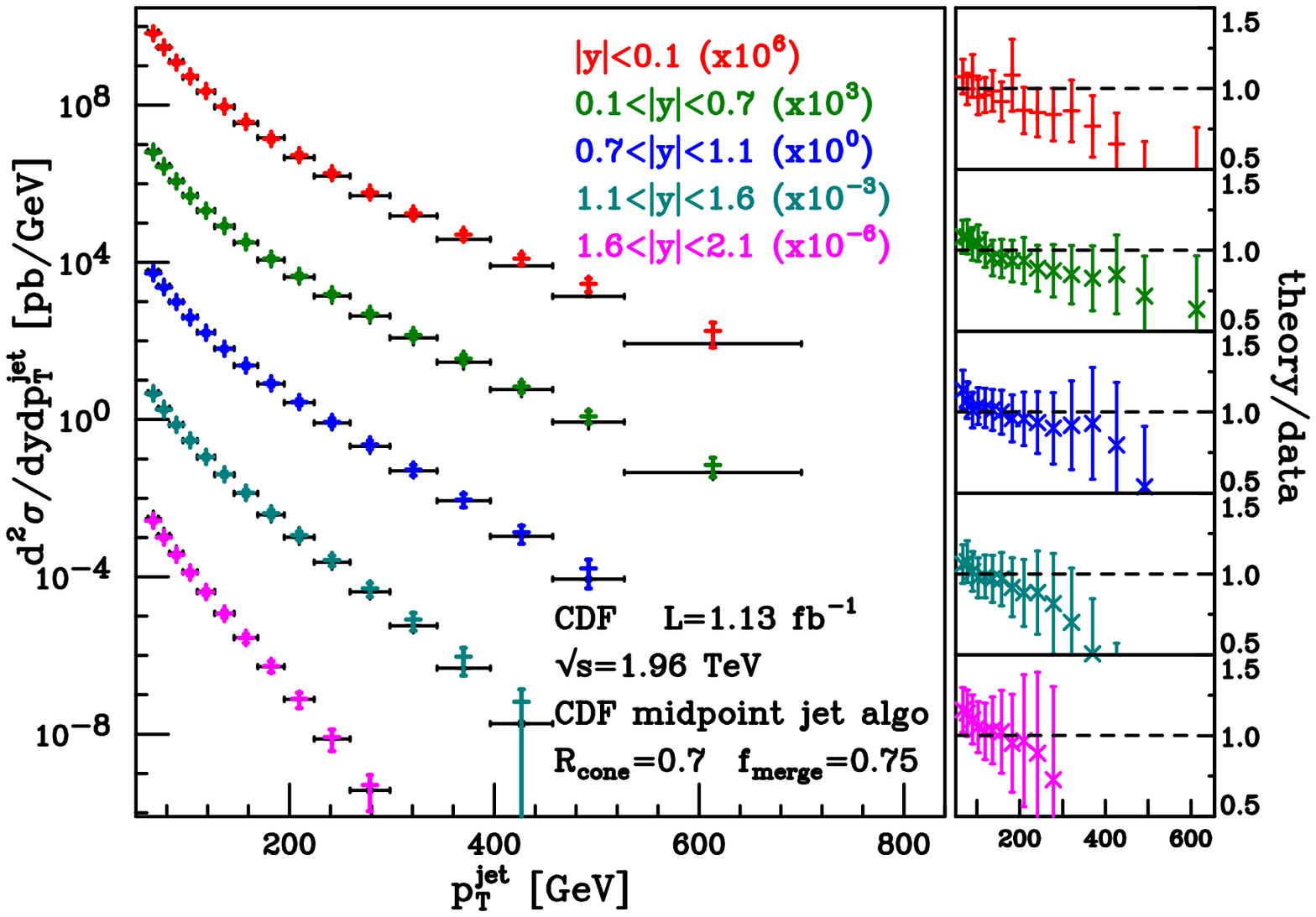,width=1.\textwidth}
\par\end{centering}
\caption{\label{fig:CDF_dptdy_midalgo_plots} Predictions and experimental
  results for the double-differential inclusive jet cross section as a
  function of the transverse momentum of the jet $\pt^{\rm jet}$, for
  different bins of jet rapidity, $y$, as measured by the CDF Collaboration,
  using the cone-based midpoint jet algorithm.  Black lines are the
  \POWHEG+\PYTHIA{} results (error bars are drawn too, even if almost
  invisible on the plot scale), while coloured bars are the experimental
  data (with errors represented as vertical bars)~\cite{Aaltonen:2008eq}.
  Data are shown from top to bottom in order of increasing rapidity.}
\end{figure}

\begin{figure}
\begin{centering}
\epsfig{file=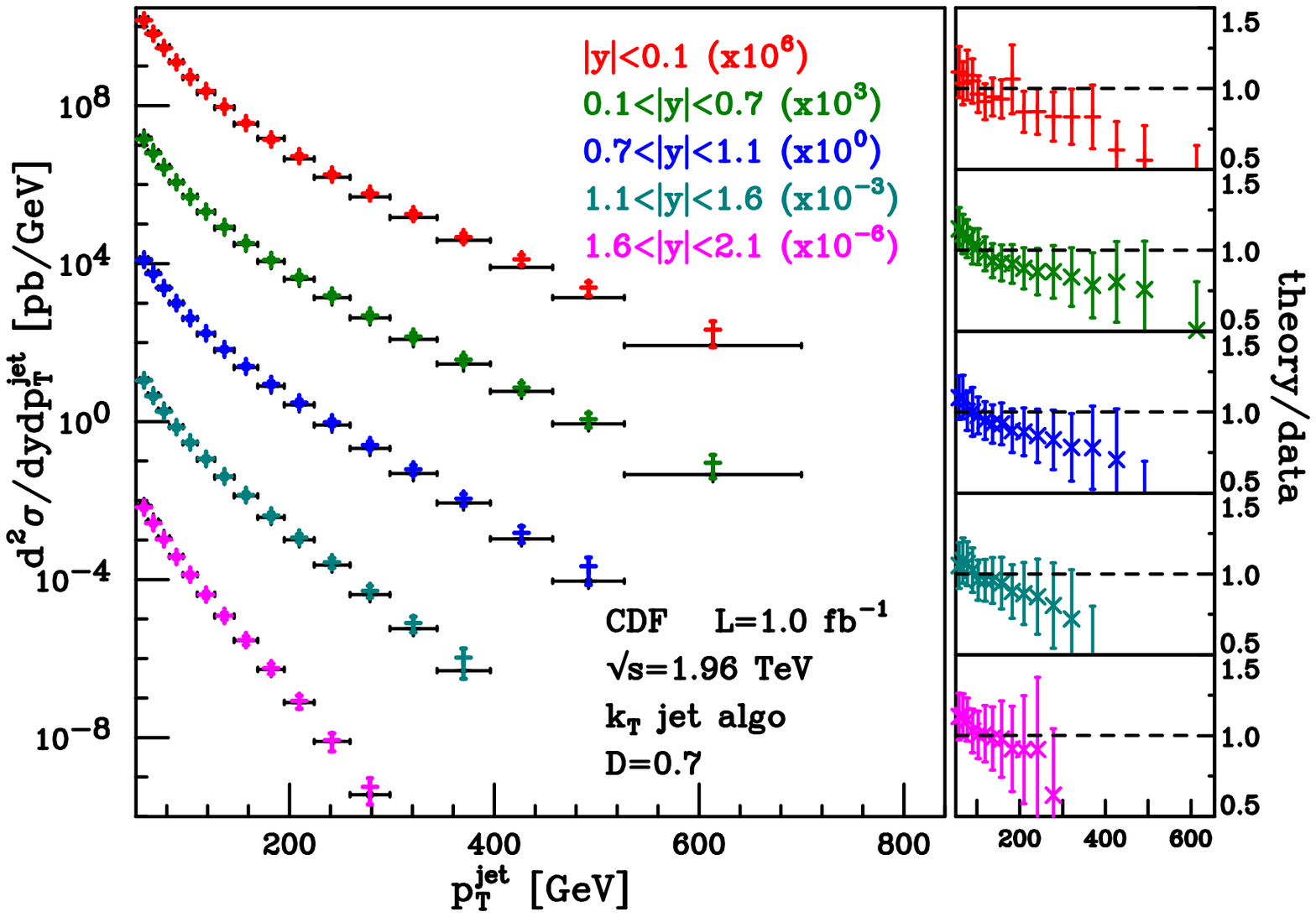,width=1.\textwidth}
\par\end{centering}
\caption{\label{fig:CDF_dptdy_ktalgo_plots} Predictions and experimental
  results for the double-differential inclusive jet cross section as a
  function of the transverse momentum of the jet $\pt^{\rm jet}$, for
  different bins of jet rapidity, $y$, as measured by the CDF Collaboration,
  using the $\kt$ jet algorithm.  Black lines are the \POWHEG+\PYTHIA{}
  results (error bars are drawn too, even if almost invisible on the plot
  scale), while coloured bars are the experimental data (with errors
  represented as vertical bars)~\cite{Abulencia:2007ez}. Data are shown from
  top to bottom in order of increasing rapidity. }
\end{figure}

In figs.~\ref{fig:CDF_dptdy_midalgo_plots}
and~\ref{fig:CDF_dptdy_ktalgo_plots} we plot the double-differential
inclusive jet cross section as a function of the transverse momentum of the
jet $\pt^{\rm jet}$, for different bins of jet rapidity $y$, as measured by
the CDF Collaboration. 
Jets in fig.~\ref{fig:CDF_dptdy_midalgo_plots} are reconstructed using the
cone-based CDF midpoint jet algorithm~\cite{Blazey:2000qt}, with a jet radius
parameter $R=0.7$ and overlapping fraction $f=0.75$. After jet clustering,
only jets with
\begin{equation}
62\ \mbox{GeV} < \pt^{\rm jet} < 700\ \mbox{GeV} \quad 
\mbox{and}\quad |y^{\rm jet}|<2.1
\end{equation}
are kept. \POWHEG{} results are shown as black lines, while the experimental
data~\cite{Aaltonen:2008eq} are drawn as coloured bars, with errors obtained
by summing in quadrature the statistical and the systematic errors.

In fig.~\ref{fig:CDF_dptdy_ktalgo_plots} we compare the \POWHEG{} predictions
with results from the analysis of ref.~\cite{Abulencia:2007ez}.  Jets are
recombined using the $\kt$ algorithm~\cite{Ellis:1993tq,Catani:1993hr}, where
the jet-size parameter $D$ has been set equal to $0.7$. Only jets with
\begin{equation}
54\ \mbox{GeV} < \pt^{\rm jet} < 700\ \mbox{GeV} \quad 
\mbox{and}\quad |y^{\rm jet}|<2.1\,,
\end{equation}
are kept.

We remark that, in refs.~\cite{Aaltonen:2008eq} and~\cite{Abulencia:2007ez},
data were compared to NLO calculations corrected for the parton-to-hadron
correction factors. Here, the measured values are directly compared to the
\POWHEG{} results showered by \PYTHIA{}, and the agreement is quite good, as
can be seen from the ratios theory/data in
figs.~\ref{fig:CDF_dptdy_midalgo_plots} and~\ref{fig:CDF_dptdy_ktalgo_plots}.

\begin{figure}
\begin{centering}
\epsfig{file=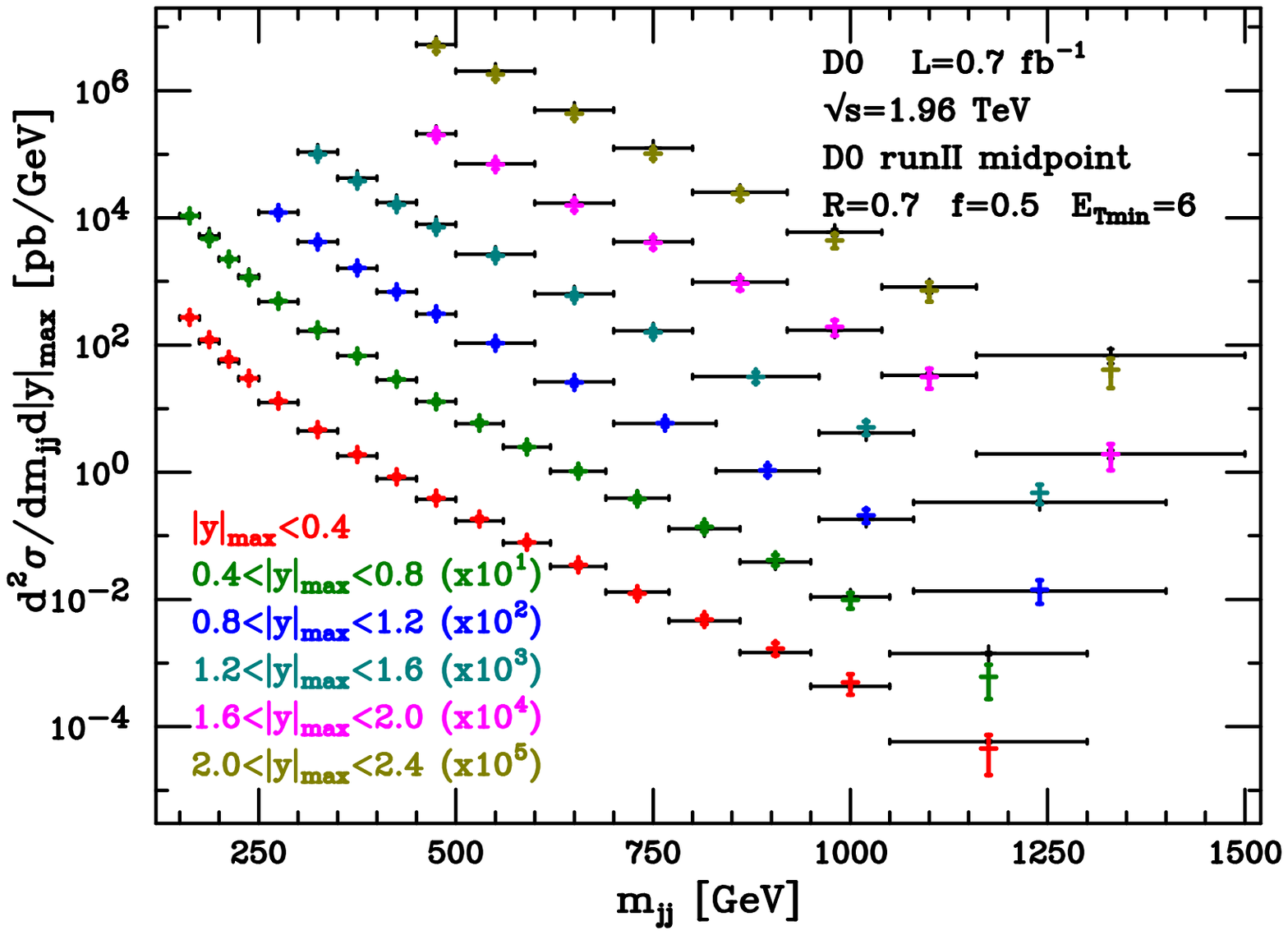,width=0.490\textwidth}
\hfill{}
\epsfig{file=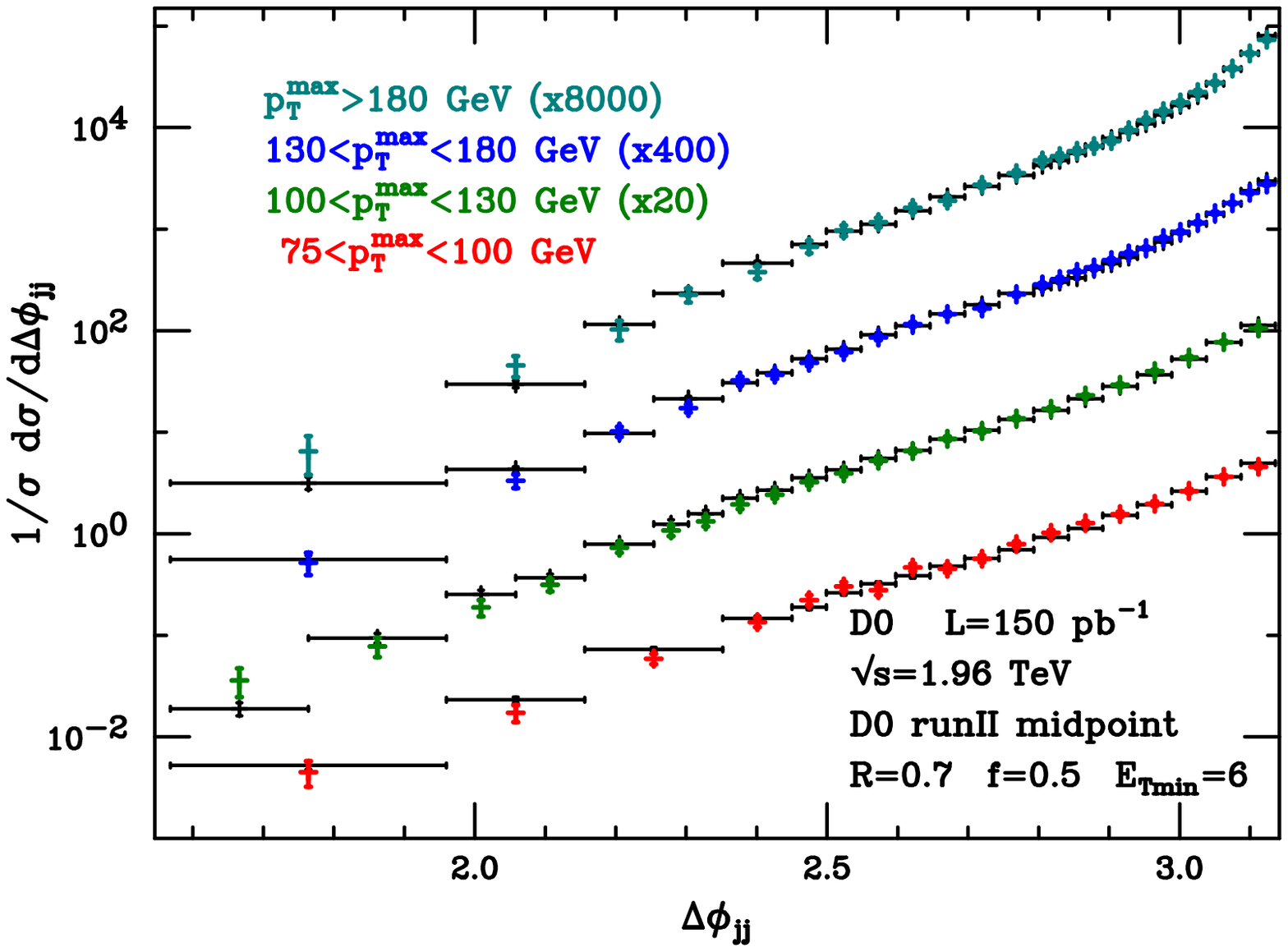,width=0.490\textwidth}
\par\end{centering}
\caption{\label{fig:D0_plots} On the left, we confront the
  \POWHEG+\PYTHIA{} results with the D0 measurement~\cite{Abazov:2010fr} of
  the double-differential dijet cross section, as a function of the invariant
  mass of the two leading jets, $m_{jj}$, for different bins of $|y|_{\max}$,
  the maximum absolute value of the rapidity of the two jets. On the
  right-hand side, we show the dijet azimuthal decorrelation,
  $\Delta\phi_{jj}$, in bins of the transverse momentum of the leading jet
  $\pt^{\max}$~\cite{Abazov:2004hm}. Black horizontal bars correspond to the
  \POWHEG{} outputs. D0 data are shown in colour, with experimental errors
  given by the vertical bars.
  Data are shown from top to bottom in order of increasing rapidity on the
  left, and of decreasing $\pt^{\max}$ on the right.}
\end{figure}
In fig.~\ref{fig:D0_plots}, we have performed a comparison between the
\POWHEG{} results and D0 data~\cite{Abazov:2010fr} for the invariant mass of
the two leading jets and for the azimuthal angle between them. For both
plots, we have used the seed-based D0 run~II midpoint cone algorithm, with
$R=0.7$, overlapping fraction $f=0.5$ and with minimum jet $\Et$ parameter
equal to 6~GeV, as in ref.~\cite{Abazov:2010fr}.  In the left plot, we
display the double-differential jet cross section as a function of the
invariant mass of the two leading jets, $m_{jj}$, for different bins of
$|y|_{\max}$, the maximum absolute value of the rapidity of the two jets.  A
minimum cut on the transverse momentum of the two leading jets is imposed,
i.e.~$\pt^{\rm jet}> 40$~GeV.  Data~\cite{Abazov:2010fr} are shown in colour
with vertical error bars, while the \POWHEG{} results are depicted in black.
Good agreement is found between data and the \POWHEG{} results over quite a
wide range of values of the dijet mass and of the rapidity intervals.

In the right plot, we show the azimuthal separation of the two hardest
jets, $\Delta\phi_{jj}$.  We require the hardest and next-to-hardest
jets to have $\pt^{\rm jet} > 75$~GeV and $\pt^{\rm jet} > 40$~GeV
respectively.  In addition, jets must have central rapidities,
i.e.~$|y_{\rm jet}|< 0.5$, as in the analysis of
ref.~\cite{Abazov:2004hm}.  Results are shown in bins of $\pt^{\max}$,
the transverse momentum of the leading jet. Notice that the results
are normalized to unity, so that the prediction power of \POWHEG{} is
less evident in this plot.  For this physical variable, the NLO
results becomes negative as $\Delta\phi_{jj}\to \pi$, i.e.~as the
third parton become soft or collinear to one of the other
twos. Instead, the \POWHEG{} curves are finite, since the Sudakov form
factor resums the leading-logarithmic divergences as this limit is
approached. On the other hand, hard jets contribute in the region
where $\Delta\phi_{jj}$ gets smaller, so that we do not expect a
perfect agreement with the \POWHEG{} curves, that at most produces 3
hard jets.

\begin{figure}
\begin{centering}
\epsfig{file=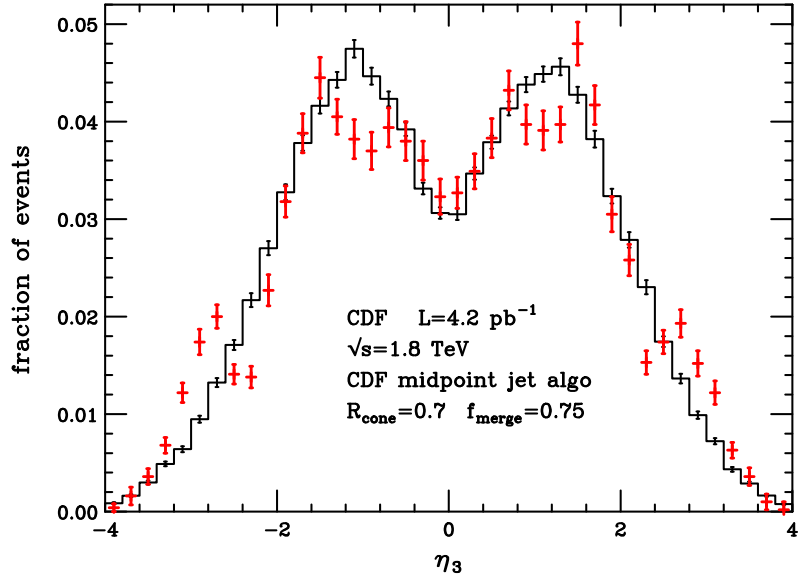,width=0.7\textwidth}
\par\end{centering}
\caption{\label{fig:CDF_coherence} The pseudorapidity spectrum of the third
  hardest jet (ordered in the transverse energy) in three-jet events,
  requiring the two leading jets to satisfy $|\eta_{1}|,|\eta_{2}|<0.7$,
  $||\phi_{1}-\phi_{2}| - \pi| < 20^\circ$, with cuts on the transverse
  energies of the leading and third jet of $E_{\sss\rm T,1}>110$~GeV and
  $E_{T,3}>10$~GeV, as in ref.~\cite{Abe:1994nj}. The red vertical bars are
  the collected data, while the \POWHEG+\PYTHIA{} result is shown as a black
  histogram.}
\end{figure}
Colour coherence effects have been observed and studied at CDF, in
ref.~\cite{Abe:1994nj}. In that paper, variables sensitive to interference
effects have been identified and measured.  The data were collected in Run~I
at 1.8~TeV, on a sample of 4.2~pb$^{-1}$.  We have generated a sample of
about 5 million weighted events with the \POWHEGBOX{}, having a Born minimum
transverse momentum of 1~GeV and a suppression factor $\ktsupp=150$~GeV. The
folding used was 1-5-1, that gave rise to a fraction of 1\% negative-weight
events.  We have applied the same jet algorithm and parameters used by other
CDF analyses.
Jets are then ordered with respect to the transverse energy, and not with
respect to the transverse momentum.  With this ordering, only events with at
least three jets are selected.  The two leading jets are required to be
central in the detector, with pseudorapidities $|\eta_{1}|,|\eta_{2}|<0.7$,
and to be back-to-back within $20^\circ$ in the transverse plane,
$||\phi_{1}-\phi_{2}| - \pi| < 20^\circ$.  The transverse energy of the first
jet and of the third jet are required to be greater than 110~GeV and 10~GeV.
In fig.~\ref{fig:CDF_coherence} we plot the CDF results (red vertical bar)
and the \POWHEG{} prediction (black histogram) for the pseudorapidity of the
third jet, $\eta_3$. The distribution is normalized to unit
area.\footnote{Note that the local distortions of the shape, such those at
  $|\eta_3|=2.5$, are due to uninstrumented regions of the detector, as
  pointed out in ref.~\cite{Abe:1994nj}.}  The colour-coherence
feature is manifest as a central dip, not present if interference effects
are not properly accounted for.

\subsection{LHC results}
The ATLAS Collaboration has recently published the first LHC data on dijet
production at $7$~TeV~\cite{Collaboration:2010wv}.  To make a comparison with
these data, we have generated a sample of about $5$ million weighted
events, with a cut on the Born transverse momentum of 1~GeV and with
$\ktsupp=200$~GeV, in order for the events to cover a region in the transverse
momentum up to 600~GeV.  We have used the folding 2-10-1, getting a fraction
of negative-weight events equal to 1.6\permil, that we have disregarded.
Jets were reconstructed using the anti-$\kt$
algorithm~\cite{Cacciari:2008gp}, which is infrared-safe at all orders.
Furthermore, it has a simple geometrical interpretation in
terms of cone-like jets. We adopted the same choice of resolution
parameters used in ref.~\cite{Collaboration:2010wv}, i.e.~$R=0.4$ and
$R=0.6$.

\begin{figure}
\begin{centering}
\epsfig{file=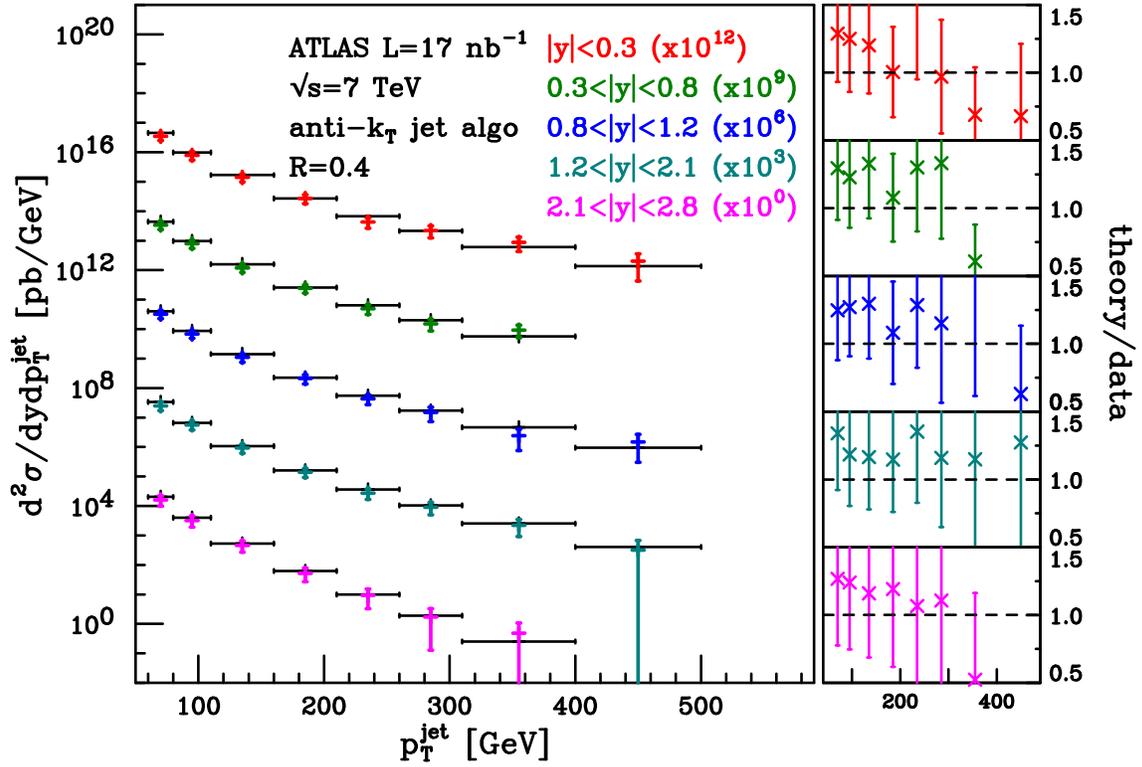,width=1.0\textwidth}
\par\end{centering}
\caption{\label{fig:ATLAS_dptdy_R04_plots} Predictions and experimental
  results for the double-differential inclusive jet cross section, as a
  function of the jet transverse momentum, $\pt^{\rm jet}$, in bins of jet
  rapidity $y$ for the jets that pass the cuts of eq.~(\ref{eq:ATLAS_cuts}).
  Black horizontal lines are the \POWHEG+\PYTHIA{} theoretical results (with
  errors, almost invisible at the scale of the plot). Coloured vertical bars
  describe the experimental data from ATLAS (systematic and statistical
  errors added in quadrature)~\cite{Collaboration:2010wv}.  Jets recombined
  using the anti-$\kt$ algorithm with $R=0.4$. Data are shown from top to
  bottom in order of increasing rapidity.}
\end{figure}

\begin{figure}
\begin{centering}
\epsfig{file=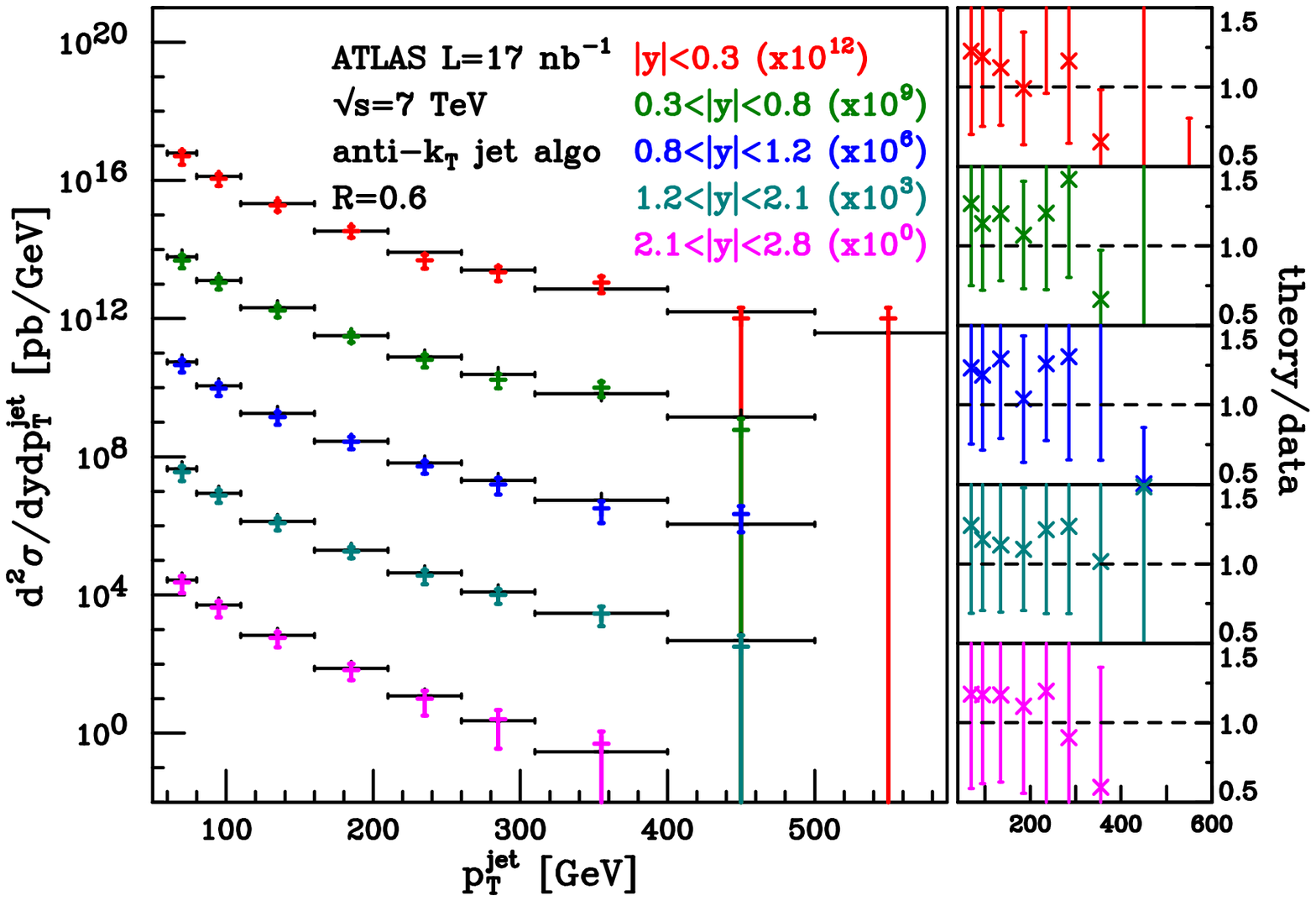,width=1.0\textwidth}
\par\end{centering}
\caption{\label{fig:ATLAS_dptdy_R06_plots} Predictions and experimental
  results for the double-differential inclusive jet cross section, as a
  function of the jet transverse momentum, $\pt^{\rm jet}$, in bins of jet
  rapidity $y$ for the jets that pass the cuts of eq.~(\ref{eq:ATLAS_cuts}).
  Black horizontal lines are the \POWHEG+\PYTHIA{} theoretical results (with
  errors, almost invisible at the scale of the plot). Coloured vertical bars
  describe the experimental data from ATLAS (systematic and statistical
  errors added in quadrature)~\cite{Collaboration:2010wv}.  Jets recombined
  using the anti-$\kt$ algorithm with $R=0.6$. Data are shown from top to
  bottom in order of increasing rapidity.}
\end{figure}

In figs.~\ref{fig:ATLAS_dptdy_R04_plots} and~\ref{fig:ATLAS_dptdy_R06_plots}
we show the inclusive double-differential cross section, as a function of the
jet transverse momentum, $\pt^{\rm jet}$, in bins of jet rapidity $y$.  In the
first figure, results for $R=0.4$ are shown, while in the second one the
value $R=0.6$ has been used.  Only  jets with
\begin{equation}
\label{eq:ATLAS_cuts}
 \pt^{\rm jet} > 60\ {\rm GeV}, \qquad  |y^{\rm jet}|<2.8
\end{equation}
are included in the plot. \POWHEG{} results are shown as black horizontal
lines, with width equal to the bin size used by ATLAS, and coloured vertical
bars correspond to the experimental data of ref.~\cite{Collaboration:2010wv},
obtained by summing in quadrature the statistical and systematic errors.  An
additional overall uncertainty of 11\%, due to the measurement of the
integrated luminosity, is also included in the error bars.  The good
agreement between theory and data is illustrated in the ratio plots on the
right hand side of the two figures.  As discussed in section~\ref{sec:R_dep},
one expects that, as the jet radius $R$ increases, more particles are
clustered in the jets that pass the cuts of eq.~(\ref{eq:ATLAS_cuts}), and
the agreement with theory is improved.

\begin{figure}
\begin{centering}
\epsfig{file=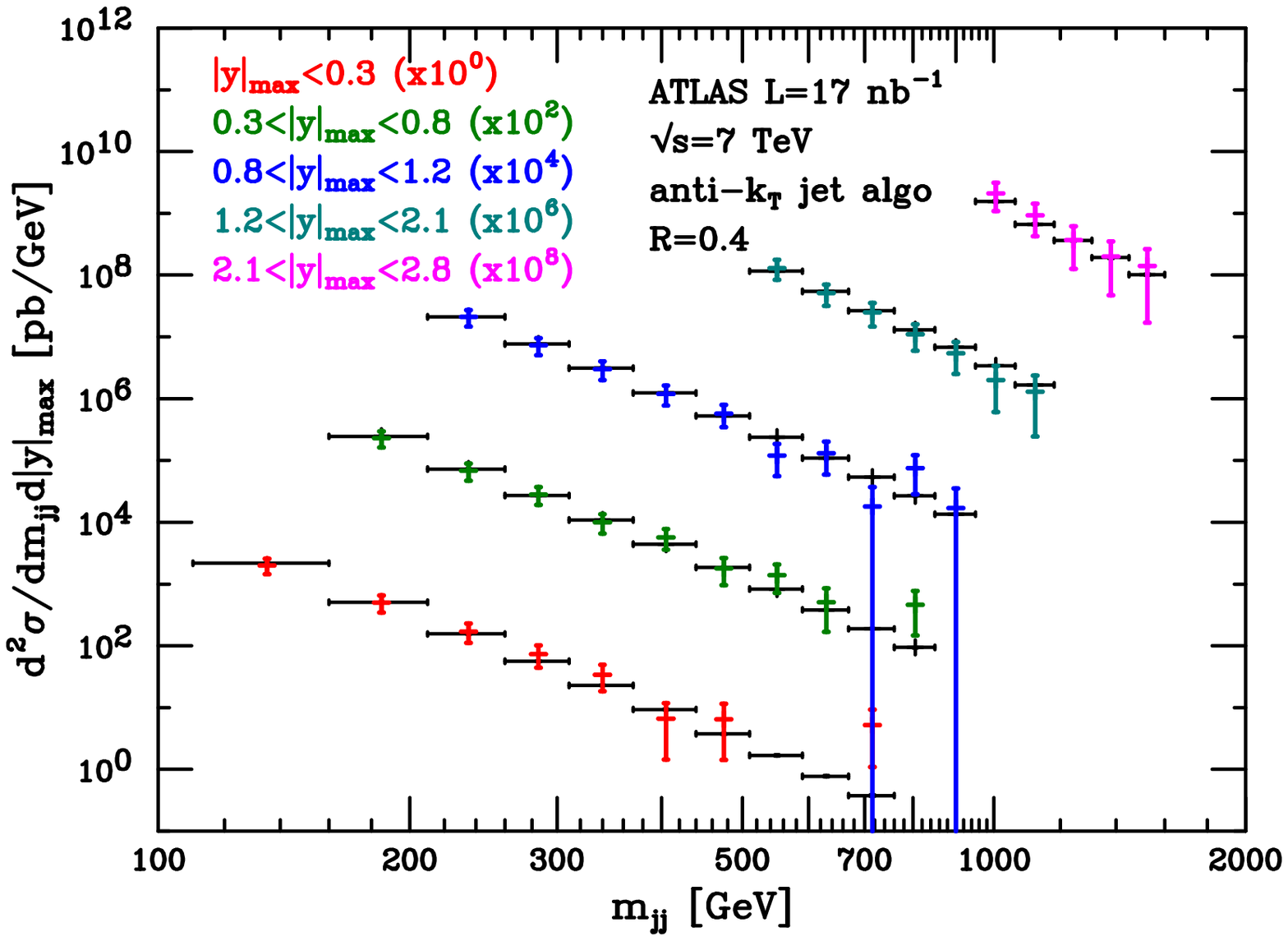,width=0.490\textwidth}
\hfill{}
\epsfig{file=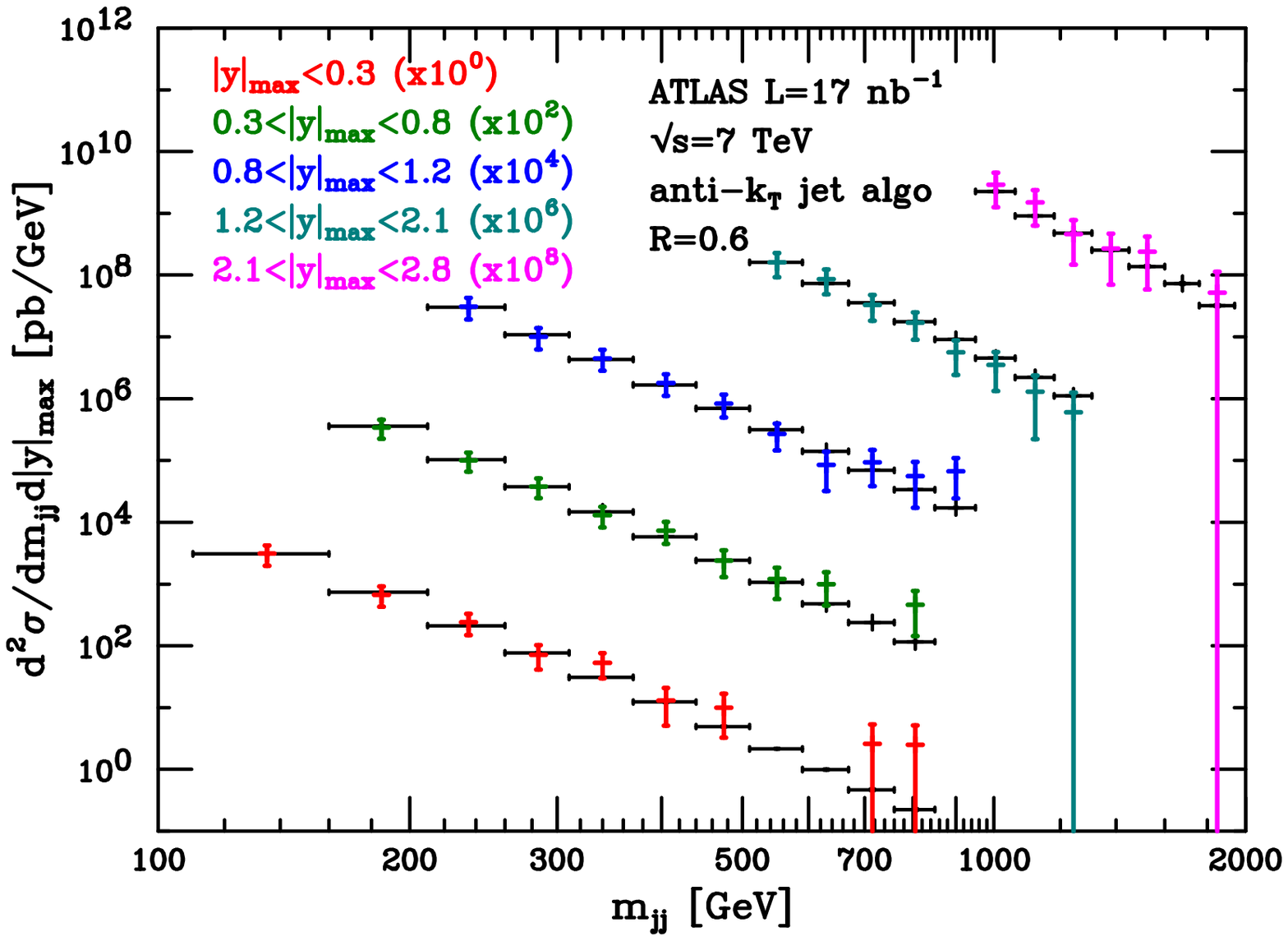,width=0.490\textwidth}
\par\end{centering}
\caption{\label{fig:ATLAS_dydmjj_plots} Predictions and experimental results
  for the double-differential inclusive jet cross section as function of the
  dijet mass $m_{jj}$ of the two leading jet pair, for different bins in
  $|y|_{\max}$, the maximum absolute rapidity of the  two jets.  The cuts
  of eq.~(\ref{eq:ATLAS_mjj_cuts}) are imposed on the two leading jets.
  Black horizontal lines are the \POWHEG+\PYTHIA{} theoretical results (with
  errors). Coloured vertical bars describe the experimental data from ATLAS
  (systematic and statistical errors added in
  quadrature)~\cite{Collaboration:2010wv}.  Jets are recombined using the
  anti-$\kt$ algorithm with $R=0.4$ in the left plot, and $R=0.6$ in the
  right one. 
  Data are shown from top to bottom in order of decreasing rapidity.}
\end{figure}
In fig.~\ref{fig:ATLAS_dydmjj_plots}, we compare the dijet
double-differential cross section, as function of the dijet mass $m_{jj}$ of
the two leading jets, for different bins in $|y|_{\max}$, the maximum
absolute rapidity of the  two jets.  The two leading jets are required to
have
\begin{equation}
\label{eq:ATLAS_mjj_cuts}
{p_{\sss\rm T,1}^{\rm jet}} > 60\ {\rm GeV}, \quad \quad 
{p_{\sss\rm T,2}^{\rm jet}} > 30\ {\rm GeV}, \quad \quad 
|y^{\rm jet}_{1}|,\ |y^{\rm jet}_{2}|<2.8\,.
\end{equation}
In the left plot, results for the anti-$\kt$ algorithm with $R=0.4$ are
shown, while on the right the value $R=0.6$ has been used.
Good agreement is found over the entire dijet mass and rapidity ranges.

\begin{figure}
\begin{centering}
\epsfig{file=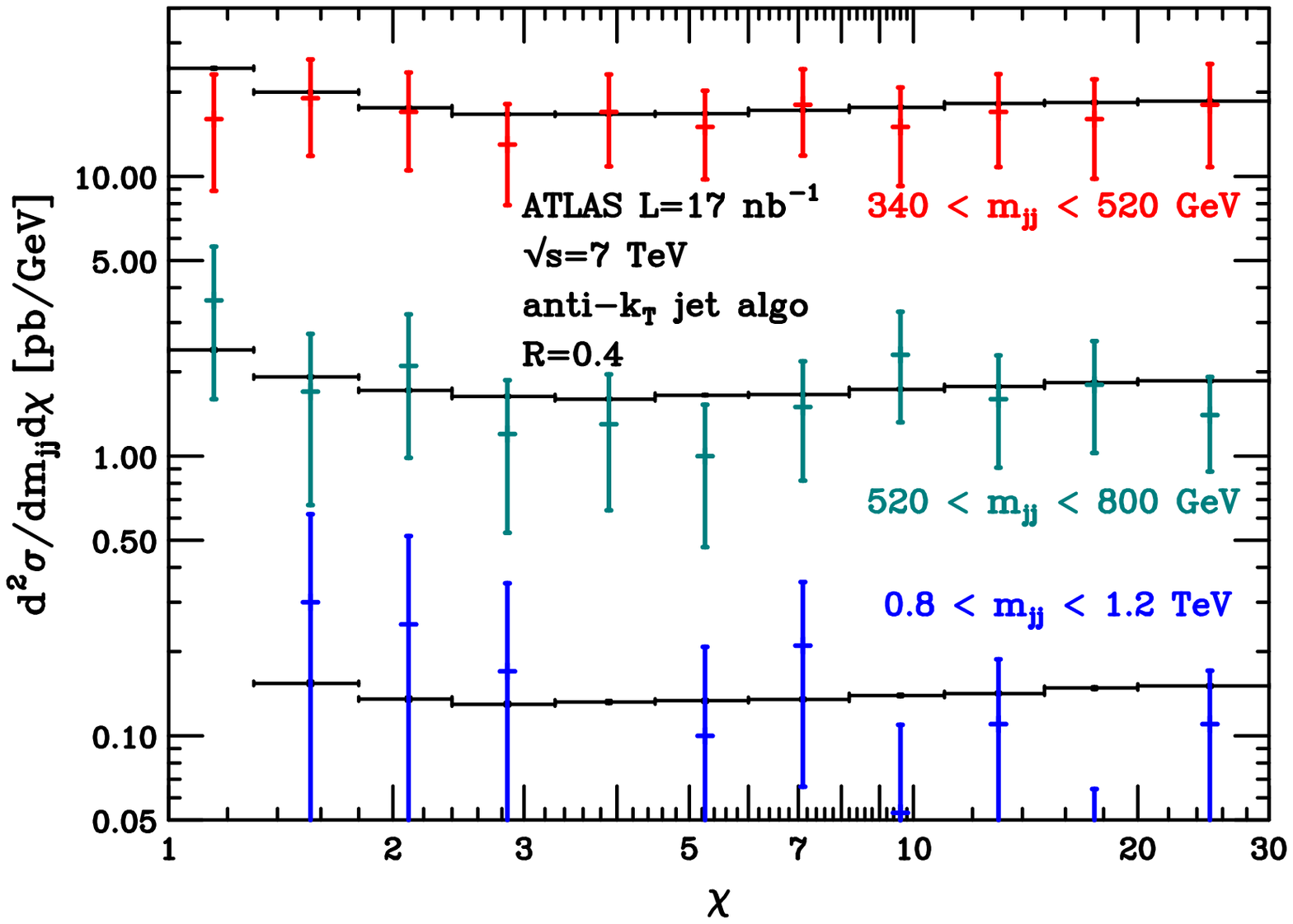,width=0.490\textwidth}
\hfill{}
\epsfig{file=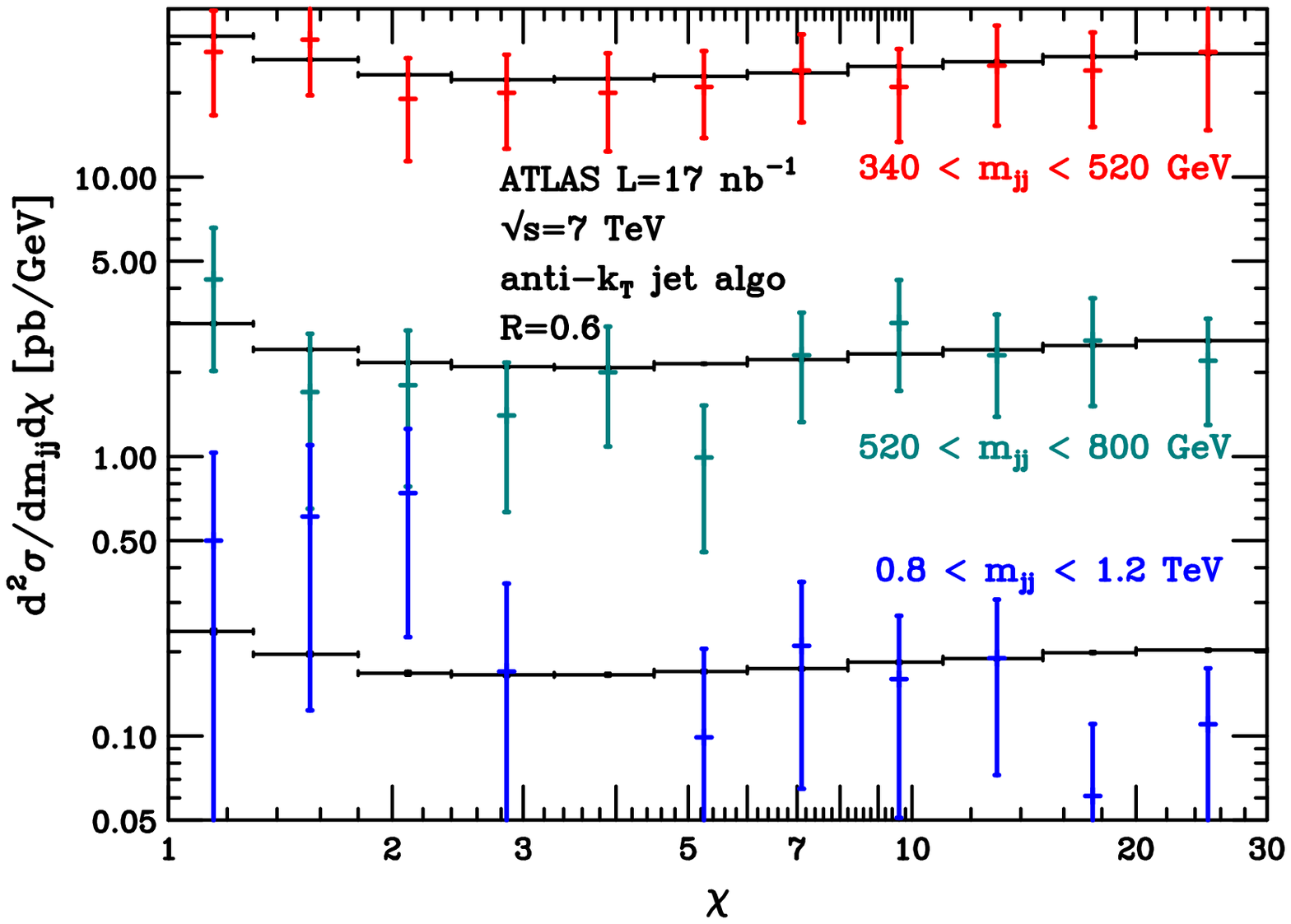,width=0.490\textwidth}
\par\end{centering}
\caption{\label{fig:ATLAS_dchidmjj_plots} Predictions and experimental
  results for the  dijet double-differential cross section as a function
  of the dijet angular variable $\chi$ (see eq.~(\ref{eq:ATLAS_chi_def})),
  for different ranges of the dijet mass $m_{jj}$ of the two leading jets.
  Jets must satisfy the cuts of eqs.~(\ref{eq:ATLAS_mjj_cuts})
  and~(\ref{eq:ATLAS_chi_cut}).  Black horizontal lines are the
  \POWHEG+\PYTHIA{} theoretical results (with errors). Coloured vertical bars
  describe the experimental data from ATLAS (systematic and statistical
  errors added in quadrature)~\cite{Collaboration:2010wv}.  Jets recombined
  using the anti-$\kt$ algorithm with $R=0.4$ in the left plot, and $R=0.6$
  on the right.}
\end{figure}
In fig.~\ref{fig:ATLAS_dchidmjj_plots}, we plot the dijet double-differential
cross section as a function of the dijet angular variable $\chi$, for
different ranges of the dijet mass $m_{jj}$ of the two leading jets. The
$\chi$ angular variable is defined in terms of the rapidities of the two
leading jets as
\begin{equation}
\label{eq:ATLAS_chi_def}
\chi = \exp (|y_1 - y_2 | )\,.
\end{equation}
Jets must pass the cuts of eq.~(\ref{eq:ATLAS_mjj_cuts}) and, in order to
reject events in which both jets are boosted in the forward or backward
directions, they must satisfy
\begin{equation}
\label{eq:ATLAS_chi_cut}
 |y_{\rm boost}|<1.1\,,
\end{equation}
where $ y_{\rm boost} = (y_1 + y_2) /2 $.  In the left-hand plot, we show results
using the anti-$\kt$ algorithm with $R=0.4$, while on the right-hand side we
show the ones for $R=0.6$.  These data are affected by quite large errors. The
\POWHEG{} results are in very good accord with data.

\section{Conclusions}
\label{sec:conclusions}
In this article we have presented a next-to-leading order parton
shower simulation of dijet hadroproduction, based on the \POWHEG{}
formalism. We have assembled our generator with the aid of the
\POWHEGBOX{} toolkit, thereby reducing the task to that of writing
a set of computer subroutines returning the Born, virtual, real
and colour-correlated Born cross sections, for given input momenta
and flavour structure, and to provide the Born-level phase space.
By contrast, the validation and phenomenological studies following
the construction of the simulation have proven subtle, with the
generator output ultimately providing us with a better understanding
of the dijet production process. 

In previous studies with next-to-leading order calculations interfaced to
parton showers (\NLOPS{}) it was found that the \NLOPS{}
predictions for inclusive observables are in accord with those of
conventional fixed-order computations. Marked differences between the
fixed-order and \NLOPS{} predictions have been essentially confined to
exclusive quantities, sensitive to the emission of soft radiation.
Conversely, in performing the same class of comparisons for inclusive
quantities in dijet production, we have found conflicts. Only
for the most inclusive observables, namely, the inclusive jet rapidity
and transverse-momentum distributions, have we found the fixed-order and
\POWHEG{} results in perfect agreement. Other observables, such as the total 
cross section subject to a single cut on the leading jet transverse
energy and, to a larger extent, the total cross section for the production
of two jets with transverse energy above a common threshold, exhibit clear
discrepancies, even for large jet radii.


We have studied these discrepancies in some detail and confirmed
that they are symptomatic of a problem in the fixed-order
computation. In regard to the cross sections with cuts on the two
hardest jets, it was already shown in ref.~\cite{Frixione:1997ks} that the NLO
real emission cross section develops large threshold terms, of the form
$\Delta \log \Delta$, $\Delta$ being the difference of the transverse-energy
cuts on the two leading jets. This large NLO correction thus spoils the
convergence of the perturbative expansion and the corresponding NLO
results display unphysical features.
We have succeeded in reaching an analogous conclusion also in the
case where a single cut is applied to the leading-$\Et$ jet.
We have used our program to assess how events with a common
underlying Born kinematics reside or migrate on either side of the $\Et$ cut
boundary, exposing in both cases
the logarithmic sensitivity of the cross sections to soft radiation.

As noted earlier, the problems of the NLO predictions in the presence of
symmetric $\Et$ cuts have been studied in the past and ways to resum these
large corrections to all orders in perturbation theory have also been
proposed~\cite{Banfi:2003jj}.  In practice, the general consensus has been to
warn against the use of symmetric cuts in experimental and theoretical dijet
studies. Having said that, let us reiterate that, even in what one might
regard as the maximally asymmetric case, in which the cut on the second jet
is set to zero, the NLO prediction is logarithmically sensitive to soft
radiation at the cut, albeit at the level of ten rather than one hundred
percent corrections to the total cross section.

Since Sudakov logarithms are always resummed in the \POWHEG{} approach,
one can expect improved agreement with data for general observables,
even the badly behaved inclusive observables mentioned above.
Furthermore, insidious cases, when these large corrections may arise
without being easily identifiable, are also handled correctly by the
\POWHEG{} generator. For example, in analysing the dijet invariant mass
distribution we have found that a small, symmetric, transverse-momentum
cut on the jets, ineffective for central jets, becomes instead effective
for the highest rapidity bins, introducing a large mismatch between the
NLO and \POWHEG{} results in that region.

The reader may wonder what in particular is special about dijet production
that gives rise to the problems that we have discussed.  In \NLOPS{}
implementations concerning the production of massive objects, the
distributions of the kinematic variables of the massive particle do not
display any pathological behaviour.  It is interesting, however, to try to
look for the effects found in dijet production also in these cases, by
building appropriate observables.  In the case of $p p \to Z/\gamma +X \to
l^+ l^- + X$, for example, we have found that, in the presence of symmetric
cuts of $45$~GeV on the transverse momentum of the leptons, the corresponding
NLO computation exhibits essentially the same ill effects discussed here in
the context of jet production. Thus, there is nothing special about dijet
production in regards to this pathological behavior. NLO predictions for
inclusive quantities in other processes can also exhibit similar weaknesses.

Finally, we have compared the results of
our program to a wide variety of Tevatron and early LHC data. Although
in these studies we have not included a full assessment of the
theoretical uncertainties, we have seen that our predictions are
in very pleasing agreement with the experimental measurements.
We believe that a more thorough study using our program can only
be performed in the framework of experimental collaborations
studying jet physics. In particular, Monte Carlo tuning of
hadronization and underlying event parameters should probably be
performed again using the \POWHEG{} dijet program, in view of the
sensitivity of jet measurements to these features.

The code of our generator can be accessed in the \POWHEGBOX{} svn
repository:\\
\centerline{\url{svn://powhegbox.mib.infn.it/trunk/POWHEG-BOX},}\\ with
username {\tt anonymous} and password {\tt anonymous}.

\acknowledgments 
We wish to thank T.~Sj\"ostrand for very useful exchanges and, in particular,
for clarifying the use of the \PYTHIA{} simulation of multiple interactions
in jet processes. We also wish to thank E.~Feng, P.~Francavilla, C.~Roda and
G.~Salam for useful correspondence.

\appendix

\section{\PYTHIA{} and \HERWIG{} settings}
\label{sec:PY_settings}
The sequence of \PYTHIA{} calls we have used in the calculation of the
results presented in section~\ref{sec:theory_bit} is the following:
\begin{verbatim}
MSTP(81)=0 ! No multiple interactions 
CALL PYINIT('USER','','',0d0)
CALL PYABEG
DO I=1,NUM_EVENTS
   CALL PYEVNT
   CALL PYANAL
ENDDO
CALL PYAEND
END
\end{verbatim}
while in  section~\ref{sec:phenomenology} the sequence was:
\begin{verbatim}
CALL PYTUNE(320)    !   Perugia 0 TUNE
CALL PYINIT('USER','','',0d0)
MSTP(86)=1
CALL PYABEG
DO I=1,NUM_EVENTS
   CALL PYEVNT
   CALL PYANAL
ENDDO
CALL PYAEND
END
\end{verbatim}
\HERWIG{} was run setting the following parameters after the call
to {\tt HWIGIN}:
\begin{verbatim}
PTRMS=2.5D0
PRSOF=0
\end{verbatim}
\providecommand{\href}[2]{#2}\begingroup\raggedright\endgroup

\end{document}